\documentclass[aps,prd,superscriptaddress,floatfix,nofootinbib,notitlepage]{revtex4-1}

\usepackage[utf8]{inputenc}
\usepackage{graphicx}
\usepackage{aas_macros}
\usepackage{bm}
\usepackage{multirow}
\usepackage{amsmath,amssymb,amsfonts}

\usepackage[colorlinks]{hyperref}
\hypersetup{
    colorlinks=true,
    linkcolor=blue,
    filecolor=blue,      
    urlcolor=blue,
    citecolor=magenta,
    pdfpagemode=FullScreen
}

\renewcommand{\vec}[1]{\bm{#1}}
\newcommand{\orcid}[1]{\hspace{1mm}\href{https://orcid.org/#1}{\includegraphics[height=0.3cm,keepaspectratio]{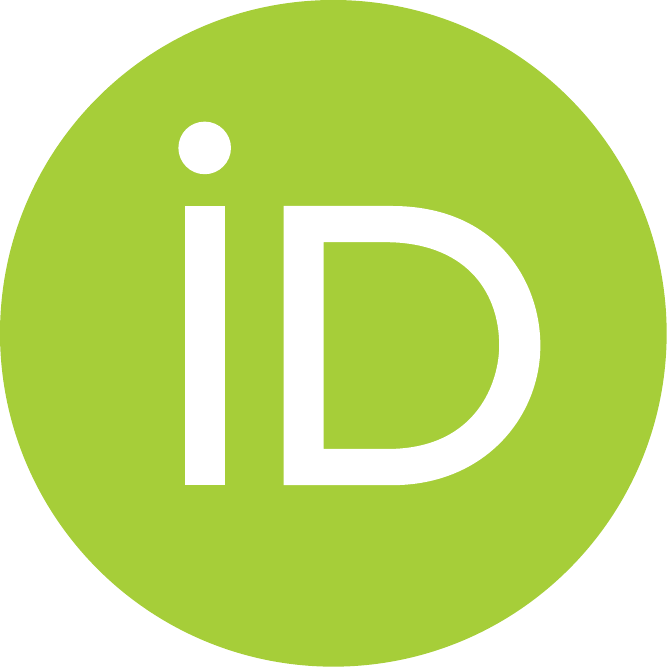}}}

\begin{document}

\title{On the stochastic nature of Galactic cosmic-ray sources}

\author{Carmelo Evoli\orcid{0000-0002-6023-5253}} 
\email{carmelo.evoli@gssi.it}
\affiliation{Gran Sasso Science Institute (GSSI), Viale Francesco Crispi 7, 67100 L'Aquila, Italy}
\affiliation{INFN-Laboratori Nazionali del Gran Sasso (LNGS),  via G. Acitelli 22, 67100 Assergi (AQ), Italy}

\author{Elena Amato\orcid{0000-0002-9881-8112}}
\email{elena.amato@inaf.it}
\affiliation{INAF-Osservatorio Astrofisico di Arcetri, Largo E. Fermi 5, 50125 Firenze, Italy}

\author{Pasquale Blasi\orcid{0000-0003-2480-599X}}
\email{pasquale.blasi@gssi.it}
\affiliation{Gran Sasso Science Institute (GSSI), Viale Francesco Crispi 7, 67100 L'Aquila, Italy}
\affiliation{INFN-Laboratori Nazionali del Gran Sasso (LNGS),  via G. Acitelli 22, 67100 Assergi (AQ), Italy}

\author{Roberto Aloisio\orcid{0000-0003-0161-5923}}
\email{roberto.aloisio@gssi.it}
\affiliation{Gran Sasso Science Institute (GSSI), Viale Francesco Crispi 7, 67100 L'Aquila, Italy}
\affiliation{INFN-Laboratori Nazionali del Gran Sasso (LNGS),  via G. Acitelli 22, 67100 Assergi (AQ), Italy}

\date{\today}

\begin{abstract}
{The precision measurements of the spectra of cosmic ray nuclei and leptons in recent years have revealed the existence of multiple features, such as the spectral break at $\sim 300$ GV rigidity seen by PAMELA and AMS-02 and more recently confirmed by DAMPE and CALET, the softening in the spectra of H and He nuclei at $\sim 10$ TV reported by DAMPE, confirming previous hints by NUCLEON and CREAM, a tiny change of slope at $\sim 40$ GeV in the electron spectrum, revealed by AMS-02, and the large spectral break at $\sim$~TeV reported by indirect (HESS, MAGIC and VERITAS) and direct (DAMPE, CALET) measurements of the total (electrons+positrons) lepton spectrum. 
In all these cases, the possibility has been suggested that these features might reflect the occasional presence of a local cosmic ray source, inducing a noticeable reshaping of the average expected spectra. All these proposals have to face the question of how likely it is for such a source to exist, a question that we address here in a quantitative way. We study the statistical properties of random distribution of sources in space and time, and the effect of the spiral structure of our Galaxy  for both the spectra of light nuclei (p and He) and leptons (electrons and positrons) in different energy regions.}
\end{abstract}

\maketitle

\section{Introduction}
\label{sec:introduction}

Despite the remarkably regular shape of the all-particle cosmic ray (CR) spectrum over many orders of magnitude in energy, recent measurements of the spectra of individual nuclear species and of leptons (electrons and positrons) have shown a wealth of unexpected features. The PAMELA~\cite{HHe.PAMELA} and AMS-02~\cite{H.AMS02,He.AMS02} experiments have independently shown that the spectra of protons and helium nuclei show a hardening at rigidity $\sim 200-300$~GV. Later AMS-02 confirmed that this feature is present also in the spectra of virtually all other nuclei in the cosmic radiation. The feature in the H and He spectra was recently confirmed also by DAMPE~\cite{H.DAMPE,He.DAMPE} and CALET~\cite{H.CALET} (to date, limited to H). 

The spectra of protons and helium as measured by DAMPE in a wider energy region than previous instruments also show a feature at rigidity $\sim 10$~TV \cite{H.DAMPE,He.DAMPE}. CALET confirmed the presence of this steepening in the proton spectrum~\cite{H.CALET}. 

These high precision data provided us with the unprecedented possibility to find a self-consistent picture of the transport of CR nuclei in the Galaxy, in terms of diffusion, advection and energy losses. The main limitation to finalize this goal is in fact in the poor knowledge we have of fragmentation cross sections of nuclei in their journey through the Galaxy, as pointed out in much current literature~\cite{Strong2007arnps,Tomassetti2015prc,Genolini2018prc,Evoli2019prd,Schroer2021prd,Korsmeier2021prd}. 

The same precision is also a powerful tool to unveil the reason for the appearance of the features discussed above. Taken at face value the hardening at few hundred GV rigidity could either result from a corresponding hardening in the spectrum of CRs injected by sources (see for instance~\cite{Ptuskin2013apj,Recchia2018mnras}), or could be due to a change in the way diffusion takes place at energies below and above the feature~\cite{Blasi2012prl,Tomassetti2012apj,Evoli2018prl}. The first class of models, associated with sources, includes the possibility that a local recent source may substantially affect the overall spectrum and cause the hardening~\cite{Kachelriess2015prl}. However, these types of models do not explain in a simple way the behavior of the secondary/primary ratios, which instead arises quite naturally in scenarios where the feature is caused by a change of diffusion regime~\cite{Genolini2017aa}. In the second classes, the reason for the different energy dependence of the diffusion coefficient below and above the break is still matter of debate: it might reflect a transition from self-generated to pre-existing turbulence~\cite{Blasi2012prl} or 
a non-trivial spatial dependence of the diffusion coefficient in the halo~\cite{Tomassetti2012apj}.

As far as the DAMPE feature at $\sim 10$ TV is concerned, there is at present no accepted explanation, although it has been proposed that it might reflect CR reacceleration by a local source~\cite{Malkov2021apj}. The results of the present work will apply also to this model.

Recent observations also changed in a dramatic way our view of CR leptons: HESS reported a measurement of the lepton spectrum up to 5 TeV~\cite{leptons.HESS}, showing substantial steepening at $\sim 900$~GeV (with $\Delta \gamma \sim 1$). CALET~\cite{leptons.CALET} and DAMPE~\cite{leptons.DAMPE} have provided the first direct measurements of the total lepton spectrum up to $\sim 5$~TeV. DAMPE largely confirmed the spectral softening at $\sim 900$~GeV, with the spectral index changing from $\sim 3.1$ to $\sim 3.9$.
The spectra of electrons~\cite{electrons.PAMELA} and positrons~\cite{positrons.PAMELA,positrons.AMS02} have been measured separately by PAMELA and AMS-02, the latter extending the measurement up to $\sim 1$~TeV. The different spectral shape of electrons and positrons, and the rising positron fraction have proven that positrons in the cosmic radiation are not only secondary products of CR interactions in the ISM. The most likely sources of electron-positron pairs are believed to be pulsar magnetospheres (see for instance~\cite{Hooper2009jcap,Grasso2009aph,Delahaye2010aa,Linden2013apj,Gaggero2013prl,DiMauro2014jcap,Yaun2015aph,Cholis2018prd,Orusa2021arxiv}). A careful assessment of the role of pulsars that escaped their parent SNR in producing the observed flux of positrons was recently presented in Ref.~\cite{Evoli2021prd}. 

The recent AMS-02 data on the electron spectrum also show evidence of an unexpected feature at $\gtrsim 40$ GeV. The collaboration suggested that this would reflect a new type of sources contributing to the flux. On the other hand, it was noted~\cite{Evoli2020prl} that the feature appears at about the same energy where a transition from Thomson to Klein-Nishina regime of ICS off UV light occurs. A careful calculation of the electron spectrum, taking into account the appropriate cross section~\cite{Fang2021chphl}, the spiral structure of the Galaxy and an updated model of the UV light, proved that a feature does appear at about the correct energy to describe the AMS-02 data. 

In virtually all cases listed above, there have been speculations concerning the role of a local recent source to the flux of CRs at the Earth~\cite{Yuksel2009prl,Fujita2009prd,Thoudam2012mnras,Yin2013prd,Boudaud2015aa,Kachelriess2015prl,LopezCoto2018prl,Fornieri2020jcap}, thereby raising the question of whether such occurrence may be considered likely enough to be plausible. Clearly the answer to this question depends on whether we are focusing on nuclei or electrons, and on the energy of the particles produced by the alleged local source. In order to address the question in a quantitative manner, here we adopt a Monte Carlo technique for the generation of the spatial localization and the temporal occurrence of the sources and a Green function formalism to describe particle transport. This approach allows us to quantify the role of fluctuations for nuclei and leptons of different energies, and even for different assumptions of the spatial distribution of sources in the Galaxy. In fact, the latter represents a crucial point in our analysis, in that we find that the role of fluctuations is very sensitive to the spiral structure of the Galaxy we live in. 

The effects of source discreteness and stochasticity on the CR flux observed at the Earth have been considered in previous literature. In particular, the author of Ref.~\cite{Lee1979apj} was the first to point out that the variance associated with random source distribution is formally infinite, as a consequence of the unavoidable presence of sources very close to the solar system and occurring at very recent times. The problem was treated by introducing either a minimum time or a minimum distance to the source, so as to make the variance finite~\cite{Blasi2012composition,Ptuskin2006adspr}. However, as noticed by~\cite{Mertsch2011jcap}, this argument is based on the hidden assumption that the flux observed at the Earth is not affected by one of such nearby recent sources. An alternative solution, as suggested by \cite{Bernard2012aa}, was to adopt a smooth density of distant sources and introduce explicitly a number of individual, nearby sources with distances and ages inferred from supernova remnant (SNR) or/and pulsar catalogues and treated as known sources. This approach has been criticized by~\cite{Mertsch2018jcap}, as source catalogues suffer from incompleteness. In particular, the completeness of catalogues drops sharply beyond a certain distance and more importantly beyond a certain age. Moreover it is not clear how to normalize the energetic input of the two contributions (distant sources and local sources), making the ratio between the two an artificial free parameter. Finally, the distinction between local and distant sources is somewhat arbitrary.

From the mathematical point of view, the divergence in the variance of fluctuations is due to a long power law tail in the probability density function (PDF) of the flux from nearby sources~\cite{Lagutin1995jetp}. However, a generalisation of the central limit theorem which applies to PDFs with power law tails and diverging second moments can make the problem tractable again~\cite{Genolini2017aa}. For nuclei, this idea was applied to find that, within homogeneous models for sources and diffusion, it is unlikely that the spectral break at 300 GeV might be due to a prominent nearby source~\cite{Genolini2017prl}. 

The situation is more critical when dealing with relativistic electrons as their galactic ``horizon'' (the region from which they can reach Earth without appreciable energy losses) shrinks significantly at high-energies, making the number of sources effectively contributing to such energies very small and the fluctuations correspondingly large.
 
As discussed in \cite{Blasi2012anisotropy}, the variance of the fluctuations in the flux also plagues the expected anisotropy, which is dominated by the most recent and closest sources. In this case, the situation is even worse in that it is the zeroth order signal (anisotropy amplitude and phase) that change wildly from one realization of the source distribution to another. Moreover, both the amplitude and the phase are deeply affected by the transport in the {\it last mile}, making the interpretation of the observations strongly dependent upon the orientation of the local magnetic field within few parsecs from the Solar system \cite{Ahlers2016prl}.

In this article we discuss in depth the statistical properties characterizing the fluxes of CR nuclei and leptons at the Earth and their fluctuations, for different assumptions on the spatial distribution of sources in the Galaxy. 

We show that the probability of a local source to affect in any appreciable way the flux of protons and helium nuclei at the Earth is negligibly small, especially if the Galactic arm structure is taken into account. Hence, from the statistical point of view, it is very unlikely that the spectral hardening at a few hundred GV or the DAMPE feature at $\gtrsim 10$ TV may be attributed to CR acceleration or reacceleration at a local source. We also comment on the implications of these findings for CR anisotropy. 

For leptons the situation is more complex in that radiative losses emphasize the role of nearby sources on the total lepton flux. For instance, we find that at 100 GeV the probability of a nearby source to contribute a flux about equal to that of all remaining sources is $\sim 0.1\%$, while this probability rises to $\sim 0.3\%$ at 1 TeV and $\sim 7\%$ at 10 TeV. All these numbers refer to the case in which the spiral structure is properly taken into account. We also checked that indeed in about 1 realization out of 10 the calculated flux of leptons at energies $\gtrsim 10$ TeV shows features that reflect the contribution of one local source. If we use a sample of local sources that we are indeed aware of, whose statistics is compatible with the number of sources in our random realizations, we confirm that the CR flux contributed by these astrophysical objects is negligible in terms of H and He, while for leptons it is very likely that Vela may show up in future observations of the lepton flux at energies $\gtrsim 10$ TeV, but may already be providing an appreciable ($\sim 20\%$) contribution to the lepton flux at $\sim 1$ TeV.

The article is organized as follows: in \S~\ref{sec:model} we describe our model for the transport of CR nuclei and leptons and for the generation of the spatial and temporal distribution of the sources. In \S~\ref{sec:results} we discuss the statistical properties describing the flux of nuclei and leptons and the related anisotropy signal. We devote \S~\ref{sec:ksources} to the description of the role of Vela and other known local sources to both the spectra of nuclei and leptons. Our conclusions are illustrated in \S~\ref{sec:conclusions}. 

\section{Model}
\label{sec:model}

Here we illustrate the main aspects of our calculations, based on the following few assumptions: primary nuclei and electrons are assumed to be accelerated at SNRs exploding in the Galaxy at a mean rate inferred from independent measurements~\cite{Dragicevich1999mnras,Rozwadowska2021newa}. Additionally, we assume that roughly 80\% of SN explosions produce a pulsar that injects in the ISM an equal number of electrons and positrons with a relatively hard spectrum up to a few hundred GeV and a cutoff at a fraction of PeV~\cite{Evoli2021prd}.

The individual SN explosions are assumed to be bursts in terms of CR injection into the ISM. As discussed in~\cite{Blasi2012composition}, introducing a time dependent injection of CRs from individual SNRs does not affect the result in an appreciable way, since the duration of the active phase of a SNR (a few tens of thousand years) is much shorter than the typical escape time of CRs from the Galaxy. Pulsars have typically a longer lifetime compared with SNRs, however as shown in~\cite{Evoli2021prd}, the burst approximation is sufficient to describe the lepton spectrum at Earth up to an energy of $\gtrsim 10$~TeV.

The diffusion coefficient and halo size to be adopted in the calculations is as inferred from standard calculations of CR transport, based for instance on weighted slab approaches to CR transport, or from more complex numerical computations (GALPROP~\cite{Moskalenko1998apja,Moskalenko1998apjb}, DRAGON~\cite{Evoli2008jcap}) once the information about secondary to primary ratios and Be/B is used \cite{Evoli2019prd,Evoli2020prd,Weinrich2020aaa,Weinrich2020aab,Korsmeier2021prd}. 

Following an approach previously exploited in~\cite{Evoli2021prd} {(and similar to the methodology adopted in several works, e.g.,~\cite{Atoyan1995prd,Delahaye2010aa,Mertsch2011jcap,Blasi2012composition,Pohl2013apj,Cholis2018prd,Manconi2020prd})}, we generate the position of SNRs in space and time in the Galaxy with a Monte Carlo technique. Each source realization is required to reproduce, on large scales, the given global spatial distribution (see below). For each type II SNR, we also generate a pulsar with an initial period drawn at random from the observed distribution and with a birth kick velocity also following the observed distribution~\cite{FaucherGiguere2006apj}. 

Since the main goal of the work presented here is to address the issue of fluctuations in high-energy CRs, we neglect advection and second-order Fermi acceleration (if present at all), that may only affect CR transport at energies below 10 GeV. 

The transport equation of the CR species $i$ reduces to the simpler form~\cite{TheBible}:
\begin{equation}\label{eq:transport}
\frac{\partial}{\partial t} n_i(t, E, \vec r) - D_i(E) \nabla^2 n_i(t, E, \vec r) 
= {\cal Q}_i(t, E, \vec r) - {\cal B}_i(t, E, \vec r),
\end{equation}
where, $n_i(\vec r,t,E) = dN/dVdE$ is the isotropic part of the differential CR spectral density.

In Eq.~\ref{eq:transport}, ${\cal Q}(\vec r,t,E) \equiv dN/dE dV dt$ is the rate of CR injection per unit volume, time and energy, $D(E)$ is the energy-dependent isotropic diffusion coefficient (assumed to be uniform in the diffusion volume), and $\mathcal{B}$ is the rate of energy losses which is different for nuclei and leptons (see~\ref{sec:nucleimodel} and \ref{sec:leptonsmodel}).

As usual, Eq.~\ref{eq:transport} is solved with the so-called free-escape boundary condition at $z=|H|$, $H$ being the height of the halo, namely $n_i(z = \pm H) = 0$. We neglect particle escape in the radial direction, a good approximation so far as the distance of the Sun to the radial boundary is larger than the size $H$ of the halo.
For the diffusion coefficient we adopt the same functional form and parameters' values as in~\cite{Evoli2021prd}, assuming for the halo size $H$ the value of $5$~kpc as deduced from the recent measurements of Be/B and Be/C ratios by AMS-02~\cite{Evoli2020prd,Weinrich2020aaa}.

The source term ${\mathcal Q}$ is expected to be a sum of discrete injection episodes in space-time (the spatial and timescale of likely accelerators being assumed much shorter than propagation length and time), whose actual positions and epochs are unknown.
Therefore, for a given ensemble of $N$ sources with positions $\vec r_s$ and ages $t_s$, ${\mathcal Q}$ can be conveniently written as:
\begin{equation}
{\mathcal Q}(t, E, \vec r) = \sum_j Q_j(E) \delta^3(\vec r_{s}^{(j)} - \vec{r}) \delta(t_{s}^{(j)} - t)
\end{equation}
where each source $j$ that belongs to the source population contributes with the injection spectrum $Q_j(E)$.

\begin{figure*}[t]
\centering
\hspace{\stretch{1}}
\includegraphics[width=0.48\textwidth]{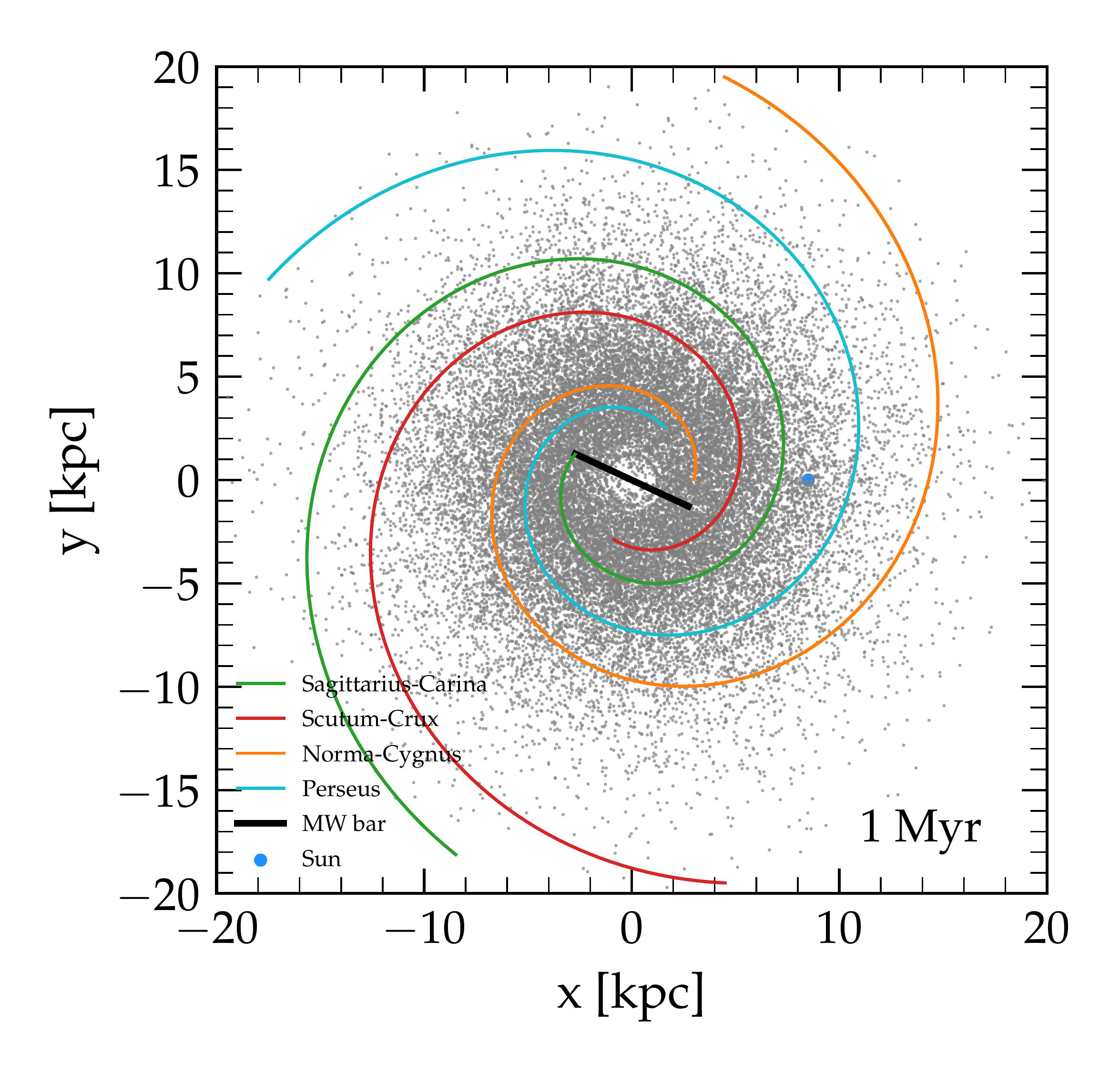}
\hspace{\stretch{1}}
\includegraphics[width=0.48\textwidth]{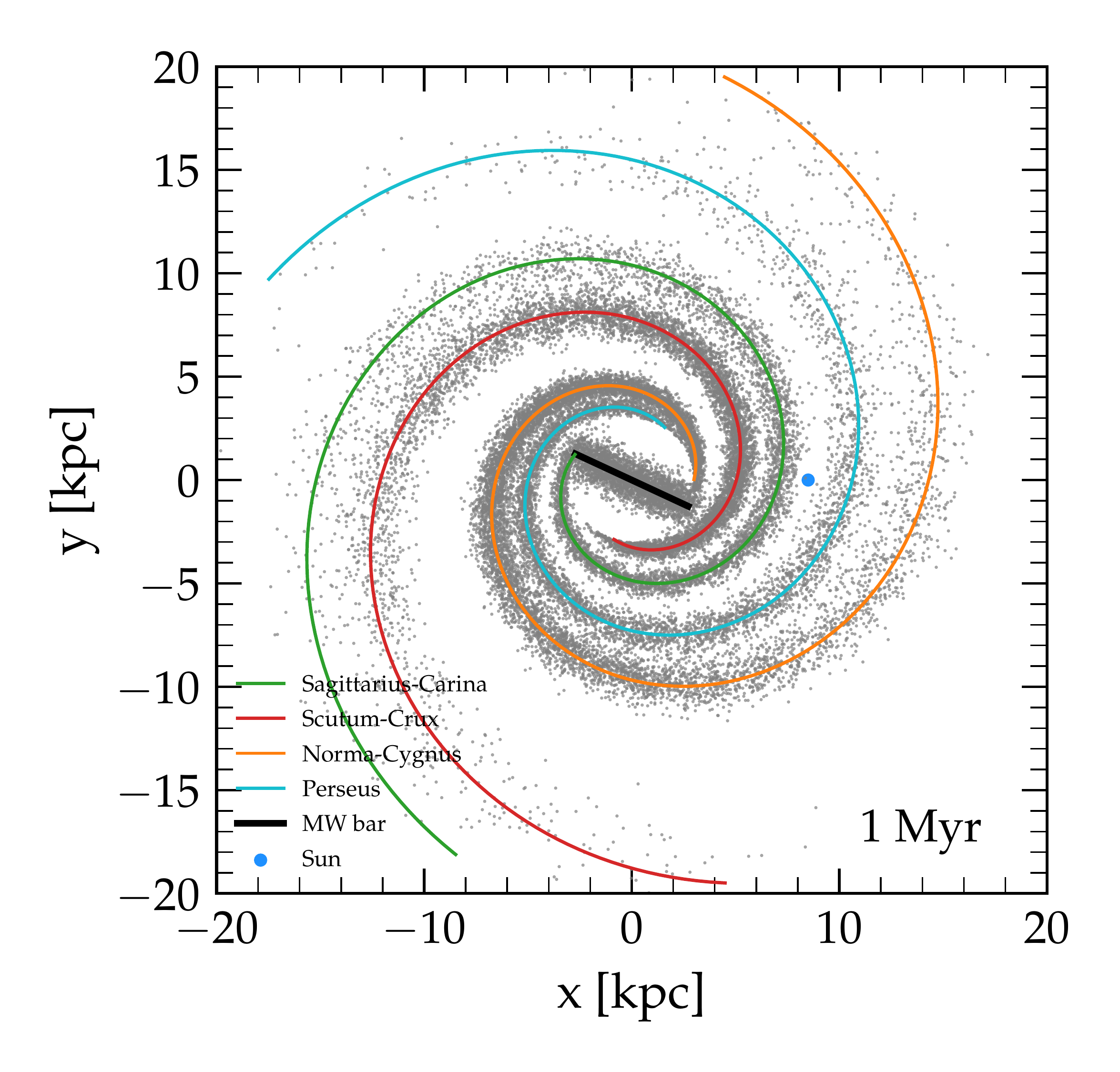}
\hspace{\stretch{1}}
\caption{The plots show the position of the explosions in the Galactic plane in a given realization and for a simulation time of 1 Myr. In the left(right) panel we show the case without(with) including the spiral pattern. In both figures we overplot the loci of the 4-arms of the Milky Way spiral structure and the position of the Sun is represented by the thick (blue) circle.}
\label{fig:galaxies}
\end{figure*}

The flux at the Earth of primary CRs of a given type at a given energy as provided by all sources (assuming bursted injection) can be written as:
\begin{equation}
\Phi_\odot(E) = \frac{c}{4\pi} \sum_j n_j(t_\odot, E, r_\odot) = \frac{c}{4\pi} \sum_j {\mathcal G}_j(t_\odot,E,\vec r_\odot \leftarrow t_{s}, E_s, \vec r_{s}) Q_j(E_s),
\end{equation}
where we have assumed that all CRs of interest are ultrarelativistic. Here ${\mathcal G}(t,E,\vec r \leftarrow t_0, E_0, \vec r_0)$ is the Green function of the transport equation, namely the solution of Eq.~\ref{eq:transport} for ${\mathcal Q} = \delta^3(\vec r - \vec r_0) \delta (t - t_0) \delta (E - E_0)$ and the same boundary conditions as for Eq.~\ref{eq:transport}.

Hereafter we adopt the notation that all quantities labelled with $\odot$ are considered as evaluated at the Earth, after summing the contributions of all relevant sources.

With this formalism, it is also easy to compute the anisotropy from discrete sources as the ratio of the diffusive particle current ($\vec J = D \vec \nabla n$) and the modulus of the isotropic current~\cite{Ptuskin2006adspr,Blasi2012anisotropy}:
\begin{equation}
\vec \delta(E) 
= \frac{\vec J_\odot(E)}{\frac 1 3 c n_\odot(E)} 
= \frac{3}{c} \frac{D(E)}{n_\odot(E)} \sum_j \vec \nabla n_j(t_\odot, E, r_\odot) .
\end{equation}

The CR flux at the Earth and its cosmic variance (fluctuations due to different realizations of the distribution of sources) can be calculated by randomly generating the location of the sources, following the probability distribution function ${\mathcal P}(\vec r)$ of finding a source between distance $\vec r$ and $\vec r + d\vec r$ from the Galactic center. The time of occurrence of the source is also drawn at random from a flat distribution, given the average Galactic SN rate $\mathcal R = 0.03~\text{event/year}$.

In this work we adopt two different models for the source distribution $\mathcal{P}$. The first one (hereafter \emph{Jelly model}) simply accounts for the average radial distribution of sources in the Galaxy, while the second model (hereafter \emph{Spiral model}) also takes into account the 4-arms spiral structure.

In both cases, the sources are assumed to be distributed in terms of Galactocentric distance $r$ as described in Ref.~\cite{Lorimer2006mnras}:
\begin{equation}
f(r) \propto \left( \frac{r}{r_\odot} \right)^a \exp \left[ -b \frac{    r - r_\odot}{r_\odot} \right],
\end{equation}
where $a = 1.9$, $b = 5.0$ and $r_\odot = 8.5$~kpc.
In the $z$-direction, the SNR distribution is assumed to be a Gaussian with scale-height $100$~pc. 

In the Jelly model, the positions of the sources are chosen by drawing at random values of $r$ and $z$ from the distributions above. 
The $x$ and $y$ coordinates are computed by generating a random angle $0 \le \phi < 2\pi$ and using the given value of $r$.
In the Spiral model, the procedure we adopt is similar to that already used in Ref.~\cite{Evoli2021prd}.

A pictorial illustration of the two scenarios is shown in Fig.~\ref{fig:galaxies} where we show the distribution of the SNRs generated in 1 Myr in the Jelly model (left panel) and in the Spiral model (right panel). 
The position of the Sun is illustrated by the thick (blue) symbol. We notice that in the Spiral model the Sun is correctly placed in the inter-arm region between Sagittarius-Carina and Perseus arms. In this model, the distance to the closest spiral arm, in the direction of the Galactic center, is about 1.5~kpc, well compatible with recent determinations, as e.g.,~\cite{NoguerasLara2021aa}.

In this approach, the CR flux behaves as a stochastic variable whose statistical properties have been the subject of several investigations~\cite{Lee1979apj,Lagutin1995jetp,Mertsch2011jcap,Bernard2012aa,Blasi2012composition,Genolini2017aa}.
It has been shown, for instance, that the probability density function (PDF) of the flux is asymmetric having a long power law tail for large fluxes which reflects the unavoidable occasional contribution of local sources. Under these conditions, the median value of the flux is not coincident with the mean value. Notably, the latter coincides with the flux predicted in standard calculations of CR transport for the smooth source density distribution. In the following, we take the median rather than the mean flux as representative of the typical flux to be expected, so that the efficiency in the injection of H and He are calculated by imposing that the median fits the data. This is not a major concern for nuclei, but as we discuss below, may become important for leptons.

Because of the long tail discussed above, these distribution functions do not have a well-defined second moment (the variance is formally divergent) and the problem of evaluating the fluctuations around the median must rely on a generalisation of the central limit theorem which applies to heavy-tailed PDFs as suggested, e.g., in~\cite{Genolini2017aa}. 
A quantitative assessment of the role of fluctuations can be obtained by defining the uncertainty intervals of the flux using the percentiles, in particular the 95\% uncertainty range of the flux  will correspond to the interval between 2.5\% and 97.5\% percentiles~(see also~\cite{Phan2021prl}).

\begin{figure*}[t]
\centering
\hspace{\stretch{1}}
\includegraphics[width=0.45\textwidth]{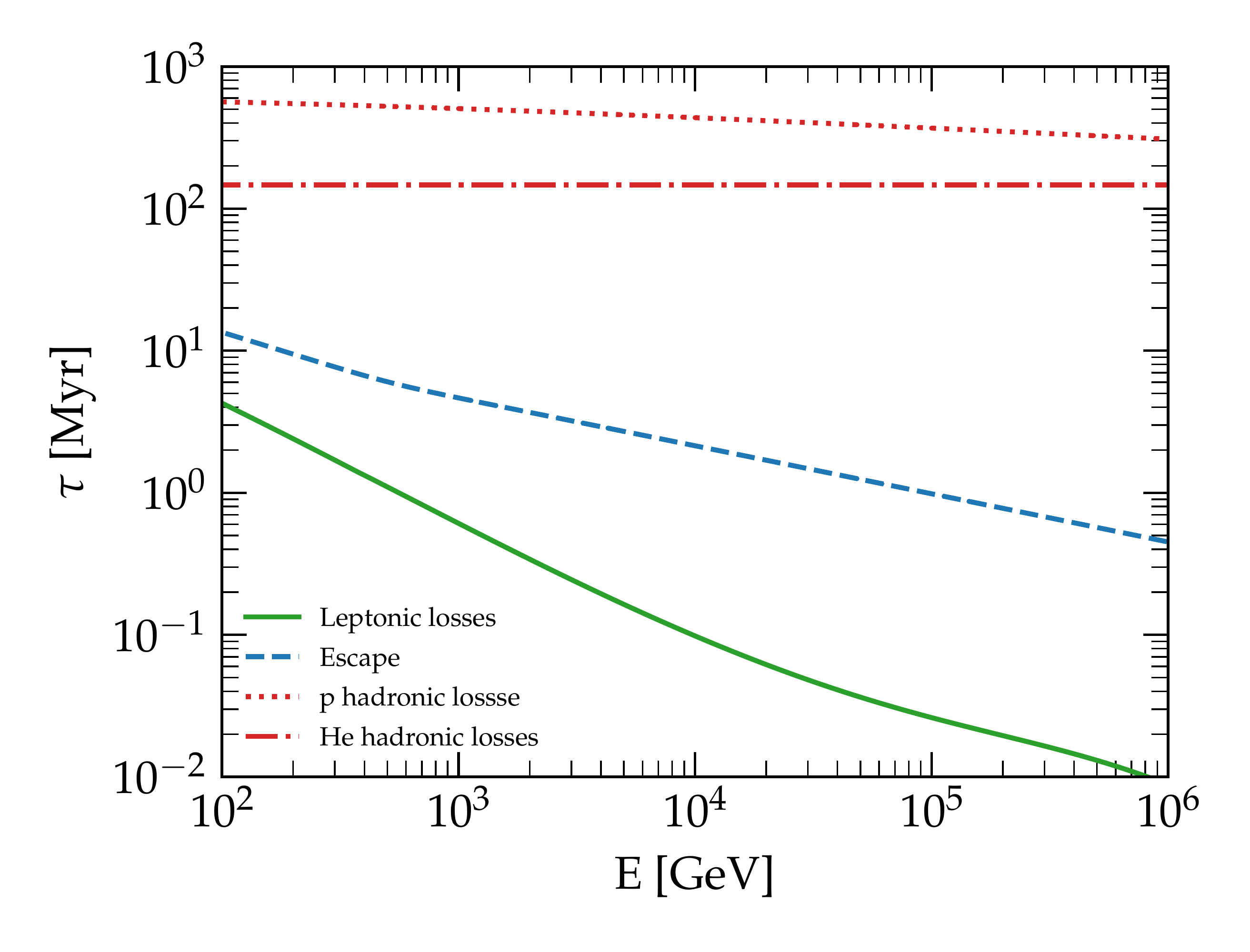}
\hspace{\stretch{1}}
\caption{Energy loss timescale for CR electrons (solid green line), protons (dotted red line) and Helium nuclei (dot-dashed red line) as a function of particle energy. The dashed blue line represents the escape timescale from the Galaxy due to diffusion assuming $H = 5$~kpc~\cite{Evoli2020prd}.}
\label{fig:losses}
\end{figure*}

\subsection{Nuclei}
\label{sec:nucleimodel}

At the energies of interest here, the transport of nuclei is dominated by diffusion and catastrophic energy losses caused by fragmentation on gas in the ISM~\cite{Moskalenko1998apja,Evoli2018jcap}:
\begin{equation}
{\cal B}_{\rm nuclei, i}(t, E, \vec r) = 
\frac{n_i(t, E, \vec r)}{\tau_{\rm sp} (E)}, 
\end{equation}
where the spallation rate is defined in terms of the mean gas density in the diffusion region $\bar{n}$ and of the cross section for spallation $\sigma_{\rm sp}$ as $\tau_{\rm sp}^{-1} = \bar{n}_{\rm gas} c \sigma_{\rm sp}$\,. The introduction of the mean density is not necessary in general, but since we use the Green function formalism here it is a useful simplification, that is accurate as long as spallation is slow, namely the spallation time scale is longer than the diffusion time. 
For H and He nuclei, that we focus on here, this is certainly the case. In this case, the mean density reads $\bar{n}_{\rm gas} \sim n_{\rm disc} (h/H)$, where the gas density in the disc of the Galaxy (with half-thickness $h = 100$~pc) is $n_{\rm disc} = 2$~cm$^{-3}$ corresponding to a mean surface density of $\sim 2.3$~mgram/cm$^2$~\cite{Ferriere2001review}.

We model the cross section for inelastic collisions of CR protons with interstellar gas by using the parametrization provided by~\cite{Kafexhiu2014prd}. For Helium, we adopt the constant value of $\sigma_{\rm sp,He} = 110$~mbarn since the  available measurements of this quantity do not manifest any deviation from a constant value above few GeV's~\cite{Coste2012aa}.  
The Green function for the transport of CR nuclear species $k$ reads~\cite{Blasi2012composition}:
\begin{multline}
{\mathcal G}^{\rm nuclei}(t,E,\vec r \leftarrow t_s, \vec r_s) =
\frac{1}{\left[ \pi \lambda_k^2(E) \right]^{3/2}} 
\exp\left[-\frac{\Delta t}{\tau_{\rm sp,k}(E)}\right]  \\
\exp\left[-\frac{(x-x_s)^2 + (y-y_s)^2}{\lambda_k^2(E)} \right]    
\sum_{n=-\infty}^{+\infty} (-1)^n \exp\left[ - \frac{(z-z_n')^2}{\lambda_k^2(E)}\right],\label{eq:greennuclei}
\end{multline}
where we accounted for the fact that the CR energy is conserved during propagation (namely, $E=E_s$) and we have defined $\Delta t = t - t_s$ and $\lambda_k = \sqrt{4 D_k(E) \Delta t}$.

The infinite summation in the Green function written in Eq.~\ref{eq:greennuclei} is introduced to satisfy the correct boundary condition at $z = \pm H$ through the image charge method and having defined $z_n' = (-1)^n z_s + 2nH$ as the $z$-coordinates of the image sources.

Concerning the injection, we assume that all sources are identical in the sense that the energy of each SN explosion is fixed and equal to $10^{51}$ erg and each of them contributes with the same spectrum for a given CR charge $Z$.
In general, the injection term reads:
\begin{equation}
Q_k(E) = q_{0,k} \left(\frac{E}{10 \, \text{GeV}}\right)^{-\gamma_k} \exp\left(-\frac{E}{Z E_{\rm max,p}}\right),
\end{equation}
where $k = \{\text{H, He}\}$, $q_{0,k}$ is a normalization, $\gamma_k$ is the slope of the injection spectrum (different for protons and Helium) and $E_{\rm max,p}$ is the maximum energy of protons.
All the parameters are chosen in such a way as to reproduce local observations as we discuss in Sec.~\ref{sec:results}. 

\subsection{Leptons}
\label{sec:leptonsmodel}

The energy losses of CR leptons are dominated by inverse Compton scattering (ICS) on the photons of the interstellar radiation fields (ISRFs) and cosmic microwave background (CMB), and synchrotron emission in the Galactic magnetic field.
For the magnetic field we assume U$_{\rm B} = 0.025$~eV~cm$^{-3}$ (corresponding to a magnetic field B$_{0} = 1$~$\mu$G), as representative of the average energy density in the halo~\cite{Ferriere2001review}. 
The ISRF is made of the CMB, whose energy density is U$_{\rm CMB} = 0.25$~eV~cm$^{-3}$~\cite{Planck2020aa} everywhere, and of a second component which is the result of emission by stars and re-processing of the starlight by dust~\cite{Porter2005,Popescu2017mnras}.
In order to model this component we adopt a fit provided by~\cite{Delahaye2010aa} which entails the sum of 5 gray-bodies, identified with an infrared (IR), an optical and 3 ultraviolet components (the temperatures and normalizations of each component are given in Table~2 of \cite{Delahaye2010aa}).

The rate of (continuous) energy losses of these two processes can be then written as:
\begin{equation}\label{eq:bE}
{\mathcal B}_{\rm leptons,i}(t, E, \vec r) = -\frac{\partial}{\partial E} \left[b(E) n_i(t, E, \vec r) \right]
\end{equation}
where
\begin{equation}
b(E) = \frac{4}{3} c \sigma_T \left[ \sum_\mu f_{\rm KN}(E, T_\mu) U_{\gamma,\mu} + U_B \right] \left(\frac{E}{m_e c^2}\right)^2 
\end{equation}
being $m_e$ the electron mass, $\sigma_T$ the Thomson scattering cross section and $U_{\gamma,\mu}$ and $U_B$ the energy densities in the form of photons of type $\mu=\{\text{ISRF},~\text{CMB}\}$ and of magnetic field respectively. 
The function $f_{\rm KN}$ effectively describes the modification to the Thomson cross section due to the Klein-Nishina corrections and we adopt the parametrization recently proposed by~\cite{Fang2021chphl}.

A comprehensive comparison of the timescales introduced so far is given in Fig.~\ref{fig:losses} where we show the time scale for energy losses of leptons and nuclei, $\tau_{\rm loss, lep}(E) = E / b(E)$ and $\tau_{\rm sp}(E)$ respectively, as a function of particle energy, compared with the timescale for diffusive escape from the Galaxy, $\tau_{\rm esc} = H^2 / 2D(E)$, assuming $H = 5$~kpc.
In this plot, as well as in the rest of the paper, we focus on energies larger than $100$~GeV, which are the ones where the role of fluctuations may at least in principle be most visible. At such energies, energy losses of leptons are quite important, while for nuclei they play a negligible role. 

The Green function of Eq.~\ref{eq:transport} for leptons can be written as follows:
\begin{equation}
{\mathcal G}^{\rm leptons} (t, E, \vec r \leftarrow t_s, E_s, \vec r_s) = \delta(\Delta t - \Delta \tau) {\mathcal G}_{\vec r} (E, \vec r \leftarrow E_s, \vec r_s),
\end{equation}
where the $\delta$-function shows that a particle injected with energy $E_s$ is observed after a time $\Delta t$ with energy $E < E_s$ only if the elapsed time corresponds to the time during which the energy of a particle decreases from $E_s$ to $E$ because of losses. 
This loss time is defined as:
\begin{equation}
\Delta \tau(E, E_s) \equiv \int_E^{E_s} \frac{dE'}{b(E')} \,.
\end{equation}

The spatial part of the Green function reads:
\begin{equation}
{\mathcal G}_{\vec r} = \frac{1}{b(E)} 
\frac{1}{\pi \lambda_e^2} \exp\left[ -\frac{(x-x_{s})^{2}+(y-y_{s})^{2}}{\lambda_e^2}\right] 
\frac{1}{\sqrt{\pi}\lambda_e} \sum_{n=-\infty}^{+\infty} (-1)^{n} \exp \left[ -\frac{(z-z'_{n})^{2}}{\lambda_e^2} \right],
\end{equation}
where $\lambda_e$ is the distance covered by a lepton under the effect of losses and diffusion, and it is defined as:
\begin{equation}
\lambda_e^2(E, E_s) \equiv 4 {\int_E^{E_s} dE' \, \frac{D(E')}{|b(E')|}}.
\end{equation}

In the following we assume that leptons comprise two components, namely primary electrons accelerated in SNRs and the electron-positron pairs produced in pulsars.

The injection spectrum of electrons from SNRs is expected to be a power law with a cutoff at an energy where losses balance acceleration. As discussed in Refs.~\cite{Zirakashvili2007aa,Blasi2010mnras} the shape of the cutoff depends on the diffusion coefficient in the acceleration region. 
For Bohm diffusion, most reasonable in the case of strong magnetic field amplification, the cutoff shape can be calculated analytically \cite{Zirakashvili2007aa,Blasi2010mnras} and the spectrum injected into the ISM can be written as:
\begin{equation}
Q_{\rm SNR}(E) = q_{0,e} \left(\frac{E}{\rm GeV}\right)^{-\gamma_e} \exp\left[-\left(\frac{E}{E_c}\right)^2\right],
\end{equation}
where the normalization $q_{0,e}$ and the injection slope of this primary component, $\gamma_e \gtrsim 2$, are chosen in such a way as to reproduce observations of the total electron spectrum at the Earth.

The cutoff energy $E_c$ is set by equating acceleration and losses timescales in the acceleration region.
For Bohm diffusion ($D_{\textrm B}$) and synchrotron losses in a magnetic field of $\sim$0.1~mG, typical conditions for the environment downstream of a SNR shock~\cite{Vink2012aarv}, the electron spectrum develops a cutoff at $E_{\rm c} \simeq 36$~TeV~\cite{Evoli2021prd}. 

Finally, we assume that electron-positron pairs are also injected into the ISM by pulsars, after these leave the parent SNR due to their birth kick velocity (see \cite{Evoli2021prd} for a detailed discussion). The source term for positrons and electrons from an individual pulsar is assumed here to be burstlike at the time when the pulsar escapes its parent remnant, $t_{\rm BS}$:
\begin{equation}\label{eq:source-pwn}
{\mathcal Q}_j(t, E, \vec r) =  Q_{\text{PWN}}(E, t_{\rm BS}) \delta^3(\vec r_{s}^{(j)} - \vec{r}) \delta (t_{s}^{(j)} + t_{\rm BS} - t)
\end{equation}
where, following~\cite{AmatoBlasi2018}, $t_{\rm BS} \simeq 56$~kyr and the spectrum $Q_{\rm PWN}(E)$ is modeled as a broken power law, with slope $\gamma_{\rm L}$ below the break $E_b$ and slope $\gamma_{\rm H}$ above:
\begin{equation}
Q_{\rm PWN}(E, t) = q_{0,e^\pm} (t) 
\exp \left[-\frac{E}{E_{\rm drop}(t)}\right] 
\times
\begin{cases}
(E / E_{\rm b})^{-\gamma_{\rm L}} & E < E_{\rm b} \\
(E / E_{\rm b})^{-\gamma_{\rm H}} & E \ge E_{\rm b}.
\label{eq:injpulsar}
\end{cases} 
\end{equation}

In most cases, observations of electromagnetic radiation from individual pulsars require $\gamma_{\rm L} \sim 1-1.9$ and $\gamma_{\rm H}\sim 2.5$. This functional form provides a good description of the emission from PWNe both within the parent SNR and in the bow shock phase (see \cite{Bykov2017ssrv} and references therein). 
The cutoff position $E_{\rm drop}$ reflects the potential drop of the pulsar at the time of leaving the parent remnant~\cite{Arons2003apj,Kotera2015jcap}.

\section{Results}
\label{sec:results}

One realization of the source distribution in the Galaxy is generated by extracting at random in time and space from the appropriate distributions (see discussion above) a number of sources sufficiently large to reach equilibrium in both the spectra of nuclei and leptons. For nuclei with energy $\sim 100$ GeV this requires that sources are produced over a time span of $>10$ Myr. Typically the statistical analyses discussed below are obtained by generating $\sim 10^5$ realizations for each model of spatial distribution of the sources (Jelly versus Spiral model). Below we discuss separately our results for nuclei and leptons. 

\subsection{Nuclei}

\begin{figure*}[t]
\centering
\hspace{\stretch{1}}
\includegraphics[width=0.9\textwidth]{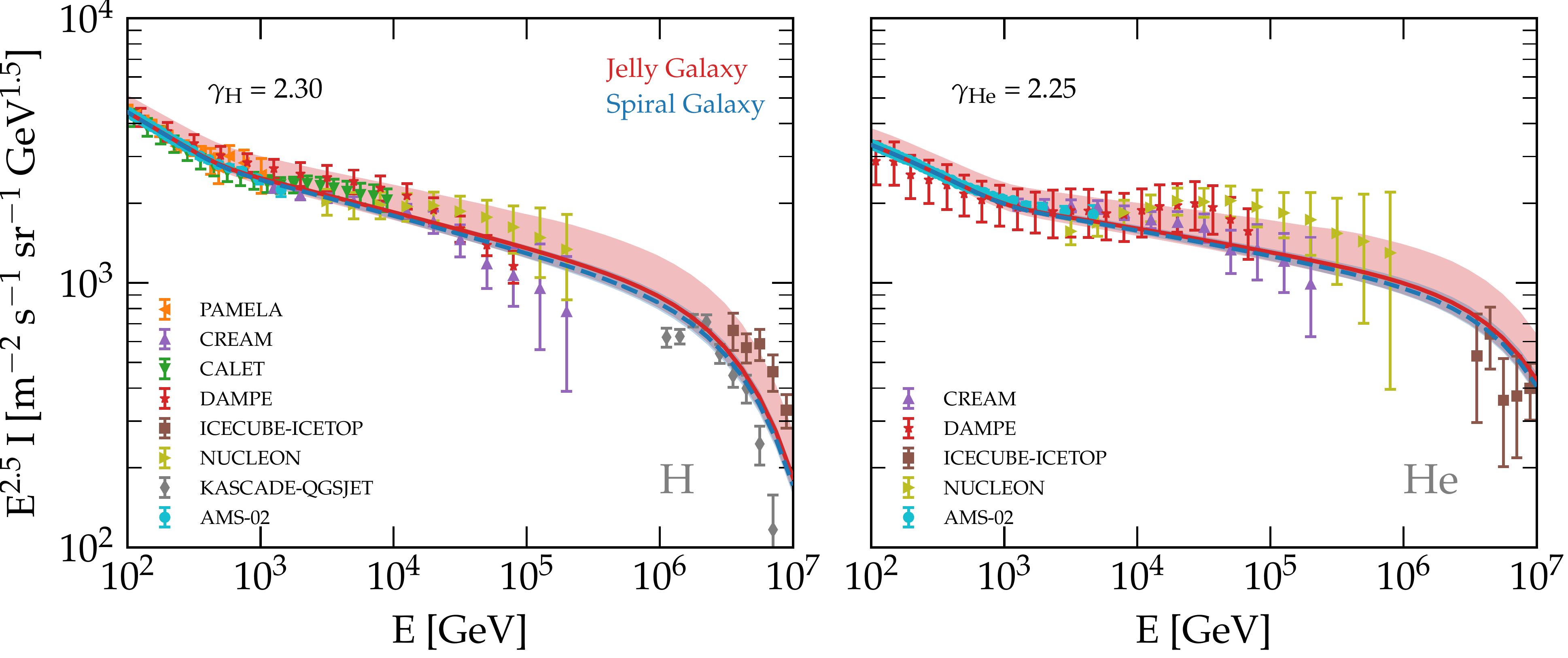}
\hspace{\stretch{1}}
\caption{The intensity of protons (left panel) and Helium nuclei (right panel) resulting from the sum of all SNRs throughout the Galaxy. Lines represent the median flux computed over $10^5$ realizations of the Milky Way in the {\it Jelly} configuration (solid red line) or {\it Spiral} configuration (dashed blue line). The shadowed areas represent the 95\% uncertainty range computed as the 2.5\% and 97.5\% percentiles of the intensity PDF. Data are show for the experiments: AMS-02~\cite{H.AMS02,He.AMS02}, CALET~\cite{H.CALET}, CREAM~\cite{HHe.CREAM}, DAMPE~\cite{H.DAMPE,He.DAMPE}, ICECUBE-ICETOP~\cite{HHe.ICECUBE-ICETOP}, KASCADE~\cite{H.KASCADE}, NUCLEON~\cite{HHe.NUCLEON} and PAMELA~\cite{HHe.PAMELA}.}
\label{fig:nuclei}
\end{figure*}

In Fig.~\ref{fig:nuclei} we compare our model predictions with data provided by recent experiments in the energy region $\gtrsim 100$ GeV for H (left panel) and He (right panel).~All the results that we obtain for H and He can be straightforwardly applied to heavier nuclei as long as diffusive escape is faster than fragmentation. This latter constraint derives from having used the Green functions which does not apply when spallation losses becomes dominant. The approximation is appropriate for protons and He nuclei in the whole energy range considered here. For heavier nuclei the minimum energy for which spallation losses can be neglected becomes gradually higher.
The normalization of the injection rates of H and He are calculated by imposing that the median flux fits the data. The fit is clearly dominated by the high precision AMS-02 data. We confirm findings of previous analyses suggesting that the source spectra of the two elements are required to be slightly different ($\gamma_{\text{H}}=2.3$ and $\gamma_{\text{He}}=2.25$). Notice that we introduced a cutoff in rigidity at $R=7$ PeV, although this does not affect our main conclusions at lower energies.
Each panel in Fig.~\ref{fig:nuclei} shows the median flux for the two models of spatial distribution of the sources, the Jelly (red solid line) and Spiral (blue dashed curve) model. 
The shadowed areas represent the 95\% uncertainty range computed as the 2.5\% and 97.5\% percentiles of the PDF of the flux. In particular, the red area refers to the Jelly model and the blue area to the Spiral model. 

One can notice that the fluctuations are visibly less pronounced in the Spiral model, as one can easily expect since the closest sources are more concentrated around the centroid of the nearest arms, hence the probability to find local recent sources around the position of the Earth is lower.

\begin{figure*}[t]
\centering
\hspace{\stretch{1}}
\includegraphics[width=0.95\textwidth]{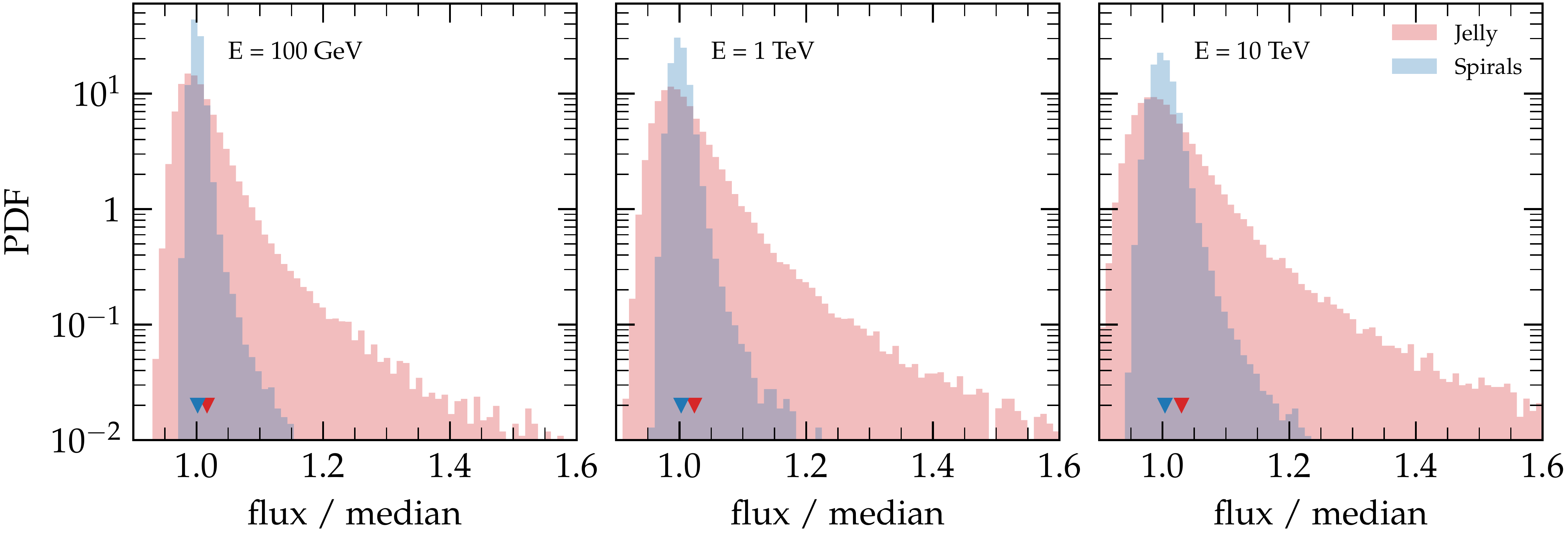}
\hspace{\stretch{1}}
\caption{The PDF of the CR flux normalized to the median value computed for three representative energies ($E = 100$~GeV, $E = 1$~TeV, $E = 10$~TeV). The Milky Way configuration dubbed {\it Jelly}({\it Spiral}) is shown as a red(blue) histogram. The triangles indicate the position of the mean flux (normalized to the median).}
\label{fig:protonspdfs}
\end{figure*}

The fact that these fluctuations are associated to the high flux tail in the PDF is well illustrated in Fig.~\ref{fig:protonspdfs}, where we show the PDF of the fluxes at three energies, 100 GeV, 1 TeV and 10 TeV, for the Jelly (red shadow) and Spiral (blue shadow) model. The tail is much more pronounced in the Jelly model, which translates into a higher probability that an individual source can contribute a sizeable fraction of the flux observed at a given energy. 
The PDFs shown in Fig.~\ref{fig:protonspdfs} illustrate in a clear way how it is much more likely to have a flux much higher than the median rather than a much lower flux, again reflecting the occasional presence of a local source. 

\begin{figure*}[t]
\centering
\hspace{\stretch{1}}
\includegraphics[width=0.95\textwidth]{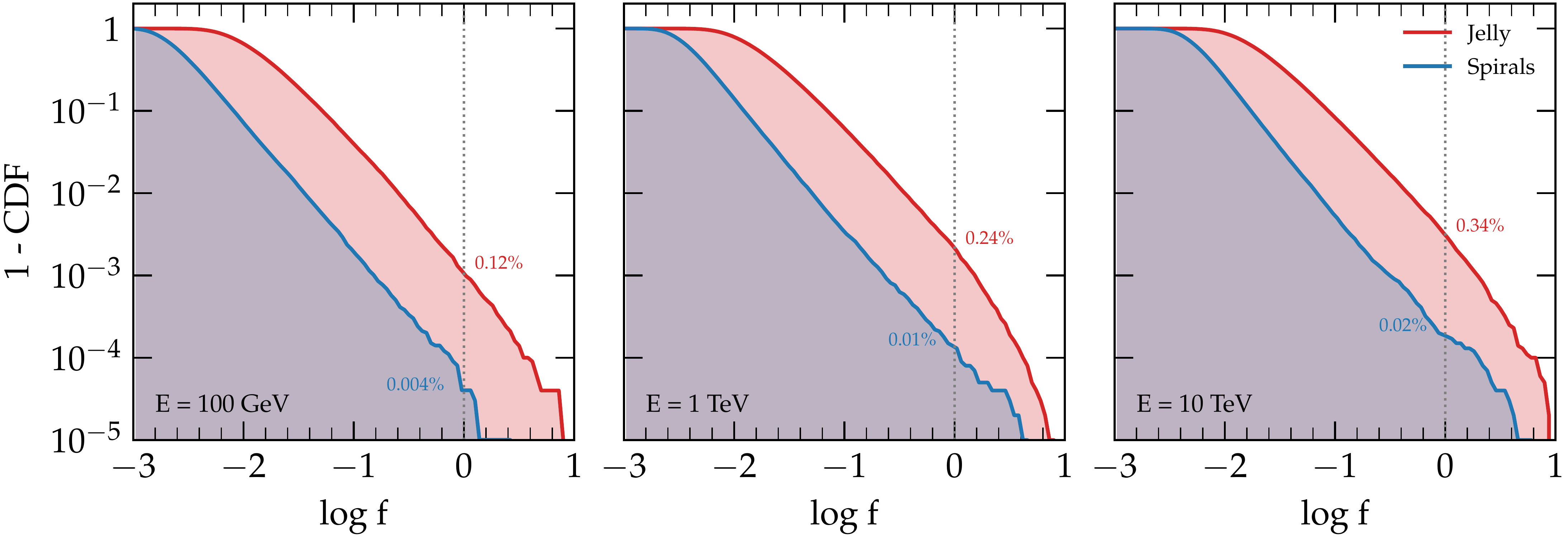}
\hspace{\stretch{1}}
\caption{The complement to the CDF of the quantity $f$ (defined in the text) for each realization is shown for the {\it Jelly} galaxies (in red) and \emph{Spirals} (in blue). A dotted vertical line marks the value $f=1$ which is the case in which a single source contributes to the local flux as much as all the others added together. The percentages reported in the plots indicate the corresponding probability. The three panels refer to three different energies as in Fig.~\ref{fig:protonspdfs}.}
\label{fig:protonssinglesource}
\end{figure*}

This way of interpreting the PDF is also shown in Fig.~\ref{fig:protonssinglesource} where we plot the complement to the {Cumulative Distribution Function, CDF}, or $1-$~CDF, as a function of the fraction $f$ of the total flux contributed by the single source providing the largest flux at the Earth (with this definition, the situation $f=1$ corresponds to the case where the most intense source produces a flux at the Earth that is as large as the sum of all other sources). The red (blue) shaded area refers to the Jelly (Spiral) model, while the three panels show the results for the same three energies as in Fig.~\ref{fig:protonspdfs}. The vertical dotted line guides the eye to the fraction $f=1$. At 100 GeV, one has $f\sim 1$ with a probability of $\sim 10^{-3}$ in the Jelly model, while for the Spiral model this probability drops to virtually zero. While increasing the energy of the arriving CRs, this probability becomes somewhat larger. At 10 TeV, one source can impact on the total flux with a probability of $\sim 2\times 10^{-3}$ in the Jelly model, while this probability drops to $10^{-4}$ in the Spiral model. The spiral structure of the Galaxy reduces the odds of finding a source exceedingly close to the Earth location. These results suggest that it is very unlikely that features in the proton (or helium) spectrum may be the result of the accidental proximity to a local source, thereby casting serious doubts on the class of models where these proximity effects are invoked as an explanation of either the spectral break at $300$ GeV or the so-called DAMPE feature at $\sim 10$ TeV. 
Our results extend previous analyses, see e.g.~\cite{Genolini2017aa}, as we included for the first time the spiral distribution of sources in the calculation of the probability of having an individual source affecting the spectrum at the Earth.

Investigation of the effect of local random sources on the observed CR anisotropy and its fluctuations have been presented by several authors~\cite{Lee1979apj,Ptuskin2006adspr,Blasi2012anisotropy}.

In fact, the anisotropy has typically two components, one associated with the global distribution of sources and the other associated with local sources. The first contribution weighs the fact that there may be more sources in one half of the sky (typically toward the Galactic center region if the detector is placed in a peripheral region of the Galaxy, as in the case of the Earth) than in the opposite direction. The second contribution reflects the stochastic nature of local sources. 

\begin{figure*}[t]
\centering
\includegraphics[width=0.9\textwidth]{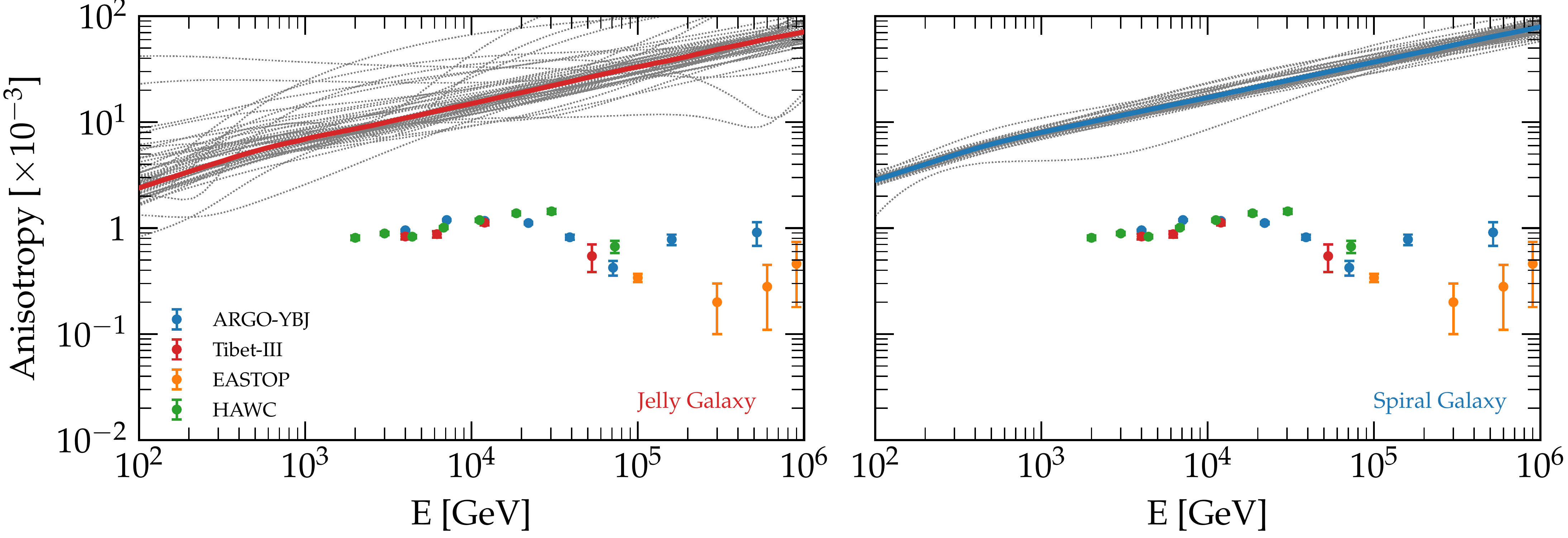}
\caption{CR anisotropy amplitude for 30 random realizations of sources in the {\it Jelly} (left) and {\it Spiral} (right) model. The solid thick line shows the median amplitude computed over $10^4$~realizations. Datasets from ARGO~\cite{anisotropy.ARGO}, EASTOP~\cite{anisotropy.EASTOP}, HAWC~\cite{anisotropy.HAWC} and TIBET-III~\cite{anisotropy.TIBET} are overplotted with dotted symbols.}
\label{fig:protonanisotropy} 
\end{figure*}

As pointed out in~\cite{Ptuskin2006adspr,Blasi2012anisotropy}, the fluctuations in the anisotropy signal (amplitude and phase) are typically very large, so that little can be learned. Moreover, since the signal is typically dominated by a handful of sources within a few hundred parsec from the Earth, both the amplitude and the phase of the anisotropy are very sensitive to the configuration of the magnetic field in the proximity of the Solar system. Since the level of magnetic perturbations expected in the ISM is at most at the level $\delta B/B\sim 1$ (but typically much smaller) the role of perpendicular diffusion in the determination of anisotropy is crucial~\cite{Ahlers2016prl}. This effect is neglected in global calculations of CR transport in that, when averaged over distances $\gg L_c\sim 10-50$ pc (the coherence scale of the Galactic field), the diffusion can be approximated as isotropic. 
With all these caveats, in Fig.~\ref{fig:protonanisotropy} we show the predicted amplitude of the anisotropy signal at the Earth location for the Jelly model (left panel) and the Spiral model (right panel) of the spatial distribution of sources, compared with observations. The median amplitude rises with energy as $\sim E^{1/3}$, which reflects the energy dependence of the diffusion coefficient above the break. 
As already pointed out in~\cite{Ptuskin2006adspr,Blasi2012anisotropy}, the anisotropy computed with these approaches is systematically larger than the observed one. In the Jelly model (similar to that discussed in \cite{Blasi2012anisotropy}) the fluctuations are rather strong, but still the median anisotropy is a factor of a few larger than the measured one. 
In the Spiral model, while the mean anisotropy does not change that much, the fluctuations are considerably smaller, as a result of the fact that the probability of having nearby sources is smaller than in the Jelly model. 

As mentioned above, and as discussed in detail in~\cite{Mertsch2015prl,Ahlers2016prl}, this might reflect the important role of anisotropic diffusion from sources in the Sun proximity, an effect that is neglected in our calculations. 

In addition to these effects related to anisotropic transport, as was pointed out by~\cite{Zirakashvili2005ijmpa}, the anisotropy signal might be severely decreased by a smaller overall diffusion coefficient in the region surrounding the solar system, while leaving the total flux basically unchanged.

\subsection{Leptons}
\label{sec:leptons}

\begin{figure*}[t]
\centering
\hspace{\stretch{1}}
\includegraphics[width=0.9\textwidth]{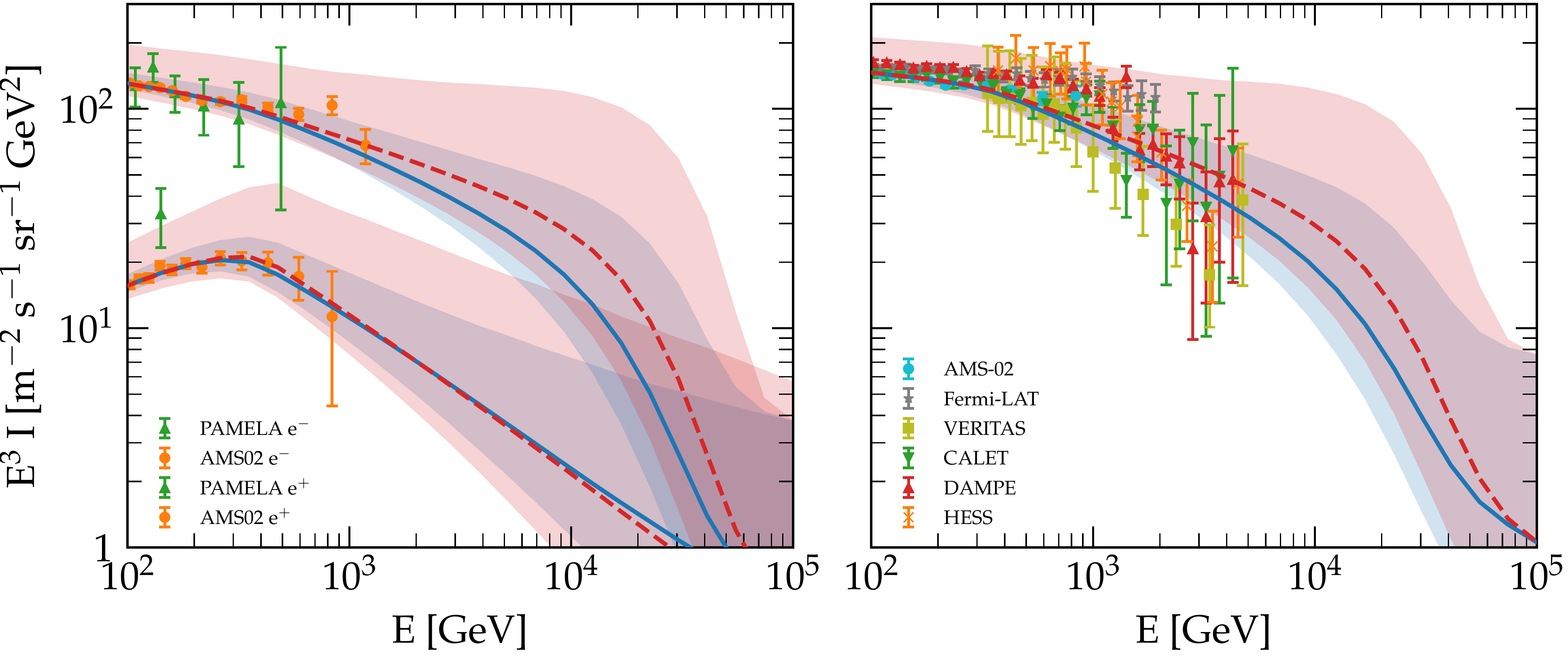}
\hspace{\stretch{1}}
\caption{The intensities of CR electrons and positrons are shown separately in the left panel, the total lepton flux is shown in the right panel. The theoretical lines include the contribution of electrons from SNRs and electrons and positrons from PWNe. Lines represent the median flux computed over $10^5$ realizations of the Milky Way in the {\it Jelly} configuration (dashed red line) or {\it Spiral} configuration (solid blue line). The shadowed areas represent the 95\% uncertainty range. Data are shown for the experiments: AMS-02~\cite{electrons.AMS02,positrons.AMS02}, CALET~\cite{leptons.CALET}, DAMPE~\cite{leptons.DAMPE}, FERMI-LAT~\cite{leptons.FERMI}, HESS~\cite{leptons.HESS}, PAMELA~\cite{electrons.PAMELA,positrons.PAMELA} and VERITAS~\cite{leptons.VERITAS}.}
\label{fig:leptons}
\end{figure*}

The transport of leptons on Galactic scales is severely affected by ICS and synchrotron losses, increasingly so at high energies. 
These losses introduce an effective horizon $l(E)\sim \sqrt{D(E)\tau_l(E)}$ at which a source can possibly be located so that CR electrons and/or positrons of energy $E$ can be received at the Earth. Since $l(E)$ decreases with energy, the number of sources that can contribute to the lepton flux at high energy decreases, and this can be easily understood to cause an increase in the flux fluctuations at high energies. 
It may be useful to remind the reader that, as discussed in~\cite{Evoli2021prd}, for the propagation parameters that best describe AMS-02 data on primary and secondary nuclei, the number of sources contributing to the local electron flux is considerably larger than reported in previous literature (e.g.,~we expect around $10^3$ sources to lie within the $\sim$~TeV lepton horizon both in the uniform case and in the case with spirals as the diffusion length is larger than the average spiral distance). 

The spectrum of electrons and positrons and the total lepton spectrum as calculated here are shown in the left and right panel of Fig.~\ref{fig:leptons} respectively, compared with a selection of available recent data. As for the case of nuclei, we show the median flux (continuous lines) and the fluctuations (shaded areas) for the Jelly case (in red) and for the Spiral model (blue). Different from the case of nuclei, the injection spectrum required to reproduce the local spectrum is different in the two setups. In particular, in the case without spirals the energy losses are less important and therefore the injection spectrum is required to be somewhat steeper ($\gamma = 2.63$) than in the case with spirals ($\gamma = 2.58$ as already found in~\cite{Evoli2021prd}). Hence, we confirm previous claims~\cite{Gaggero2013prl} that assuming a realistic distribution of sources can slightly reduce the tension between the injection slopes of protons and electrons. Nevertheless, this effect is far from sufficient to solve this problem, and it remains true that other physical effects need to be invoked. The problem is certainly non-trivial, in that the most likely option, namely that the electron spectrum may be made steeper by losses in the downstream region of a SNR shock \cite{Diesing2019prl}, requires quite extreme conditions in the late stages of the SNR evolution~\cite{Cristofari2021aa} (see also \cite{Morlino2021mnras} for a parametric treatment of the problem). As a result, the issue remains open.  

\begin{figure*}[t]
\centering
\hspace{\stretch{1}}
\includegraphics[width=0.95\textwidth]{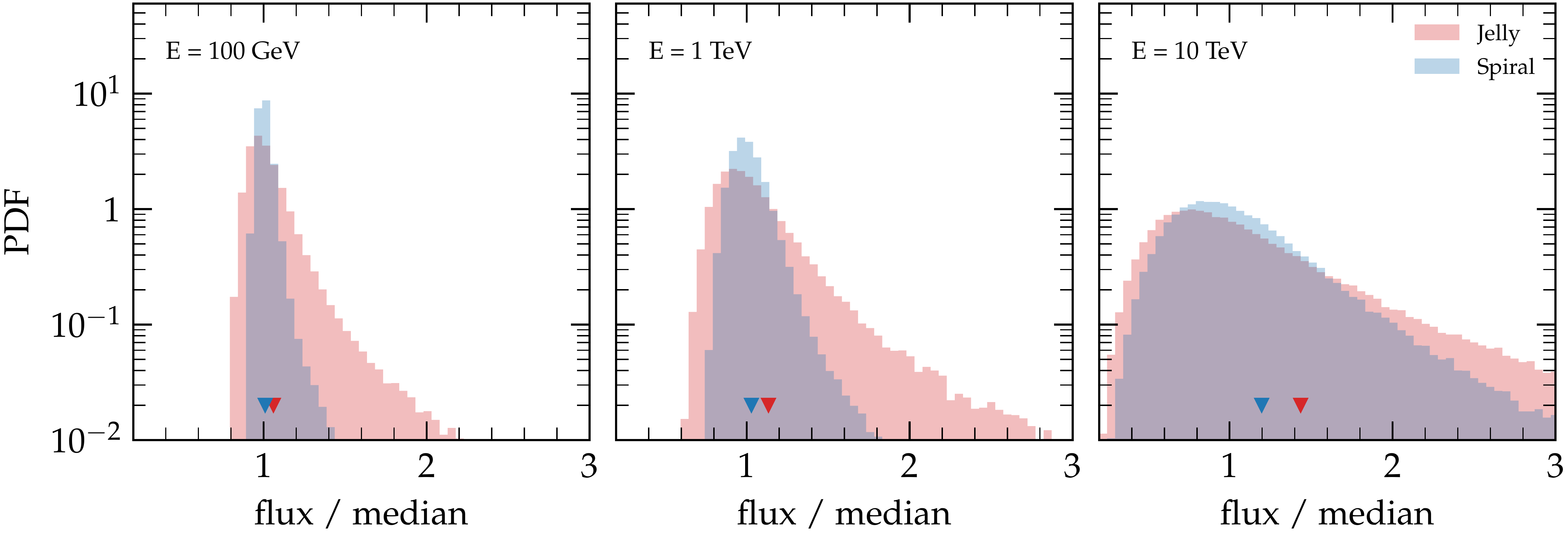}
\hspace{\stretch{1}}
\caption{The same plot as in Fig.~\ref{fig:protonspdfs} for the lepton case.}
\label{fig:leptonspdfs}
\end{figure*}

As expected, the fluctuations around the median lepton flux are very large, more so at energies $\gtrsim$ TeV, in contrast with the relatively mild fluctuations we found in the case of nuclei. However it remains true that accounting for the spiral structure of the Galaxy in terms of source distribution considerably reduces the dependence of the flux on cosmic variance. This can be appreciated in Fig.~\ref{fig:leptonspdfs}, where we show the PDF of the flux at three energies (100 GeV, 1 TeV and 10 TeV) for the Jelly (red) and Spiral (blue) model. At energies $\gtrsim 10$ TeV the cases of Jelly and Spiral distribution are not as well separated, as a result of the fact that the number of sources becomes very small in both cases. 

\begin{figure*}[t]
\centering
\hspace{\stretch{1}}
\includegraphics[width=0.95\textwidth]{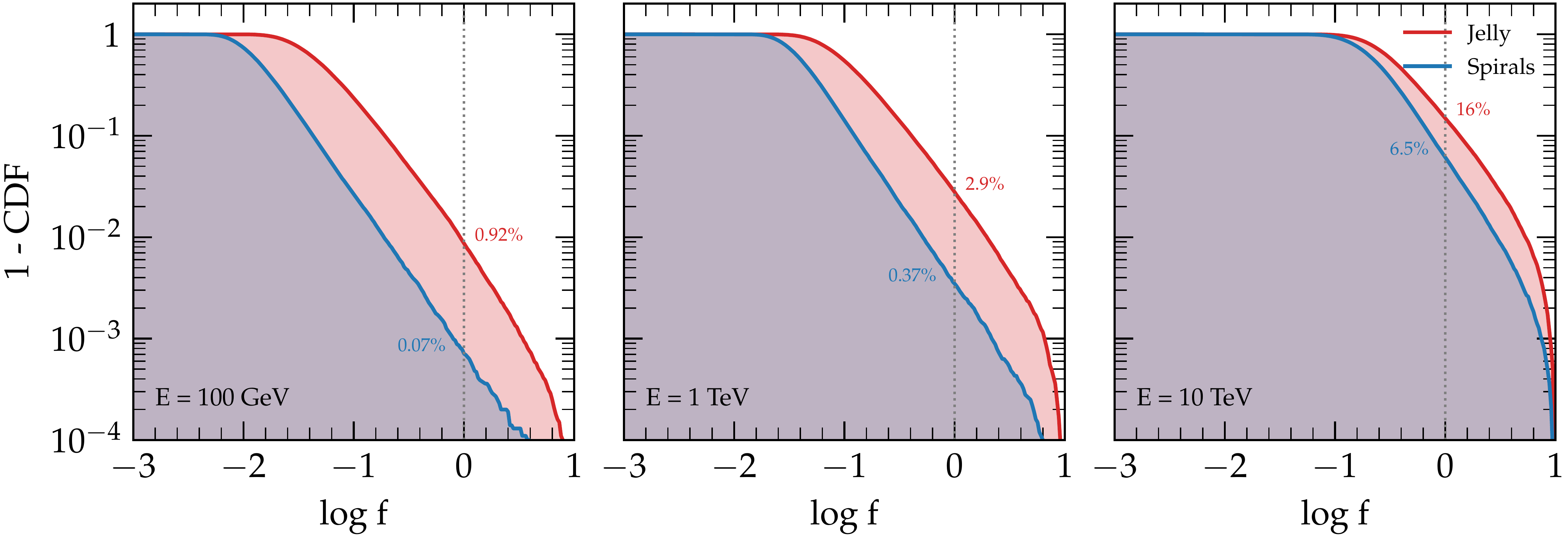}
\hspace{\stretch{1}}
\caption{The same plot as in Fig.~\ref{fig:protonssinglesource} for the lepton case.}
\label{fig:leptonsf}
\end{figure*}

We repeated here the calculation of the probability of having one source providing a contribution that is a fraction $f$ of the total flux, as plotted in Fig. \ref{fig:leptonsf}. For the Jelly model, the probability of having one source dominating the flux ($f=1$) at 100 GeV is of order 1\%, while it reduces the $\lesssim 0.1\%$ for the spiral model. These probabilities rise while increasing the lepton energy. At 10 TeV there is about $\sim 10\%$ probability of having a source contributing as much as the total flux of CR leptons, confirming that at such energies the fluctuations are expected to have a large effect. In fact, at such energies it is likely to see a local source dominating the all-lepton spectrum, a prediction that should be soon testable with data from DAMPE and CALET. 

\begin{figure*}[t]
\centering
\hspace{\stretch{1}}
\includegraphics[width=0.9\textwidth]{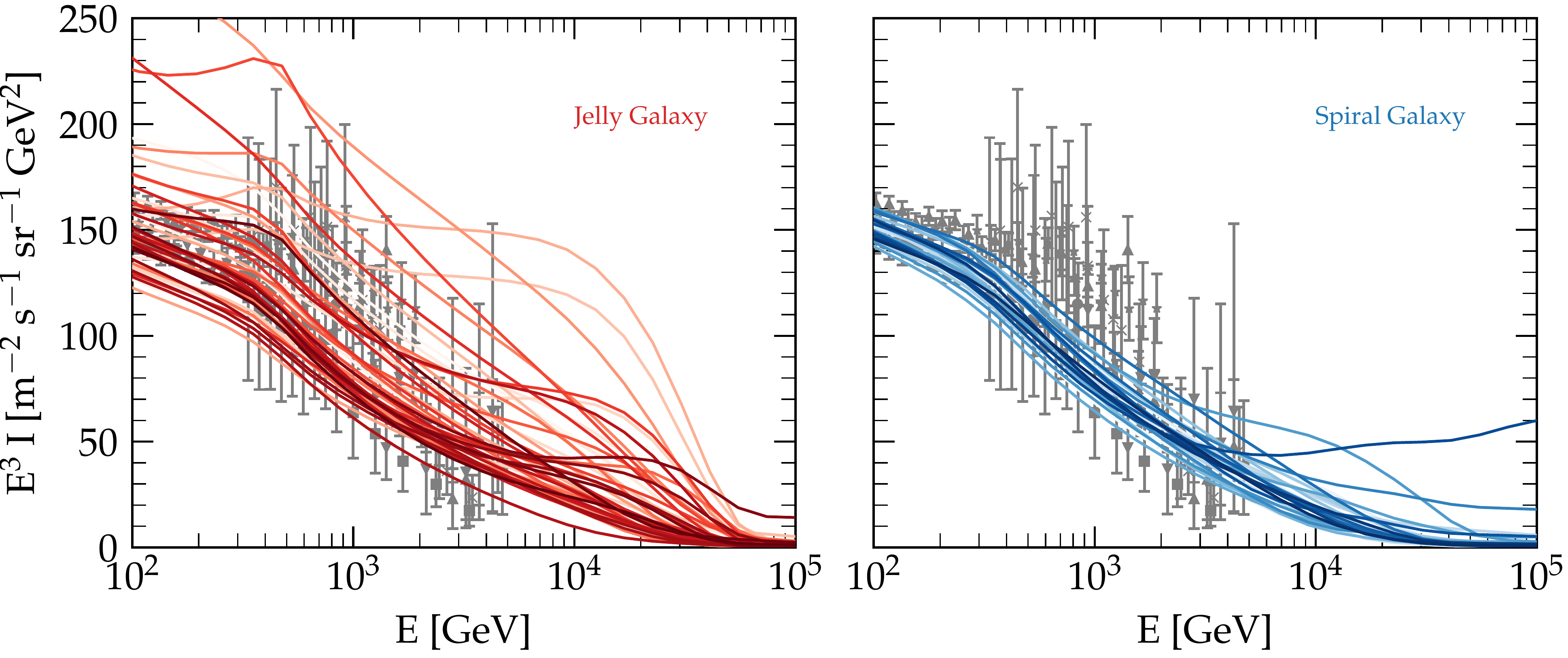}
\hspace{\stretch{1}}
\caption{The coloured lines show 30 different random realisations from the Monte Carlo simulations for the case of Jelly Galaxy (left panel, shades of red) and Spiral Galaxy (right panel, shades of blue). Datasets as in Fig.~\ref{fig:leptons} are overplotted with gray symbols.}
\label{fig:leptonsfluctuations}
\end{figure*}

This prediction is also illustrated, perhaps in a clearer way, in Fig.~\ref{fig:leptonsfluctuations}, as the lepton flux in a collection of 30 random realizations of the source distribution. The figure shows in a rather immediate way how the fluctuations are larger in the Jelly model. However, even in the Spiral model, which should be considered more realistic, the fluctuations at $\gtrsim 10$ TeV are prominent and some local sources are bound to dominate. It is impossible at the present time to predict in advance how the spectrum at such high energies should behave, in that it depends on the location and stage of the relevant sources. 

\begin{figure*}[t]
\centering
\hspace{\stretch{1}}
\includegraphics[width=0.9\textwidth]{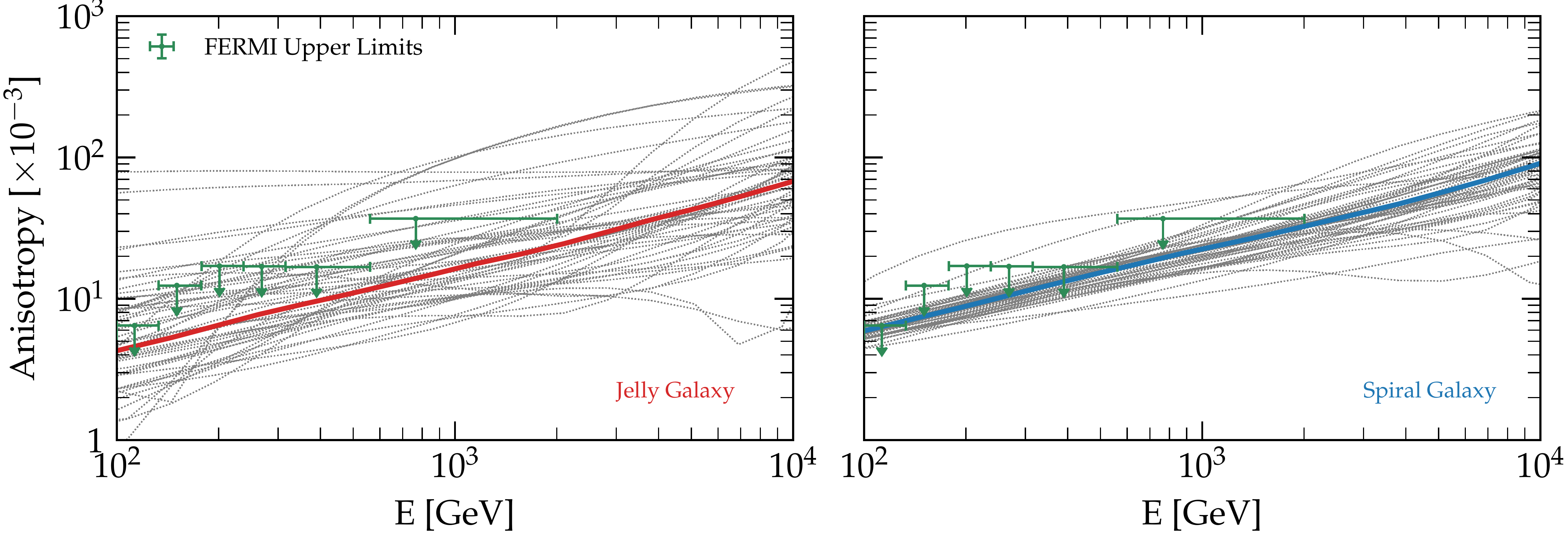}
\hspace{\stretch{1}}
\caption{Lepton anisotropy amplitude for 30 random realizations of sources in the {\it Jelly} (left) and {\it Spiral} (right) model. The solid thick line shows the median amplitude computed over $10^4$~realizations. Data points show the 95\% CL upper limits on the dipole anisotropy from~\cite{Ackermann2010prd}.}
\label{fig:leptonsanisotropy}
\end{figure*}

The anisotropy in the lepton signal has the same peculiarities already discussed for the case of nuclei, but made more severe by the larger fluctuations. In Fig. \ref{fig:leptonsanisotropy} we show the anisotropy as a function of energy for some realizations of the source distribution in the Jelly (left panel) and Spiral model (right panel). In the same plot we show the upper limits imposed by FERMI-LAT observations~\cite{Ackermann2010prd}. Most realizations, especially in the Spiral model, are compatible with these upper limits. As for nuclei, the actual pattern of anisotropy reflects the specific distribution (in space and time) of the most recent local sources. For leptons, as discussed above, not only the anisotropy but also the flux at energies $\gtrsim$ few TeV is dominated by such proximity effects. 

\subsection{Local known sources}
\label{sec:ksources}

The statistical considerations illustrated above need to face the fact that we live in a specific realization of our Galaxy and we know some ``real'' sources (SNRs) that satisfy our definition of ``local'' sources as introduced in the previous section; in particular, the Vela SNR is at a distance of $\sim 300$ pc and its age is around $\sim 20$ kyr, making it a very good candidate source of high energy leptons~\cite{Kobayashi2004apj,Ahlers2016prl,Manconi2017jcap}. 

\begin{figure*}[t]
\centering
\hspace{\stretch{1}}
\includegraphics[width=0.45\textwidth]{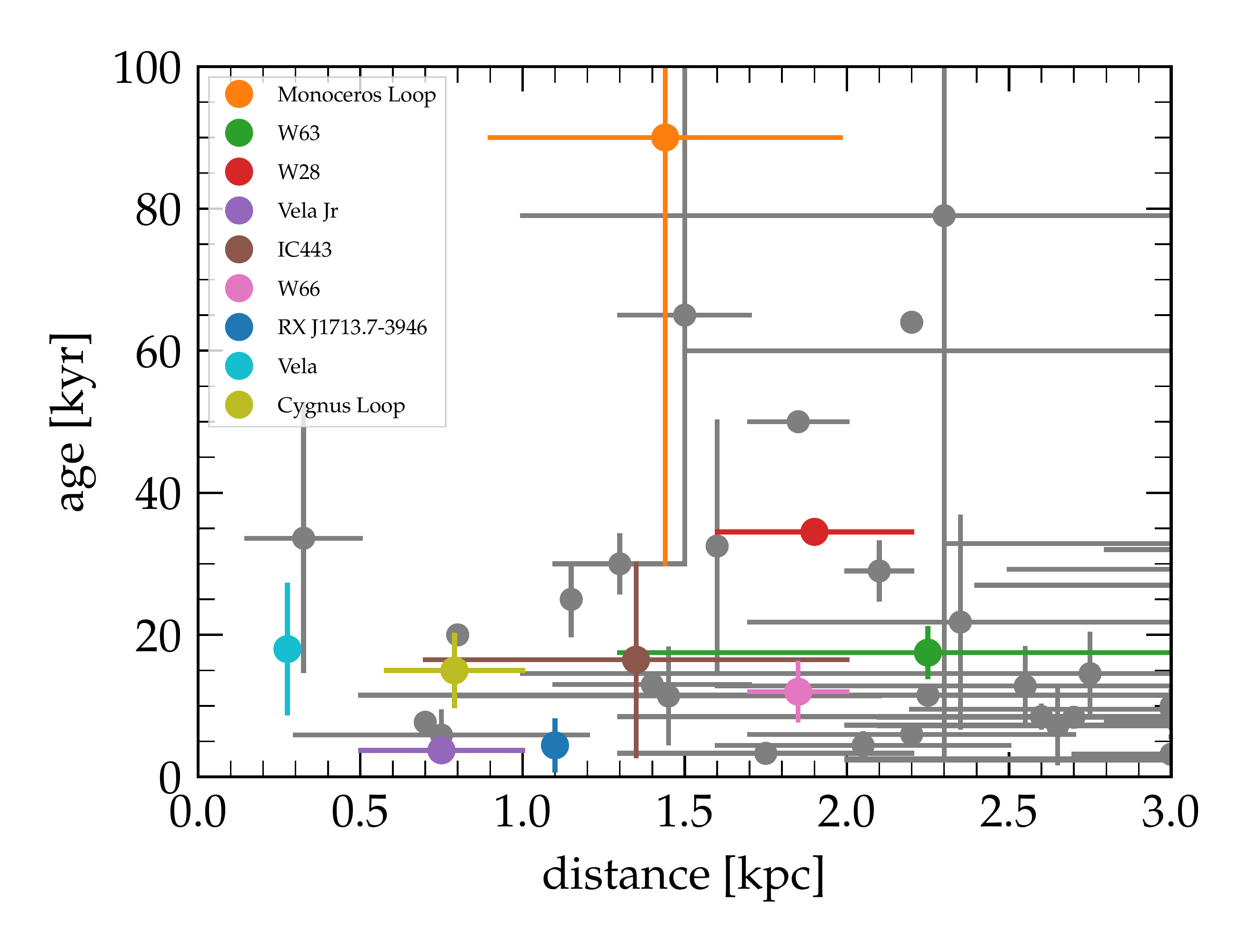}
\hspace{\stretch{1}}
\caption{The ages and distances of the SNRs within a distance of 3 kpc from the Sun selected from the ATNF catalogue~\cite{ATNF.cat}. The error bars show the reported uncertainty both in distance and age. Specific objects are highlighted with different colors as in legend.}
\label{fig:protoncat}
\end{figure*}

\begin{figure*}[t]
\centering
\hspace{\stretch{1}}
\includegraphics[width=0.45\textwidth]{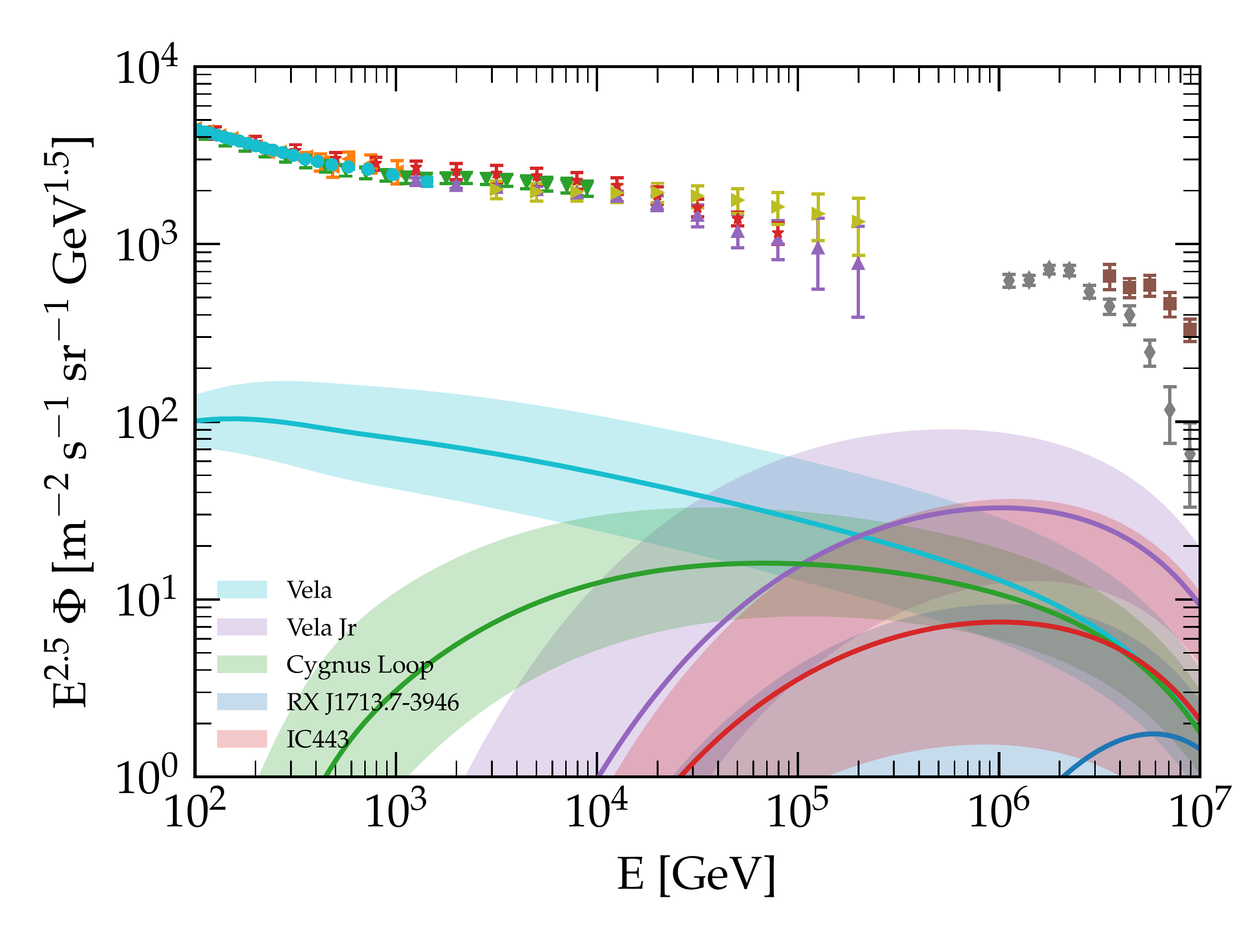}
\hspace{\stretch{1}}
\includegraphics[width=0.45\textwidth]{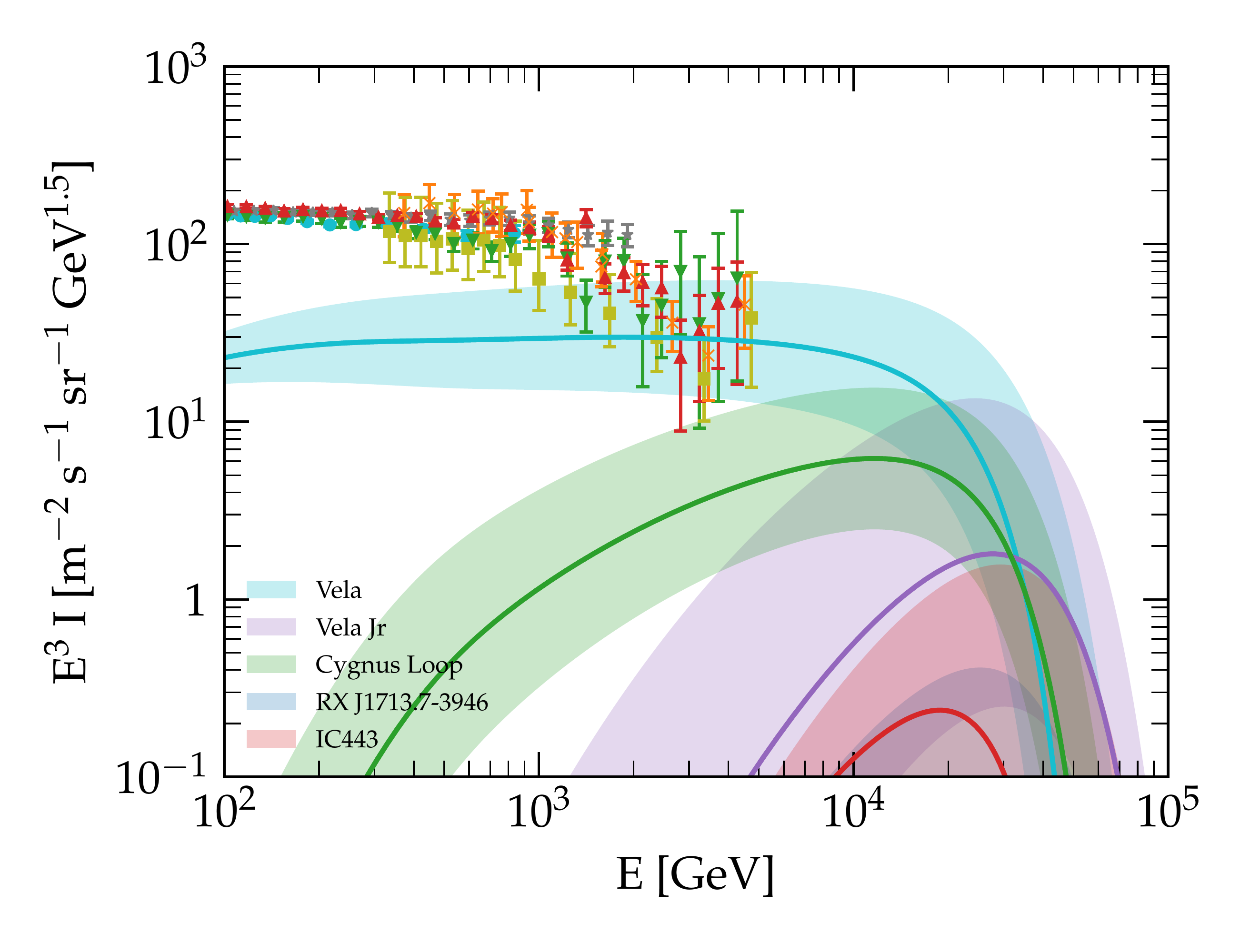}
\hspace{\stretch{1}}
\caption{Left panel: Prediction for the proton flux at the Earth from individual (known) nearby sources assuming the same efficiency and parameters as for the Spiral model in Fig.~\ref{fig:nuclei}. Right panel: the same for the lepton case.}
\label{fig:leptoncat}
\end{figure*}

In Fig.~\ref{fig:protoncat} we show a collection of ages and distances of SNRs within 3 kpc from the Sun, selected from the ATNF catalog~\cite{ATNF.cat}. In Fig.~\ref{fig:leptoncat} (left panel) we show the spectra of protons produced by sources with the same characteristics as the sources in Fig.~\ref{fig:protoncat},  assuming that they are all identical in terms of the spectrum that they inject into the ISM. Hence the difference in the spectra at the earth only derives from the different distances and ages of the sources. One should notice how these local sources provide a very small contribution to the total proton spectrum, thereby confirming once more that the spectrum of protons is always dominated, at all energies of interest here, by the contribution of numerous distant sources rather than a few local sources. 
Notice also that, at odds with what has been done in much previous work on the subject of local sources, here we do not attribute to these astrophysical objects any peculiar properties. They are in all respects just local SNRs, identical to the distant ones in their ability to produce CRs. 

The situation illustrated above changes completely for leptons: in the right panel of Fig.~\ref{fig:leptoncat} we show the spectrum of leptons at the Earth as due to the same few local SNR sources, assumed to be all identical, and compared with the observed lepton spectrum. In this case one can see that even the spectrum of leptons from Vela alone is sufficient to account for most electrons measured at energies $\gtrsim 5$ TeV. 
Below 10 TeV, the contribution of all galactic pulsars to the total lepton flux is on average much smaller than $10$\%, so that it is even less probable that a single source prevails over all others. For instance, we have checked that the flux expected from a pulsar like Geminga (with a distance of $\sim 250$~pc and age of $\sim 300$~kyr~\cite{Faherty2007apss}) is orders of magnitude smaller than the observed flux, if we assume that the injection spectrum and the efficiency are the same as that of the galactic pulsar population.

Clearly these statements should be taken with a grain of salt, in that all sources here were treated as bursts, while for local sources the duration of the injection can be comparable with the propagation time and/or the loss times, so that the time dependence of the injection may become an important factor in shaping the spectrum observed at the Earth. Qualitatively one may expect, in scenarios where higher energy particles escape in earlier phases, that the observed spectra may turn out to be somewhat harder than shown in Fig.~\ref{fig:leptoncat}, due to this phenomenon. But we stress once more that these considerations are related to details of the sources that are very hard to handle at this time. 

Moreover, for nearby SNRs, the flux received at the Earth may be affected by details of the diffusive transport: for instance, since Vela is located at a distance from Earth that corresponds to only a few coherence lengths of the Galactic magnetic field, it is plausible that anisotropy in the diffusive transport plays an important role. These effects are not taken into account here, since this would require the knowledge of details of the magnetic field (both ordered and turbulent) that are not available. 

What can be stated with reasonable certainty is that the local sources should not produce appreciable modifications of the spectrum of nuclei, while they should considerably shape the spectrum of leptons at high energy, to the point that above $\sim 10$ TeV, a dominating source, presumably Vela, might be the main contributor and produce considerable spectral modifications to be measured with DAMPE and CALET. 

\section{Conclusions}
\label{sec:conclusions}

In this work we focused our attention on the role of local sources in shaping the spectrum of both nuclei (here we only considered H and He) and leptons, in order to assess the statistical likelihood of models in which spectral features are attributed to the occasional contribution of such local sources. The main starting point here is that the variance of the fluctuations of the flux at the Earth is formally divergent~\cite{Lee1979apj}, which reflects the fact that in a truly random distribution of sources, one source can potentially be located at arbitrarily small distances from the Earth, thereby providing an arbitrarily large contribution to the flux. Hence it is of the utmost importance to investigate the statistical properties of the flux, seen as a statistical variable, and its fluctuations. 

To do so, we adopted the PDF of the flux as our statistical indicator and investigated the implications of the long high flux tail that is typical of the PDF \cite{Lagutin1995jetp}. A generalisation of the central limit theorem which applies to PDFs with power law tails and diverging second moments makes the problem statistically well posed~\cite{Genolini2017aa}. 

We proved that the shape of the PDF is very sensitive to the spatial distribution of sources: our calculations were repeated in two cases, that of a Jelly model, in which the radial distribution of sources reflects the observed distribution of SNRs and pulsars, and the Spiral model in which not only the radial distribution but also the arm structure is taken into account. It can be easily understood that the former model leads to a higher chance of having nearby sources, while most local sources in the Spiral model are bound to be in the closest arm. This difference reflects in the shape of the tail of the PDF of fluxes. 

Moreover, the PDF is very different for nuclei (for which losses are of marginal importance) and leptons, that lose energy through ICS and synchrotron radiation quite efficiently and more so at high energies. 

We can reliably conclude that the local sources are very unlikely to affect the spectrum of CR nuclei at all energies of relevance for us, meaning that statistically it is very improbable that the hardening at a few hundred GV and the DAMPE feature at $\gtrsim 10$ TV may be due to local sources. In fact, in the Jelly model of source distribution, the probability to have one source dominating the flux at 100 GeV is only $0.1\%$, dropping to $\sim 4\times 10^{-5}$ when the spiral structure is taken into account. At 10 TeV, these probabilities becomes $\sim 0.3\%$ and $\sim 0.02\%$ respectively. 

As expected, the situation is more interesting in the case of leptons. For the transport parameters derived from the measurements of B/C and Be/B \cite{Evoli2020prd}, the transport of electrons is always dominated by energy losses at $E\gtrsim 10$ GeV \cite{Evoli2021prd}, hence the sources of these particles are bound to be located at distances appreciably smaller than the halo size $H$. As a result, the flux of leptons (electrons plus positrons) is affected by much larger fluctuations compared to the spectra of nuclei. The fluctuations become particularly large for energies $\gtrsim$~few hundred GeV for the Jelly model and above $\sim$TeV for the Spiral model. 

Nevertheless, the probability of one source dominating the lepton flux at 1 TeV is only $\sim 3\%$ in the Jelly model and $0.4\%$ when spirals are taken into account. Hence the spectral break at $\sim 1$~TeV, measured by Cherenkov telescopes~\cite{leptons.HESS,leptons.VERITAS} and direct experiments~\cite{leptons.CALET,leptons.DAMPE} is unlikely to be due to local sources.
On the other hand, at $\gtrsim 10$ TeV, the diffusion-loss time for electrons becomes comparable with the typical distance between sources, hence the chances that a single source may dominate the flux increases correspondingly. In fact, the probability for that to happen is about $16\%$ in the Jelly model and $\sim 7\%$ in the Spiral model. 

We conclude that it is not only possible but actually probable that above 10 TeV the flux of leptons may be receiving a substantial contribution from a local source and that such contribution should become even more visible when measurements will be available, presumably with DAMPE and CALET, at even higher energies. We showed this by using a sample of a few tens of realizations of source distributions and plotting the corresponding flux, that in a few of such realizations showed anomalous behaviour at $E\gtrsim 10$ TeV. 

The number of local recent sources that we are actually aware of from astronomical observations is in perfect accord with the statistical properties of our Spiral model. Hence we also tried to check the expected contribution of such known local sources to the spectrum of CR nuclei and leptons. We confirm once again that the flux of H and He contributed by Vela and the other local sources in the catalogs is negligible at all energies of relevance. On the other hand, Vela is expected to contribute a flux of leptons that basically saturates the lepton flux at $E\gtrsim 10$ TeV.

As all calculations trying to describe a complex phenomenon such as the production and transport of CRs, also our calculations have limitations and caveats, which we briefly discuss below. 

The first point we want to flag concerns our assumption that the sources are treated as bursts, while in general the injection of particles occurs in a time dependent manner. The importance of this limitation is different for nuclei and leptons. For nuclei, as we discussed above, at all energies the flux is dominated by the contribution of many distant sources, and the assumption of burst injection is inconsequential. For leptons, at energies above a few hundred GeV the contribution of local sources becomes important. For such sources, it is possible that the duration of the injection period (tens of thousand of years for SNRs and hundreds of thousands of years for pulsars) may become comparable with the diffusion time from the source to Earth, in which case the burst assumption may perform more poorly. On the other hand, accounting in a reliable way for the time dependent injection of CR leptons in the ISM is all but trivial and a few comments are worth making here. 

First, electrons are expected to be leaving the acceleration region of a SNR when their maximum energy is not dominated by losses, which happens in the late stages of the SNR evolution. This point has been made clearly in a recent work \cite{Cristofari2021aa}. The spectrum of electrons is dominated by the contribution of particles leaving the remnant at the very end of the SNR evolution. In this case, the burst assumption may in fact be justified, although the details of this phenomenon are unknown and did not receive much attention in the existing literature. In fact, as pointed out in~\cite{Diesing2019prl,Cristofari2021aa}, the very spectrum of electrons is expected to be shaped by synchrotron energy losses in these late stages, and this phenomenon is poorly known as well. 

For electrons and positrons from pulsars, the phenomenology is even more complex: as long as the pulsar is inside the parent SNR, the pairs are subject to quite extreme conditions in terms of energy losses, and these particles are not expected to escape into the ISM, at least not at high energies. Injection of pairs into the ISM is expected to take place when the pulsar leaves the remnant and becomes a bow shock nebula~\cite{AmatoBlasi2018}, at a time that depends on the birth kick velocity of the pulsar. The pulsar luminosity in the form of pairs decreases in time~\cite{AmatoBlasi2018,Evoli2021prd}, so that most energy is liberated when the pulsar escapes the remnant. In this sense the assumption of burst emission is not expected to be too bad, as discussed in~\cite{Evoli2021prd}. Moreover the lepton flux in the TeV region is still expected to be dominated by the contribution of electrons from SNRs rather than a local pulsar.  

The second point to keep in mind in assessing the reliability of the calculations presented in this work and in previous articles is that the spectrum of leptons and nuclei actually released by SNRs and pulsars is not well established. In fact, as discussed in \cite{Cristofari2019mnras,Cristofari2021aa}, the spectrum of CR protons (and nuclei in general), when calculated according to our knowledge of CR acceleration in SNRs, is not a pure power law, as a result of the fact that there are at least two contributions, that of CRs escaping at the maximum energy at any given time and that of particles leaving the SNR at the end of the evolution. Moreover, different types of SNRs have different maximum energy. The sources that are able to reach PeV energies are not common type Ia and II supernovae, but rather very energetic events that may work in different way. In fact, as mentioned in \cite{Cristofari2019mnras}, the feature in the protons and He spectrum at $\gtrsim 10$ TeV might reflect these phenomena. 
As mentioned above, all these issues become even more severe for electrons injected into the ISM by SNRs, because of the importance of energy losses in the late stages of the SNR evolution~\cite{Diesing2019prl,Cristofari2021aa}. 

A third caveat to keep in mind is that in our calculations we assumed that all SNRs have the same intrinsic energy. As mentioned earlier, it is well known that SNRs of different types may have different rates and different kinetic energy in the form of ejecta. One may speculate that accounting for such effects may lead to an increase in the level of fluctuations, especially at the highest energies, which might only be contributed by rare but very energetic SNRs. 

The last caveat to keep in mind is related to our description of CR transport from nearby sources. All such sources are located inside the disc and at typical distances of order hundreds of pc from the solar system. It is possible and in fact expected that CR transport on such scales may be regulated by effects that are hard to account for in a global description such as the one presented here and used in much literature. There are several reasons for this statement: first, inside the disc the large fraction of neutral gas is expected to lead to substantial ion-neutral damping, and hence to a larger diffusion coefficient. This effect is negligible on global CR properties, such as B/C and Be/B, but can affect CR transport from local sources. Second, diffusive transport from a local source may be very sensitive to the magnetic field structure in our cosmic neighborhood. This point has been made very clearly in~\cite{Ahlers2016prl} as it affects the amplitude and phase of the CR anisotropy. In terms of flux at the Earth, the local conditions may be important because we may not be magnetically connected in the same way to local sources, hence for some of them perpendicular diffusion may be more important than parallel diffusion. In the conditions that we believe are appropriate for the ISM, the level of turbulence is low, $\delta B/B\lesssim 1$, hence the perpendicular diffusion coefficient is much smaller than the parallel one. Moreover all the quantities depend rather sensibly upon the type of turbulence responsible for CR scattering and on the level of anisotropy in the turbulent cascade (at least for Alfv\'enic turbulence). Finally, if the region around the Sun has been affected by recent (over the last tens of million years) activity, it is plausible that the local diffusion coefficient may be quite smaller than the one on Galactic scales. As discussed by~\cite{Zirakashvili2005ijmpa}, the existence of a bubble with suppressed diffusion around the Sun would not lead to a major change in the flux of CR nuclei at the Earth but might affect the anisotropy and the flux of leptons from nearby sources. 

\begin{acknowledgments}
C.~E.~acknowledges the European Commission for support under the H2020-MSCA-IF-2016 action, Grant No.~751311 GRAPES 8211 Galactic cosmic-RAy Propagation: An Extensive Study.
We are also grateful to the LNGS Computing and Network Service for computing resources and support on U-LITE cluster at Laboratori Nazionali del Gran Sasso of INFN.
\end{acknowledgments}

\bibliographystyle{myapsrev4-2}
\bibliography{2021-stochastic-sources.bib}

\begin{thebibliography}{107}%
\makeatletter
\providecommand \@ifxundefined [1]{%
 \@ifx{#1\undefined}
}%
\providecommand \@ifnum [1]{%
 \ifnum #1\expandafter \@firstoftwo
 \else \expandafter \@secondoftwo
 \fi
}%
\providecommand \@ifx [1]{%
 \ifx #1\expandafter \@firstoftwo
 \else \expandafter \@secondoftwo
 \fi
}%
\providecommand \natexlab [1]{#1}%
\providecommand \enquote  [1]{``#1''}%
\providecommand \bibnamefont  [1]{#1}%
\providecommand \bibfnamefont [1]{#1}%
\providecommand \citenamefont [1]{#1}%
\providecommand \href@noop [0]{\@secondoftwo}%
\providecommand \href [0]{\begingroup \@sanitize@url \@href}%
\providecommand \@href[1]{\@@startlink{#1}\@@href}%
\providecommand \@@href[1]{\endgroup#1\@@endlink}%
\providecommand \@sanitize@url [0]{\catcode `\\12\catcode `\$12\catcode
  `\&12\catcode `\#12\catcode `\^12\catcode `\_12\catcode `\%12\relax}%
\providecommand \@@startlink[1]{}%
\providecommand \@@endlink[0]{}%
\providecommand \url  [0]{\begingroup\@sanitize@url \@url }%
\providecommand \@url [1]{\endgroup\@href {#1}{\urlprefix }}%
\providecommand \urlprefix  [0]{URL }%
\providecommand \Eprint [0]{\href }%
\providecommand \doibase [0]{https://doi.org/}%
\providecommand \selectlanguage [0]{\@gobble}%
\providecommand \bibinfo  [0]{\@secondoftwo}%
\providecommand \bibfield  [0]{\@secondoftwo}%
\providecommand \translation [1]{[#1]}%
\providecommand \BibitemOpen [0]{}%
\providecommand \bibitemStop [0]{}%
\providecommand \bibitemNoStop [0]{.\EOS\space}%
\providecommand \EOS [0]{\spacefactor3000\relax}%
\providecommand \BibitemShut  [1]{\csname bibitem#1\endcsname}%
\let\auto@bib@innerbib\@empty
\bibitem [{\citenamefont {{Adriani}}\ \emph
  {et~al.}(2011{\natexlab{a}})\citenamefont {{Adriani}}, \citenamefont
  {{Barbarino}}, \citenamefont {{Bazilevskaya}}, \citenamefont {{Bellotti}},
  \citenamefont {{Boezio}}, \citenamefont {{Bogomolov}}, \citenamefont
  {{Bonechi}}, \citenamefont {{Bongi}}, \citenamefont {{Bonvicini}},
  \citenamefont {{Borisov}}, \citenamefont {{Bottai}}, \citenamefont {{Bruno}},
  \citenamefont {{Cafagna}}, \citenamefont {{Campana}}, \citenamefont
  {{Carbone}}, \citenamefont {{Carlson}}, \citenamefont {{Casolino}},
  \citenamefont {{Castellini}}, \citenamefont {{Consiglio}}, \citenamefont {{De
  Pascale}}, \citenamefont {{De Santis}}, \citenamefont {{De Simone}},
  \citenamefont {{Di Felice}}, \citenamefont {{Galper}}, \citenamefont
  {{Gillard}}, \citenamefont {{Grishantseva}}, \citenamefont {{Jerse}},
  \citenamefont {{Karelin}}, \citenamefont {{Koldashov}}, \citenamefont
  {{Krutkov}}, \citenamefont {{Kvashnin}}, \citenamefont {{Leonov}},
  \citenamefont {{Malakhov}}, \citenamefont {{Malvezzi}}, \citenamefont
  {{Marcelli}}, \citenamefont {{Mayorov}}, \citenamefont {{Menn}},
  \citenamefont {{Mikhailov}}, \citenamefont {{Mocchiutti}}, \citenamefont
  {{Monaco}}, \citenamefont {{Mori}}, \citenamefont {{Nikonov}}, \citenamefont
  {{Osteria}}, \citenamefont {{Palma}}, \citenamefont {{Papini}}, \citenamefont
  {{Pearce}}, \citenamefont {{Picozza}}, \citenamefont {{Pizzolotto}},
  \citenamefont {{Ricci}}, \citenamefont {{Ricciarini}}, \citenamefont
  {{Rossetto}}, \citenamefont {{Sarkar}}, \citenamefont {{Simon}},
  \citenamefont {{Sparvoli}}, \citenamefont {{Spillantini}}, \citenamefont
  {{Stozhkov}}, \citenamefont {{Vacchi}}, \citenamefont {{Vannuccini}},
  \citenamefont {{Vasilyev}}, \citenamefont {{Voronov}}, \citenamefont
  {{Yurkin}}, \citenamefont {{Wu}}, \citenamefont {{Zampa}}, \citenamefont
  {{Zampa}},\ and\ \citenamefont {{Zverev}}}]{HHe.PAMELA}%
  \BibitemOpen
  \bibfield  {author} {\bibinfo {author} {O.~{Adriani}}, \bibinfo {author}
  {G.~C. {Barbarino}}, \bibinfo {author} {G.~A. {Bazilevskaya}} et~al.,\ }\href
  {https://doi.org/10.1126/science.1199172} {\bibfield  {journal} {\bibinfo
  {journal} {Science}\ }\textbf {\bibinfo {volume} {332}},\ \bibinfo {pages}
  {69} (\bibinfo {year} {2011}{\natexlab{a}})},\ \Eprint
  {https://arxiv.org/abs/1103.4055} {arXiv:1103.4055 [astro-ph.HE]}
  \BibitemShut {NoStop}%
\bibitem [{\citenamefont {{Aguilar}}\ \emph {et~al.}(2015)\citenamefont
  {{Aguilar}}, \citenamefont {{Aisa}}, \citenamefont {{Alpat}}, \citenamefont
  {{Alvino}}, \citenamefont {{Ambrosi}}, \citenamefont {{Andeen}},
  \citenamefont {{Arruda}}, \citenamefont {{Attig}}, \citenamefont
  {{Azzarello}}, \citenamefont {{Bachlechner}},\ and\ \citenamefont
  {et~al.}}]{H.AMS02}%
  \BibitemOpen
  \bibfield  {author} {\bibinfo {author} {M.~{Aguilar}}, \bibinfo {author}
  {D.~{Aisa}}, \bibinfo {author} {B.~{Alpat}} et~al.,\ }\href
  {https://doi.org/10.1103/PhysRevLett.114.171103} {\bibfield  {journal}
  {\bibinfo  {journal} {\prl}\ }\textbf {\bibinfo {volume} {114}},\ \bibinfo
  {eid} {171103} (\bibinfo {year} {2015})}\BibitemShut {NoStop}%
\bibitem [{\citenamefont {{Aguilar}}\ \emph {et~al.}(2021)\citenamefont
  {{Aguilar}}, \citenamefont {{Ali Cavasonza}}, \citenamefont {{Ambrosi}},
  \citenamefont {{Arruda}}, \citenamefont {{Attig}}, \citenamefont {{Barao}},
  \citenamefont {{Barrin}}, \citenamefont {{Bartoloni}}, \citenamefont
  {{Ba{\c{s}}e{\u{g}}mez-du Pree}}, \citenamefont {{Bates}},\ and\
  \citenamefont {et~al.}}]{He.AMS02}%
  \BibitemOpen
  \bibfield  {author} {\bibinfo {author} {M.~{Aguilar}}, \bibinfo {author}
  {L.~{Ali Cavasonza}}, \bibinfo {author} {G.~{Ambrosi}} et~al.,\ }\href
  {https://doi.org/10.1016/j.physrep.2020.09.003} {\bibfield  {journal}
  {\bibinfo  {journal} {\physrep}\ }\textbf {\bibinfo {volume} {894}},\
  \bibinfo {pages} {1} (\bibinfo {year} {2021})}\BibitemShut {NoStop}%
\bibitem [{\citenamefont {{An}}\ \emph {et~al.}(2019)\citenamefont {{An}},
  \citenamefont {{Asfandiyarov}}, \citenamefont {{Azzarello}}, \citenamefont
  {{Bernardini}}, \citenamefont {{Bi}}, \citenamefont {{Cai}}, \citenamefont
  {{Chang}}, \citenamefont {{Chen}}, \citenamefont {{Chen}}, \citenamefont
  {{Chen}}, \citenamefont {{Chen}}, \citenamefont {{Cui}}, \citenamefont
  {{Cui}}, \citenamefont {{Dai}}, \citenamefont {{D'Amone}}, \citenamefont {{De
  Benedittis}}, \citenamefont {{De Mitri}}, \citenamefont {{Di Santo}},
  \citenamefont {{Ding}}, \citenamefont {{Dong}}, \citenamefont {{Dong}},
  \citenamefont {{Dong}}, \citenamefont {{Donvito}}, \citenamefont {{Droz}},
  \citenamefont {{Duan}}, \citenamefont {{Duan}}, \citenamefont {{D'Urso}},
  \citenamefont {{Fan}}, \citenamefont {{Fan}}, \citenamefont {{Fang}},
  \citenamefont {{Feng}}, \citenamefont {{Feng}}, \citenamefont {{Fusco}},
  \citenamefont {{Gallo}}, \citenamefont {{Gan}}, \citenamefont {{Gao}},
  \citenamefont {{Gargano}}, \citenamefont {{Gong}}, \citenamefont {{Gong}},
  \citenamefont {{Guo}}, \citenamefont {{Guo}}, \citenamefont {{Guo}},
  \citenamefont {{Han}}, \citenamefont {{Hu}}, \citenamefont {{Huang}},
  \citenamefont {{Huang}}, \citenamefont {{Huang}}, \citenamefont {{Ionica}},
  \citenamefont {{Jiang}}, \citenamefont {{Jin}}, \citenamefont {{Kong}},
  \citenamefont {{Lei}}, \citenamefont {{Li}}, \citenamefont {{Li}},
  \citenamefont {{Li}}, \citenamefont {{Li}}, \citenamefont {{Li}},
  \citenamefont {{Liang}}, \citenamefont {{Liang}}, \citenamefont {{Liao}},
  \citenamefont {{Liu}}, \citenamefont {{Liu}}, \citenamefont {{Liu}},
  \citenamefont {{Liu}}, \citenamefont {{Liu}}, \citenamefont {{Liu}},
  \citenamefont {{Loparco}}, \citenamefont {{Luo}}, \citenamefont {{Ma}},
  \citenamefont {{Ma}}, \citenamefont {{Ma}}, \citenamefont {{Ma}},
  \citenamefont {{Ma}}, \citenamefont {{Marsella}}, \citenamefont
  {{Mazziotta}}, \citenamefont {{Mo}}, \citenamefont {{Niu}}, \citenamefont
  {{Pan}}, \citenamefont {{Peng}}, \citenamefont {{Peng}}, \citenamefont
  {{Qiao}}, \citenamefont {{Rao}}, \citenamefont {{Salinas}}, \citenamefont
  {{Shang}}, \citenamefont {{Shen}}, \citenamefont {{Shen}}, \citenamefont
  {{Shen}}, \citenamefont {{Song}}, \citenamefont {{Su}}, \citenamefont {{Su}},
  \citenamefont {{Sun}}, \citenamefont {{Surdo}}, \citenamefont {{Teng}},
  \citenamefont {{Tykhonov}}, \citenamefont {{Vitillo}}, \citenamefont
  {{Wang}}, \citenamefont {{Wang}}, \citenamefont {{Wang}}, \citenamefont
  {{Wang}}, \citenamefont {{Wang}}, \citenamefont {{Wang}}, \citenamefont
  {{Wang}}, \citenamefont {{Wang}}, \citenamefont {{Wang}}, \citenamefont
  {{Wang}}, \citenamefont {{Wang}}, \citenamefont {{Wang}}, \citenamefont
  {{Wang}}, \citenamefont {{Wei}}, \citenamefont {{Wei}}, \citenamefont
  {{Wei}}, \citenamefont {{Wen}}, \citenamefont {{Wu}}, \citenamefont {{Wu}},
  \citenamefont {{Wu}}, \citenamefont {{Wu}}, \citenamefont {{Wu}},
  \citenamefont {{Xi}}, \citenamefont {{Xia}}, \citenamefont {{Xu}},
  \citenamefont {{Xu}}, \citenamefont {{Xu}}, \citenamefont {{Xu}},
  \citenamefont {{Xue}}, \citenamefont {{Yang}}, \citenamefont {{Yang}},
  \citenamefont {{Yang}}, \citenamefont {{Yang}}, \citenamefont {{Yao}},
  \citenamefont {{Yu}}, \citenamefont {{Yuan}}, \citenamefont {{Yue}},
  \citenamefont {{Zang}}, \citenamefont {{Zhang}}, \citenamefont {{Zhang}},
  \citenamefont {{Zhang}}, \citenamefont {{Zhang}}, \citenamefont {{Zhang}},
  \citenamefont {{Zhang}}, \citenamefont {{Zhang}}, \citenamefont {{Zhang}},
  \citenamefont {{Zhang}}, \citenamefont {{Zhang}}, \citenamefont {{Zhang}},
  \citenamefont {{Zhang}}, \citenamefont {{Zhang}}, \citenamefont {{Zhao}},
  \citenamefont {{Zhao}}, \citenamefont {{Zhao}}, \citenamefont {{Zhou}},
  \citenamefont {{Zhou}}, \citenamefont {{Zhu}}, \citenamefont {{Zhu}},\ and\
  \citenamefont {{Zimmer}}}]{H.DAMPE}%
  \BibitemOpen
  \bibfield  {author} {\bibinfo {author} {Q.~{An}}, \bibinfo {author}
  {R.~{Asfandiyarov}}, \bibinfo {author} {P.~{Azzarello}} et~al.,\ }\href
  {https://doi.org/10.1126/sciadv.aax3793} {\bibfield  {journal} {\bibinfo
  {journal} {Science Advances}\ }\textbf {\bibinfo {volume} {5}},\ \bibinfo
  {pages} {eaax3793} (\bibinfo {year} {2019})},\ \Eprint
  {https://arxiv.org/abs/1909.12860} {arXiv:1909.12860 [astro-ph.HE]}
  \BibitemShut {NoStop}%
\bibitem [{\citenamefont {{Alemanno}}\ \emph {et~al.}(2021)\citenamefont
  {{Alemanno}}, \citenamefont {{An}}, \citenamefont {{Azzarello}},
  \citenamefont {{Barbato}}, \citenamefont {{Bernardini}}, \citenamefont
  {{Bi}}, \citenamefont {{Cai}}, \citenamefont {{Catanzani}}, \citenamefont
  {{Chang}}, \citenamefont {{Chen}}, \citenamefont {{Chen}}, \citenamefont
  {{Chen}}, \citenamefont {{Cui}}, \citenamefont {{Cui}}, \citenamefont
  {{Cui}}, \citenamefont {{Dai}}, \citenamefont {{D'Amone}}, \citenamefont {{de
  Benedittis}}, \citenamefont {{de Mitri}}, \citenamefont {{de Palma}},
  \citenamefont {{Deliyergiyev}}, \citenamefont {{di Santo}}, \citenamefont
  {{Dong}}, \citenamefont {{Dong}}, \citenamefont {{Donvito}}, \citenamefont
  {{Droz}}, \citenamefont {{Duan}}, \citenamefont {{Duan}}, \citenamefont
  {{D'Urso}}, \citenamefont {{Fan}}, \citenamefont {{Fan}}, \citenamefont
  {{Fang}}, \citenamefont {{Fang}}, \citenamefont {{Feng}}, \citenamefont
  {{Feng}}, \citenamefont {{Fusco}}, \citenamefont {{Gao}}, \citenamefont
  {{Gargano}}, \citenamefont {{Gong}}, \citenamefont {{Gong}}, \citenamefont
  {{Guo}}, \citenamefont {{Guo}}, \citenamefont {{Guo}}, \citenamefont {{Han}},
  \citenamefont {{Hu}}, \citenamefont {{Huang}}, \citenamefont {{Huang}},
  \citenamefont {{Huang}}, \citenamefont {{Ionica}}, \citenamefont {{Jiang}},
  \citenamefont {{Kong}}, \citenamefont {{Kotenko}}, \citenamefont
  {{Kyratzis}}, \citenamefont {{Lei}}, \citenamefont {{Li}}, \citenamefont
  {{Li}}, \citenamefont {{Li}}, \citenamefont {{Li}}, \citenamefont {{Liang}},
  \citenamefont {{Liu}}, \citenamefont {{Liu}}, \citenamefont {{Liu}},
  \citenamefont {{Liu}}, \citenamefont {{Liu}}, \citenamefont {{Liu}},
  \citenamefont {{Loparco}}, \citenamefont {{Luo}}, \citenamefont {{Ma}},
  \citenamefont {{Ma}}, \citenamefont {{Ma}}, \citenamefont {{Ma}},
  \citenamefont {{Marsella}}, \citenamefont {{Mazziotta}}, \citenamefont
  {{Mo}}, \citenamefont {{Niu}}, \citenamefont {{Pan}}, \citenamefont
  {{Parenti}}, \citenamefont {{Peng}}, \citenamefont {{Peng}}, \citenamefont
  {{Perrina}}, \citenamefont {{Qiao}}, \citenamefont {{Rao}}, \citenamefont
  {{Ruina}}, \citenamefont {{Salinas}}, \citenamefont {{Shang}}, \citenamefont
  {{Shen}}, \citenamefont {{Shen}}, \citenamefont {{Shen}}, \citenamefont
  {{Silveri}}, \citenamefont {{Song}}, \citenamefont {{Stolpovskiy}},
  \citenamefont {{Su}}, \citenamefont {{Su}}, \citenamefont {{Sun}},
  \citenamefont {{Surdo}}, \citenamefont {{Teng}}, \citenamefont {{Tykhonov}},
  \citenamefont {{Wang}}, \citenamefont {{Wang}}, \citenamefont {{Wang}},
  \citenamefont {{Wang}}, \citenamefont {{Wang}}, \citenamefont {{Wang}},
  \citenamefont {{Wang}}, \citenamefont {{Wang}}, \citenamefont {{Wang}},
  \citenamefont {{Wei}}, \citenamefont {{Wei}}, \citenamefont {{Wei}},
  \citenamefont {{Wen}}, \citenamefont {{Wu}}, \citenamefont {{Wu}},
  \citenamefont {{Wu}}, \citenamefont {{Wu}}, \citenamefont {{Wu}},
  \citenamefont {{Xia}}, \citenamefont {{Xu}}, \citenamefont {{Xu}},
  \citenamefont {{Xu}}, \citenamefont {{Xu}}, \citenamefont {{Xue}},
  \citenamefont {{Yang}}, \citenamefont {{Yang}}, \citenamefont {{Yang}},
  \citenamefont {{Yao}}, \citenamefont {{Yu}}, \citenamefont {{Yuan}},
  \citenamefont {{Yuan}}, \citenamefont {{Yue}}, \citenamefont {{Zang}},
  \citenamefont {{Zhang}}, \citenamefont {{Zhang}}, \citenamefont {{Zhang}},
  \citenamefont {{Zhang}}, \citenamefont {{Zhang}}, \citenamefont {{Zhang}},
  \citenamefont {{Zhang}}, \citenamefont {{Zhang}}, \citenamefont {{Zhang}},
  \citenamefont {{Zhang}}, \citenamefont {{Zhao}}, \citenamefont {{Zhao}},
  \citenamefont {{Zhao}}, \citenamefont {{Zhou}}, \citenamefont {{Zhu}},\ and\
  \citenamefont {{Dampe Collaboration}}}]{He.DAMPE}%
  \BibitemOpen
  \bibfield  {author} {\bibinfo {author} {F.~{Alemanno}}, \bibinfo {author}
  {Q.~{An}}, \bibinfo {author} {P.~{Azzarello}} et~al.,\ }\href
  {https://doi.org/10.1103/PhysRevLett.126.201102} {\bibfield  {journal}
  {\bibinfo  {journal} {\prl}\ }\textbf {\bibinfo {volume} {126}},\ \bibinfo
  {eid} {201102} (\bibinfo {year} {2021})},\ \Eprint
  {https://arxiv.org/abs/2105.09073} {arXiv:2105.09073 [astro-ph.HE]}
  \BibitemShut {NoStop}%
\bibitem [{\citenamefont {{Adriani}}\ \emph {et~al.}(2019)\citenamefont
  {{Adriani}}, \citenamefont {{Akaike}}, \citenamefont {{Asano}}, \citenamefont
  {{Asaoka}}, \citenamefont {{Bagliesi}}, \citenamefont {{Berti}},
  \citenamefont {{Bigongiari}}, \citenamefont {{Binns}}, \citenamefont
  {{Bonechi}}, \citenamefont {{Bongi}}, \citenamefont {{Brogi}}, \citenamefont
  {{Bruno}}, \citenamefont {{Buckley}}, \citenamefont {{Cannady}},
  \citenamefont {{Castellini}}, \citenamefont {{Checchia}}, \citenamefont
  {{Cherry}}, \citenamefont {{Collazuol}}, \citenamefont {{di Felice}},
  \citenamefont {{Ebisawa}}, \citenamefont {{Fuke}}, \citenamefont {{Guzik}},
  \citenamefont {{Hams}}, \citenamefont {{Hasebe}}, \citenamefont {{Hibino}},
  \citenamefont {{Ichimura}}, \citenamefont {{Ioka}}, \citenamefont
  {{Ishizaki}}, \citenamefont {{Israel}}, \citenamefont {{Kasahara}},
  \citenamefont {{Kataoka}}, \citenamefont {{Kataoka}}, \citenamefont
  {{Katayose}}, \citenamefont {{Kato}}, \citenamefont {{Kawanaka}},
  \citenamefont {{Kawakubo}}, \citenamefont {{Kohri}}, \citenamefont
  {{Krawczynski}}, \citenamefont {{Krizmanic}}, \citenamefont {{Lomtadze}},
  \citenamefont {{Maestro}}, \citenamefont {{Marrocchesi}}, \citenamefont
  {{Messineo}}, \citenamefont {{Mitchell}}, \citenamefont {{Miyake}},
  \citenamefont {{Moiseev}}, \citenamefont {{Mori}}, \citenamefont {{Mori}},
  \citenamefont {{Mori}}, \citenamefont {{Motz}}, \citenamefont {{Munakata}},
  \citenamefont {{Murakami}}, \citenamefont {{Nakahira}}, \citenamefont
  {{Nishimura}}, \citenamefont {{de Nolfo}}, \citenamefont {{Okuno}},
  \citenamefont {{Ormes}}, \citenamefont {{Ozawa}}, \citenamefont {{Pacini}},
  \citenamefont {{Palma}}, \citenamefont {{Papini}}, \citenamefont
  {{Penacchioni}}, \citenamefont {{Rauch}}, \citenamefont {{Ricciarini}},
  \citenamefont {{Sakai}}, \citenamefont {{Sakamoto}}, \citenamefont
  {{Sasaki}}, \citenamefont {{Shimizu}}, \citenamefont {{Shiomi}},
  \citenamefont {{Sparvoli}}, \citenamefont {{Spillantini}}, \citenamefont
  {{Stolzi}}, \citenamefont {{Suh}}, \citenamefont {{Sulaj}}, \citenamefont
  {{Takahashi}}, \citenamefont {{Takayanagi}}, \citenamefont {{Takita}},
  \citenamefont {{Tamura}}, \citenamefont {{Terasawa}}, \citenamefont
  {{Tomida}}, \citenamefont {{Torii}}, \citenamefont {{Tsunesada}},
  \citenamefont {{Uchihori}}, \citenamefont {{Ueno}}, \citenamefont
  {{Vannuccini}}, \citenamefont {{Wefel}}, \citenamefont {{Yamaoka}},
  \citenamefont {{Yanagita}}, \citenamefont {{Yoshida}}, \citenamefont
  {{Yoshida}},\ and\ \citenamefont {{Calet Collaboration}}}]{H.CALET}%
  \BibitemOpen
  \bibfield  {author} {\bibinfo {author} {O.~{Adriani}}, \bibinfo {author}
  {Y.~{Akaike}}, \bibinfo {author} {K.~{Asano}} et~al.,\ }\href
  {https://doi.org/10.1103/PhysRevLett.122.181102} {\bibfield  {journal}
  {\bibinfo  {journal} {\prl}\ }\textbf {\bibinfo {volume} {122}},\ \bibinfo
  {eid} {181102} (\bibinfo {year} {2019})},\ \Eprint
  {https://arxiv.org/abs/1905.04229} {arXiv:1905.04229 [astro-ph.HE]}
  \BibitemShut {NoStop}%
\bibitem [{\citenamefont {{Strong}}\ \emph {et~al.}(2007)\citenamefont
  {{Strong}}, \citenamefont {{Moskalenko}},\ and\ \citenamefont
  {{Ptuskin}}}]{Strong2007arnps}%
  \BibitemOpen
  \bibfield  {author} {\bibinfo {author} {A.~W. {Strong}}, \bibinfo {author}
  {I.~V. {Moskalenko}}\ and\ \bibinfo {author} {V.~S. {Ptuskin}},\ }\href
  {https://doi.org/10.1146/annurev.nucl.57.090506.123011} {\bibfield  {journal}
  {\bibinfo  {journal} {Annual Review of Nuclear and Particle Science}\
  }\textbf {\bibinfo {volume} {57}},\ \bibinfo {pages} {285} (\bibinfo {year}
  {2007})},\ \Eprint {https://arxiv.org/abs/astro-ph/0701517}
  {arXiv:astro-ph/0701517 [astro-ph]} \BibitemShut {NoStop}%
\bibitem [{\citenamefont {{Tomassetti}}(2015)}]{Tomassetti2015prc}%
  \BibitemOpen
  \bibfield  {author} {\bibinfo {author} {N.~{Tomassetti}},\ }\href
  {https://doi.org/10.1103/PhysRevC.92.045808} {\bibfield  {journal} {\bibinfo
  {journal} {\prc}\ }\textbf {\bibinfo {volume} {92}},\ \bibinfo {eid} {045808}
  (\bibinfo {year} {2015})},\ \Eprint {https://arxiv.org/abs/1509.05776}
  {arXiv:1509.05776 [astro-ph.HE]} \BibitemShut {NoStop}%
\bibitem [{\citenamefont {{G{\'e}nolini}}\ \emph {et~al.}(2018)\citenamefont
  {{G{\'e}nolini}}, \citenamefont {{Maurin}}, \citenamefont {{Moskalenko}},\
  and\ \citenamefont {{Unger}}}]{Genolini2018prc}%
  \BibitemOpen
  \bibfield  {author} {\bibinfo {author} {Y.~{G{\'e}nolini}}, \bibinfo {author}
  {D.~{Maurin}}, \bibinfo {author} {I.~V. {Moskalenko}} et~al.,\ }\href
  {https://doi.org/10.1103/PhysRevC.98.034611} {\bibfield  {journal} {\bibinfo
  {journal} {\prc}\ }\textbf {\bibinfo {volume} {98}},\ \bibinfo {eid} {034611}
  (\bibinfo {year} {2018})},\ \Eprint {https://arxiv.org/abs/1803.04686}
  {arXiv:1803.04686 [astro-ph.HE]} \BibitemShut {NoStop}%
\bibitem [{\citenamefont {{Evoli}}\ \emph {et~al.}(2019)\citenamefont
  {{Evoli}}, \citenamefont {{Aloisio}},\ and\ \citenamefont
  {{Blasi}}}]{Evoli2019prd}%
  \BibitemOpen
  \bibfield  {author} {\bibinfo {author} {C.~{Evoli}}, \bibinfo {author}
  {R.~{Aloisio}}\ and\ \bibinfo {author} {P.~{Blasi}},\ }\href
  {https://doi.org/10.1103/PhysRevD.99.103023} {\bibfield  {journal} {\bibinfo
  {journal} {\prd}\ }\textbf {\bibinfo {volume} {99}},\ \bibinfo {eid} {103023}
  (\bibinfo {year} {2019})},\ \Eprint {https://arxiv.org/abs/1904.10220}
  {arXiv:1904.10220 [astro-ph.HE]} \BibitemShut {NoStop}%
\bibitem [{\citenamefont {{Schroer}}\ \emph {et~al.}(2021)\citenamefont
  {{Schroer}}, \citenamefont {{Evoli}},\ and\ \citenamefont
  {{Blasi}}}]{Schroer2021prd}%
  \BibitemOpen
  \bibfield  {author} {\bibinfo {author} {B.~{Schroer}}, \bibinfo {author}
  {C.~{Evoli}}\ and\ \bibinfo {author} {P.~{Blasi}},\ }\href
  {https://doi.org/10.1103/PhysRevD.103.123010} {\bibfield  {journal} {\bibinfo
   {journal} {\prd}\ }\textbf {\bibinfo {volume} {103}},\ \bibinfo {eid}
  {123010} (\bibinfo {year} {2021})},\ \Eprint
  {https://arxiv.org/abs/2102.12576} {arXiv:2102.12576 [astro-ph.HE]}
  \BibitemShut {NoStop}%
\bibitem [{\citenamefont {{Korsmeier}}\ and\ \citenamefont
  {{Cuoco}}(2021)}]{Korsmeier2021prd}%
  \BibitemOpen
  \bibfield  {author} {\bibinfo {author} {M.~{Korsmeier}}\ and\ \bibinfo
  {author} {A.~{Cuoco}},\ }\href {https://doi.org/10.1103/PhysRevD.103.103016}
  {\bibfield  {journal} {\bibinfo  {journal} {\prd}\ }\textbf {\bibinfo
  {volume} {103}},\ \bibinfo {eid} {103016} (\bibinfo {year} {2021})},\ \Eprint
  {https://arxiv.org/abs/2103.09824} {arXiv:2103.09824 [astro-ph.HE]}
  \BibitemShut {NoStop}%
\bibitem [{\citenamefont {{Ptuskin}}\ \emph {et~al.}(2013)\citenamefont
  {{Ptuskin}}, \citenamefont {{Zirakashvili}},\ and\ \citenamefont
  {{Seo}}}]{Ptuskin2013apj}%
  \BibitemOpen
  \bibfield  {author} {\bibinfo {author} {V.~{Ptuskin}}, \bibinfo {author}
  {V.~{Zirakashvili}}\ and\ \bibinfo {author} {E.-S. {Seo}},\ }\href
  {https://doi.org/10.1088/0004-637X/763/1/47} {\bibfield  {journal} {\bibinfo
  {journal} {\apj}\ }\textbf {\bibinfo {volume} {763}},\ \bibinfo {eid} {47}
  (\bibinfo {year} {2013})},\ \Eprint {https://arxiv.org/abs/1212.0381}
  {arXiv:1212.0381 [astro-ph.HE]} \BibitemShut {NoStop}%
\bibitem [{\citenamefont {{Recchia}}\ and\ \citenamefont
  {{Gabici}}(2018)}]{Recchia2018mnras}%
  \BibitemOpen
  \bibfield  {author} {\bibinfo {author} {S.~{Recchia}}\ and\ \bibinfo {author}
  {S.~{Gabici}},\ }\href {https://doi.org/10.1093/mnrasl/slx191} {\bibfield
  {journal} {\bibinfo  {journal} {\mnras}\ }\textbf {\bibinfo {volume} {474}},\
  \bibinfo {pages} {L42} (\bibinfo {year} {2018})},\ \Eprint
  {https://arxiv.org/abs/1710.01111} {arXiv:1710.01111 [astro-ph.HE]}
  \BibitemShut {NoStop}%
\bibitem [{\citenamefont {{Blasi}}\ \emph {et~al.}(2012)\citenamefont
  {{Blasi}}, \citenamefont {{Amato}},\ and\ \citenamefont
  {{Serpico}}}]{Blasi2012prl}%
  \BibitemOpen
  \bibfield  {author} {\bibinfo {author} {P.~{Blasi}}, \bibinfo {author}
  {E.~{Amato}}\ and\ \bibinfo {author} {P.~D. {Serpico}},\ }\href
  {https://doi.org/10.1103/PhysRevLett.109.061101} {\bibfield  {journal}
  {\bibinfo  {journal} {\prl}\ }\textbf {\bibinfo {volume} {109}},\ \bibinfo
  {eid} {061101} (\bibinfo {year} {2012})},\ \Eprint
  {https://arxiv.org/abs/1207.3706} {arXiv:1207.3706 [astro-ph.HE]}
  \BibitemShut {NoStop}%
\bibitem [{\citenamefont {{Tomassetti}}(2012)}]{Tomassetti2012apj}%
  \BibitemOpen
  \bibfield  {author} {\bibinfo {author} {N.~{Tomassetti}},\ }\href
  {https://doi.org/10.1088/2041-8205/752/1/L13} {\bibfield  {journal} {\bibinfo
   {journal} {\apjl}\ }\textbf {\bibinfo {volume} {752}},\ \bibinfo {eid} {L13}
  (\bibinfo {year} {2012})},\ \Eprint {https://arxiv.org/abs/1204.4492}
  {arXiv:1204.4492 [astro-ph.HE]} \BibitemShut {NoStop}%
\bibitem [{\citenamefont {{Evoli}}\ \emph
  {et~al.}(2018{\natexlab{a}})\citenamefont {{Evoli}}, \citenamefont {{Blasi}},
  \citenamefont {{Morlino}},\ and\ \citenamefont {{Aloisio}}}]{Evoli2018prl}%
  \BibitemOpen
  \bibfield  {author} {\bibinfo {author} {C.~{Evoli}}, \bibinfo {author}
  {P.~{Blasi}}, \bibinfo {author} {G.~{Morlino}} et~al.,\ }\href
  {https://doi.org/10.1103/PhysRevLett.121.021102} {\bibfield  {journal}
  {\bibinfo  {journal} {\prl}\ }\textbf {\bibinfo {volume} {121}},\ \bibinfo
  {eid} {021102} (\bibinfo {year} {2018}{\natexlab{a}})},\ \Eprint
  {https://arxiv.org/abs/1806.04153} {arXiv:1806.04153 [astro-ph.HE]}
  \BibitemShut {NoStop}%
\bibitem [{\citenamefont {{Kachelrie{\ss}}}\ \emph {et~al.}(2015)\citenamefont
  {{Kachelrie{\ss}}}, \citenamefont {{Neronov}},\ and\ \citenamefont
  {{Semikoz}}}]{Kachelriess2015prl}%
  \BibitemOpen
  \bibfield  {author} {\bibinfo {author} {M.~{Kachelrie{\ss}}}, \bibinfo
  {author} {A.~{Neronov}}\ and\ \bibinfo {author} {D.~V. {Semikoz}},\ }\href
  {https://doi.org/10.1103/PhysRevLett.115.181103} {\bibfield  {journal}
  {\bibinfo  {journal} {\prl}\ }\textbf {\bibinfo {volume} {115}},\ \bibinfo
  {eid} {181103} (\bibinfo {year} {2015})},\ \Eprint
  {https://arxiv.org/abs/1504.06472} {arXiv:1504.06472 [astro-ph.HE]}
  \BibitemShut {NoStop}%
\bibitem [{\citenamefont {{Genolini}}\ \emph {et~al.}(2017)\citenamefont
  {{Genolini}}, \citenamefont {{Salati}}, \citenamefont {{Serpico}},\ and\
  \citenamefont {{Taillet}}}]{Genolini2017aa}%
  \BibitemOpen
  \bibfield  {author} {\bibinfo {author} {Y.~{Genolini}}, \bibinfo {author}
  {P.~{Salati}}, \bibinfo {author} {P.~D. {Serpico}} et~al.,\ }\href
  {https://doi.org/10.1051/0004-6361/201629903} {\bibfield  {journal} {\bibinfo
   {journal} {\aap}\ }\textbf {\bibinfo {volume} {600}},\ \bibinfo {eid} {A68}
  (\bibinfo {year} {2017})},\ \Eprint {https://arxiv.org/abs/1610.02010}
  {arXiv:1610.02010 [astro-ph.HE]} \BibitemShut {NoStop}%
\bibitem [{\citenamefont {{Malkov}}\ and\ \citenamefont
  {{Moskalenko}}(2021)}]{Malkov2021apj}%
  \BibitemOpen
  \bibfield  {author} {\bibinfo {author} {M.~A. {Malkov}}\ and\ \bibinfo
  {author} {I.~V. {Moskalenko}},\ }\href
  {https://doi.org/10.3847/1538-4357/abe855} {\bibfield  {journal} {\bibinfo
  {journal} {\apj}\ }\textbf {\bibinfo {volume} {911}},\ \bibinfo {eid} {151}
  (\bibinfo {year} {2021})},\ \Eprint {https://arxiv.org/abs/2010.02826}
  {arXiv:2010.02826 [astro-ph.HE]} \BibitemShut {NoStop}%
\bibitem [{\citenamefont {{Aharonian}}\ \emph {et~al.}(2009)\citenamefont
  {{Aharonian}}, \citenamefont {{Akhperjanian}}, \citenamefont {{Anton}},
  \citenamefont {{Barres de Almeida}}, \citenamefont {{Bazer-Bachi}},
  \citenamefont {{Becherini}}, \citenamefont {{Behera}}, \citenamefont
  {{Bernl{\"o}hr}}, \citenamefont {{Bochow}}, \citenamefont {{Boisson}},
  \citenamefont {{Bolmont}}, \citenamefont {{Borrel}}, \citenamefont
  {{Brucker}}, \citenamefont {{Brun}}, \citenamefont {{Brun}}, \citenamefont
  {{B{\"u}hler}}, \citenamefont {{Bulik}}, \citenamefont {{B{\"u}sching}},
  \citenamefont {{Boutelier}}, \citenamefont {{Chadwick}}, \citenamefont
  {{Charbonnier}}, \citenamefont {{Chaves}}, \citenamefont {{Cheesebrough}},
  \citenamefont {{Chounet}}, \citenamefont {{Clapson}}, \citenamefont
  {{Coignet}}, \citenamefont {{Dalton}}, \citenamefont {{Daniel}},
  \citenamefont {{Davids}}, \citenamefont {{Degrange}}, \citenamefont {{Deil}},
  \citenamefont {{Dickinson}}, \citenamefont {{Djannati-Ata{\"\i}}},
  \citenamefont {{Domainko}}, \citenamefont {{O'C. Drury}}, \citenamefont
  {{Dubois}}, \citenamefont {{Dubus}}, \citenamefont {{Dyks}}, \citenamefont
  {{Dyrda}}, \citenamefont {{Egberts}}, \citenamefont {{Emmanoulopoulos}},
  \citenamefont {{Espigat}}, \citenamefont {{Farnier}}, \citenamefont
  {{Feinstein}}, \citenamefont {{Fiasson}}, \citenamefont {{F{\"o}rster}},
  \citenamefont {{Fontaine}}, \citenamefont {{F{\"u}{\ss}ling}}, \citenamefont
  {{Gabici}}, \citenamefont {{Gallant}}, \citenamefont {{G{\'e}rard}},
  \citenamefont {{Gerbig}}, \citenamefont {{Giebels}}, \citenamefont
  {{Glicenstein}}, \citenamefont {{Gl{\"u}ck}}, \citenamefont {{Goret}},
  \citenamefont {{G{\"o}ring}}, \citenamefont {{Hauser}}, \citenamefont
  {{Hauser}}, \citenamefont {{Heinz}}, \citenamefont {{Heinzelmann}},
  \citenamefont {{Henri}}, \citenamefont {{Hermann}}, \citenamefont {{Hinton}},
  \citenamefont {{Hoffmann}}, \citenamefont {{Hofmann}}, \citenamefont
  {{Holleran}}, \citenamefont {{Hoppe}}, \citenamefont {{Horns}}, \citenamefont
  {{Jacholkowska}}, \citenamefont {{de Jager}}, \citenamefont {{Jahn}},
  \citenamefont {{Jung}}, \citenamefont {{Katarzy{\'n}ski}}, \citenamefont
  {{Katz}}, \citenamefont {{Kaufmann}}, \citenamefont {{Kendziorra}},
  \citenamefont {{Kerschhaggl}}, \citenamefont {{Khangulyan}}, \citenamefont
  {{Kh{\'e}lifi}}, \citenamefont {{Keogh}}, \citenamefont {{Klu{\'z}niak}},
  \citenamefont {{Kneiske}}, \citenamefont {{Komin}}, \citenamefont {{Kosack}},
  \citenamefont {{Kossakowski}}, \citenamefont {{Lamanna}}, \citenamefont
  {{Lenain}}, \citenamefont {{Lohse}}, \citenamefont {{Marandon}},
  \citenamefont {{Martin}}, \citenamefont {{Martineau-Huynh}}, \citenamefont
  {{Marcowith}}, \citenamefont {{Masbou}}, \citenamefont {{Maurin}},
  \citenamefont {{McComb}}, \citenamefont {{Medina}}, \citenamefont
  {{Moderski}}, \citenamefont {{Moulin}}, \citenamefont {{Naumann-Godo}},
  \citenamefont {{de Naurois}}, \citenamefont {{Nedbal}}, \citenamefont
  {{Nekrassov}}, \citenamefont {{Nicholas}}, \citenamefont {{Niemiec}},
  \citenamefont {{Nolan}}, \citenamefont {{Ohm}}, \citenamefont {{Olive}},
  \citenamefont {{de O{\~n}a Wilhelmi}}, \citenamefont {{Orford}},
  \citenamefont {{Ostrowski}}, \citenamefont {{Panter}}, \citenamefont {{Paz
  Arribas}}, \citenamefont {{Pedaletti}}, \citenamefont {{Pelletier}},
  \citenamefont {{Petrucci}}, \citenamefont {{Pita}}, \citenamefont
  {{P{\"u}hlhofer}}, \citenamefont {{Punch}}, \citenamefont {{Quirrenbach}},
  \citenamefont {{Raubenheimer}}, \citenamefont {{Raue}}, \citenamefont
  {{Rayner}}, \citenamefont {{Reimer}}, \citenamefont {{Renaud}}, \citenamefont
  {{Rieger}}, \citenamefont {{Ripken}}, \citenamefont {{Rob}}, \citenamefont
  {{Rosier-Lees}}, \citenamefont {{Rowell}}, \citenamefont {{Rudak}},
  \citenamefont {{Rulten}}, \citenamefont {{Ruppel}}, \citenamefont
  {{Sahakian}}, \citenamefont {{Santangelo}}, \citenamefont {{Schlickeiser}},
  \citenamefont {{Sch{\"o}ck}}, \citenamefont {{Schr{\"o}der}}, \citenamefont
  {{Schwanke}}, \citenamefont {{Schwarzburg}}, \citenamefont {{Schwemmer}},
  \citenamefont {{Shalchi}}, \citenamefont {{Sikora}}, \citenamefont
  {{Skilton}}, \citenamefont {{Sol}}, \citenamefont {{Spangler}}, \citenamefont
  {{Stawarz}}, \citenamefont {{Steenkamp}}, \citenamefont {{Stegmann}},
  \citenamefont {{Stinzing}}, \citenamefont {{Superina}}, \citenamefont
  {{Szostek}}, \citenamefont {{Tam}}, \citenamefont {{Tavernet}}, \citenamefont
  {{Terrier}}, \citenamefont {{Tibolla}}, \citenamefont {{Tluczykont}},
  \citenamefont {{van Eldik}}, \citenamefont {{Vasileiadis}}, \citenamefont
  {{Venter}}, \citenamefont {{Venter}}, \citenamefont {{Vialle}}, \citenamefont
  {{Vincent}}, \citenamefont {{Vivier}}, \citenamefont {{V{\"o}lk}},
  \citenamefont {{Volpe}}, \citenamefont {{Wagner}}, \citenamefont {{Ward}},
  \citenamefont {{Zdziarski}},\ and\ \citenamefont {{Zech}}}]{leptons.HESS}%
  \BibitemOpen
  \bibfield  {author} {\bibinfo {author} {F.~{Aharonian}}, \bibinfo {author}
  {A.~G. {Akhperjanian}}, \bibinfo {author} {G.~{Anton}} et~al.,\ }\href
  {https://doi.org/10.1051/0004-6361/200913323} {\bibfield  {journal} {\bibinfo
   {journal} {\aap}\ }\textbf {\bibinfo {volume} {508}},\ \bibinfo {pages}
  {561} (\bibinfo {year} {2009})},\ \Eprint {https://arxiv.org/abs/0905.0105}
  {arXiv:0905.0105 [astro-ph.HE]} \BibitemShut {NoStop}%
\bibitem [{\citenamefont {{Adriani}}\ \emph {et~al.}(2018)\citenamefont
  {{Adriani}}, \citenamefont {{Akaike}}, \citenamefont {{Asano}}, \citenamefont
  {{Asaoka}}, \citenamefont {{Bagliesi}}, \citenamefont {{Berti}},
  \citenamefont {{Bigongiari}}, \citenamefont {{Binns}}, \citenamefont
  {{Bonechi}}, \citenamefont {{Bongi}}, \citenamefont {{Brogi}}, \citenamefont
  {{Buckley}}, \citenamefont {{Cannady}}, \citenamefont {{Castellini}},
  \citenamefont {{Checchia}}, \citenamefont {{Cherry}}, \citenamefont
  {{Collazuol}}, \citenamefont {{di Felice}}, \citenamefont {{Ebisawa}},
  \citenamefont {{Fuke}}, \citenamefont {{Guzik}}, \citenamefont {{Hams}},
  \citenamefont {{Hareyama}}, \citenamefont {{Hasebe}}, \citenamefont
  {{Hibino}}, \citenamefont {{Ichimura}}, \citenamefont {{Ioka}}, \citenamefont
  {{Ishizaki}}, \citenamefont {{Israel}}, \citenamefont {{Kasahara}},
  \citenamefont {{Kataoka}}, \citenamefont {{Kataoka}}, \citenamefont
  {{Katayose}}, \citenamefont {{Kato}}, \citenamefont {{Kawanaka}},
  \citenamefont {{Kawakubo}}, \citenamefont {{Kohri}}, \citenamefont
  {{Krawczynski}}, \citenamefont {{Krizmanic}}, \citenamefont {{Lomtadze}},
  \citenamefont {{Maestro}}, \citenamefont {{Marrocchesi}}, \citenamefont
  {{Messineo}}, \citenamefont {{Mitchell}}, \citenamefont {{Miyake}},
  \citenamefont {{Moiseev}}, \citenamefont {{Mori}}, \citenamefont {{Mori}},
  \citenamefont {{Mori}}, \citenamefont {{Motz}}, \citenamefont {{Munakata}},
  \citenamefont {{Murakami}}, \citenamefont {{Nakahira}}, \citenamefont
  {{Nishimura}}, \citenamefont {{de Nolfo}}, \citenamefont {{Okuno}},
  \citenamefont {{Ormes}}, \citenamefont {{Ozawa}}, \citenamefont {{Pacini}},
  \citenamefont {{Palma}}, \citenamefont {{Papini}}, \citenamefont
  {{Penacchioni}}, \citenamefont {{Rauch}}, \citenamefont {{Ricciarini}},
  \citenamefont {{Sakai}}, \citenamefont {{Sakamoto}}, \citenamefont
  {{Sasaki}}, \citenamefont {{Shimizu}}, \citenamefont {{Shiomi}},
  \citenamefont {{Sparvoli}}, \citenamefont {{Spillantini}}, \citenamefont
  {{Stolzi}}, \citenamefont {{Suh}}, \citenamefont {{Sulaj}}, \citenamefont
  {{Takahashi}}, \citenamefont {{Takayanagi}}, \citenamefont {{Takita}},
  \citenamefont {{Tamura}}, \citenamefont {{Tateyama}}, \citenamefont
  {{Terasawa}}, \citenamefont {{Tomida}}, \citenamefont {{Torii}},
  \citenamefont {{Tsunesada}}, \citenamefont {{Uchihori}}, \citenamefont
  {{Ueno}}, \citenamefont {{Vannuccini}}, \citenamefont {{Wefel}},
  \citenamefont {{Yamaoka}}, \citenamefont {{Yanagita}}, \citenamefont
  {{Yoshida}}, \citenamefont {{Yoshida}},\ and\ \citenamefont {{Calet
  Collaboration}}}]{leptons.CALET}%
  \BibitemOpen
  \bibfield  {author} {\bibinfo {author} {O.~{Adriani}}, \bibinfo {author}
  {Y.~{Akaike}}, \bibinfo {author} {K.~{Asano}} et~al.,\ }\href
  {https://doi.org/10.1103/PhysRevLett.120.261102} {\bibfield  {journal}
  {\bibinfo  {journal} {\prl}\ }\textbf {\bibinfo {volume} {120}},\ \bibinfo
  {eid} {261102} (\bibinfo {year} {2018})},\ \Eprint
  {https://arxiv.org/abs/1806.09728} {arXiv:1806.09728 [astro-ph.HE]}
  \BibitemShut {NoStop}%
\bibitem [{\citenamefont {{DAMPE Collaboration}}\ \emph
  {et~al.}(2017)\citenamefont {{DAMPE Collaboration}}, \citenamefont
  {{Ambrosi}}, \citenamefont {{An}}, \citenamefont {{Asfandiyarov}},
  \citenamefont {{Azzarello}}, \citenamefont {{Bernardini}}, \citenamefont
  {{Bertucci}}, \citenamefont {{Cai}}, \citenamefont {{Chang}}, \citenamefont
  {{Chen}}, \citenamefont {{Chen}}, \citenamefont {{Chen}}, \citenamefont
  {{Chen}}, \citenamefont {{Cui}}, \citenamefont {{Cui}}, \citenamefont
  {{D'Amone}}, \citenamefont {{de Benedittis}}, \citenamefont {{De Mitri}},
  \citenamefont {{di Santo}}, \citenamefont {{Dong}}, \citenamefont {{Dong}},
  \citenamefont {{Dong}}, \citenamefont {{Dong}}, \citenamefont {{Donvito}},
  \citenamefont {{Droz}}, \citenamefont {{Duan}}, \citenamefont {{Duan}},
  \citenamefont {{Duranti}}, \citenamefont {{D'Urso}}, \citenamefont {{Fan}},
  \citenamefont {{Fan}}, \citenamefont {{Fang}}, \citenamefont {{Feng}},
  \citenamefont {{Feng}}, \citenamefont {{Fusco}}, \citenamefont {{Gallo}},
  \citenamefont {{Gan}}, \citenamefont {{Gao}}, \citenamefont {{Gao}},
  \citenamefont {{Gargano}}, \citenamefont {{Garrappa}}, \citenamefont
  {{Gong}}, \citenamefont {{Gong}}, \citenamefont {{Guo}}, \citenamefont
  {{Guo}}, \citenamefont {{Hu}}, \citenamefont {{Huang}}, \citenamefont
  {{Huang}}, \citenamefont {{Ionica}}, \citenamefont {{Jiang}}, \citenamefont
  {{Jiang}}, \citenamefont {{Jin}}, \citenamefont {{Kong}}, \citenamefont
  {{Lei}}, \citenamefont {{Li}}, \citenamefont {{Li}}, \citenamefont {{Li}},
  \citenamefont {{Li}}, \citenamefont {{Liang}}, \citenamefont {{Liang}},
  \citenamefont {{Liao}}, \citenamefont {{Liu}}, \citenamefont {{Liu}},
  \citenamefont {{Liu}}, \citenamefont {{Liu}}, \citenamefont {{Liu}},
  \citenamefont {{Loparco}}, \citenamefont {{Ma}}, \citenamefont {{Ma}},
  \citenamefont {{Ma}}, \citenamefont {{Ma}}, \citenamefont {{Ma}},
  \citenamefont {{Ma}}, \citenamefont {{Marsella}}, \citenamefont
  {{Mazziotta}}, \citenamefont {{Mo}}, \citenamefont {{Niu}}, \citenamefont
  {{Peng}}, \citenamefont {{Peng}}, \citenamefont {{Qiao}}, \citenamefont
  {{Rao}}, \citenamefont {{Salinas}}, \citenamefont {{Shang}}, \citenamefont
  {{H. Shen}}, \citenamefont {{Shen}}, \citenamefont {{Shen}}, \citenamefont
  {{Song}}, \citenamefont {{Su}}, \citenamefont {{Su}}, \citenamefont {{Sun}},
  \citenamefont {{Surdo}}, \citenamefont {{Teng}}, \citenamefont {{Tian}},
  \citenamefont {{Tykhonov}}, \citenamefont {{Vagelli}}, \citenamefont
  {{Vitillo}}, \citenamefont {{Wang}}, \citenamefont {{Wang}}, \citenamefont
  {{Wang}}, \citenamefont {{Wang}}, \citenamefont {{Wang}}, \citenamefont
  {{Wang}}, \citenamefont {{Wang}}, \citenamefont {{Wang}}, \citenamefont
  {{Wang}}, \citenamefont {{Wang}}, \citenamefont {{Wang}}, \citenamefont
  {{Wang}}, \citenamefont {{Wen}}, \citenamefont {{Wang}}, \citenamefont
  {{Wei}}, \citenamefont {{Wei}}, \citenamefont {{Wei}}, \citenamefont {{Wu}},
  \citenamefont {{Wu}}, \citenamefont {{Wu}}, \citenamefont {{Wu}},
  \citenamefont {{Wu}}, \citenamefont {{Xi}}, \citenamefont {{Xia}},
  \citenamefont {{Xin}}, \citenamefont {{Xu}}, \citenamefont {{Xu}},
  \citenamefont {{Xu}}, \citenamefont {{Xue}}, \citenamefont {{Yang}},
  \citenamefont {{Yang}}, \citenamefont {{Yang}}, \citenamefont {{Yang}},
  \citenamefont {{Yao}}, \citenamefont {{Yu}}, \citenamefont {{Yuan}},
  \citenamefont {{Yue}}, \citenamefont {{Zang}}, \citenamefont {{Zhang}},
  \citenamefont {{Zhang}}, \citenamefont {{Zhang}}, \citenamefont {{Zhang}},
  \citenamefont {{Zhang}}, \citenamefont {{Zhang}}, \citenamefont {{Zhang}},
  \citenamefont {{Zhang}}, \citenamefont {{Zhang}}, \citenamefont {{Zhang}},
  \citenamefont {{Zhang}}, \citenamefont {{Zhang}}, \citenamefont {{Zhang}},
  \citenamefont {{Zhang}}, \citenamefont {{Zhang}}, \citenamefont {{Zhang}},
  \citenamefont {{Zhang}}, \citenamefont {{Zhao}}, \citenamefont {{Zhao}},
  \citenamefont {{Zhao}}, \citenamefont {{Zhou}}, \citenamefont {{Zhou}},
  \citenamefont {{Zhu}}, \citenamefont {{Zhu}},\ and\ \citenamefont
  {{Zimmer}}}]{leptons.DAMPE}%
  \BibitemOpen
  \bibfield  {author} {\bibinfo {author} {{DAMPE Collaboration}}, \bibinfo
  {author} {G.~{Ambrosi}}, \bibinfo {author} {Q.~{An}} et~al.,\ }\href
  {https://doi.org/10.1038/nature24475} {\bibfield  {journal} {\bibinfo
  {journal} {\nat}\ }\textbf {\bibinfo {volume} {552}},\ \bibinfo {pages} {63}
  (\bibinfo {year} {2017})},\ \Eprint {https://arxiv.org/abs/1711.10981}
  {arXiv:1711.10981 [astro-ph.HE]} \BibitemShut {NoStop}%
\bibitem [{\citenamefont {{Adriani}}\ \emph
  {et~al.}(2011{\natexlab{b}})\citenamefont {{Adriani}}, \citenamefont
  {{Barbarino}}, \citenamefont {{Bazilevskaya}}, \citenamefont {{Bellotti}},
  \citenamefont {{Boezio}}, \citenamefont {{Bogomolov}}, \citenamefont
  {{Bongi}}, \citenamefont {{Bonvicini}}, \citenamefont {{Borisov}},
  \citenamefont {{Bottai}}, \citenamefont {{Bruno}}, \citenamefont {{Cafagna}},
  \citenamefont {{Campana}}, \citenamefont {{Carbone}}, \citenamefont
  {{Carlson}}, \citenamefont {{Casolino}}, \citenamefont {{Castellini}},
  \citenamefont {{Consiglio}}, \citenamefont {{de Pascale}}, \citenamefont {{de
  Santis}}, \citenamefont {{de Simone}}, \citenamefont {{di Felice}},
  \citenamefont {{Galper}}, \citenamefont {{Gillard}}, \citenamefont
  {{Grishantseva}}, \citenamefont {{Jerse}}, \citenamefont {{Karelin}},
  \citenamefont {{Koldashov}}, \citenamefont {{Krutkov}}, \citenamefont
  {{Kvashnin}}, \citenamefont {{Leonov}}, \citenamefont {{Malakhov}},
  \citenamefont {{Malvezzi}}, \citenamefont {{Marcelli}}, \citenamefont
  {{Mayorov}}, \citenamefont {{Menn}}, \citenamefont {{Mikhailov}},
  \citenamefont {{Mocchiutti}}, \citenamefont {{Monaco}}, \citenamefont
  {{Mori}}, \citenamefont {{Nikonov}}, \citenamefont {{Osteria}}, \citenamefont
  {{Palma}}, \citenamefont {{Papini}}, \citenamefont {{Pearce}}, \citenamefont
  {{Picozza}}, \citenamefont {{Pizzolotto}}, \citenamefont {{Ricci}},
  \citenamefont {{Ricciarini}}, \citenamefont {{Rossetto}}, \citenamefont
  {{Sarkar}}, \citenamefont {{Simon}}, \citenamefont {{Sparvoli}},
  \citenamefont {{Spillantini}}, \citenamefont {{Stochaj}}, \citenamefont
  {{Stockton}}, \citenamefont {{Stozhkov}}, \citenamefont {{Vacchi}},
  \citenamefont {{Vannuccini}}, \citenamefont {{Vasilyev}}, \citenamefont
  {{Voronov}}, \citenamefont {{Wu}}, \citenamefont {{Yurkin}}, \citenamefont
  {{Zampa}}, \citenamefont {{Zampa}},\ and\ \citenamefont
  {{Zverev}}}]{electrons.PAMELA}%
  \BibitemOpen
  \bibfield  {author} {\bibinfo {author} {O.~{Adriani}}, \bibinfo {author}
  {G.~C. {Barbarino}}, \bibinfo {author} {G.~A. {Bazilevskaya}} et~al.,\ }\href
  {https://doi.org/10.1103/PhysRevLett.106.201101} {\bibfield  {journal}
  {\bibinfo  {journal} {\prl}\ }\textbf {\bibinfo {volume} {106}},\ \bibinfo
  {eid} {201101} (\bibinfo {year} {2011}{\natexlab{b}})},\ \Eprint
  {https://arxiv.org/abs/1103.2880} {arXiv:1103.2880 [astro-ph.HE]}
  \BibitemShut {NoStop}%
\bibitem [{\citenamefont {{Adriani}}\ \emph {et~al.}(2013)\citenamefont
  {{Adriani}}, \citenamefont {{Barbarino}}, \citenamefont {{Bazilevskaya}},
  \citenamefont {{Bellotti}}, \citenamefont {{Bianco}}, \citenamefont
  {{Boezio}}, \citenamefont {{Bogomolov}}, \citenamefont {{Bongi}},
  \citenamefont {{Bonvicini}}, \citenamefont {{Bottai}}, \citenamefont
  {{Bruno}}, \citenamefont {{Cafagna}}, \citenamefont {{Campana}},
  \citenamefont {{Carbone}}, \citenamefont {{Carlson}}, \citenamefont
  {{Casolino}}, \citenamefont {{Castellini}}, \citenamefont {{De Donato}},
  \citenamefont {{De Santis}}, \citenamefont {{De Simone}}, \citenamefont {{Di
  Felice}}, \citenamefont {{Formato}}, \citenamefont {{Galper}}, \citenamefont
  {{Karelin}}, \citenamefont {{Koldashov}}, \citenamefont {{Koldobskiy}},
  \citenamefont {{Krutkov}}, \citenamefont {{Kvashnin}}, \citenamefont
  {{Leonov}}, \citenamefont {{Malakhov}}, \citenamefont {{Marcelli}},
  \citenamefont {{Martucci}}, \citenamefont {{Mayorov}}, \citenamefont
  {{Menn}}, \citenamefont {{Merg{\'e}}}, \citenamefont {{Mikhailov}},
  \citenamefont {{Mocchiutti}}, \citenamefont {{Monaco}}, \citenamefont
  {{Mori}}, \citenamefont {{Munini}}, \citenamefont {{Osteria}}, \citenamefont
  {{Palma}}, \citenamefont {{Papini}}, \citenamefont {{Pearce}}, \citenamefont
  {{Picozza}}, \citenamefont {{Pizzolotto}}, \citenamefont {{Ricci}},
  \citenamefont {{Ricciarini}}, \citenamefont {{Rossetto}}, \citenamefont
  {{Sarkar}}, \citenamefont {{Scotti}}, \citenamefont {{Simon}}, \citenamefont
  {{Sparvoli}}, \citenamefont {{Spillantini}}, \citenamefont {{Stochaj}},
  \citenamefont {{Stockton}}, \citenamefont {{Stozhkov}}, \citenamefont
  {{Vacchi}}, \citenamefont {{Vannuccini}}, \citenamefont {{Vasilyev}},
  \citenamefont {{Voronov}}, \citenamefont {{Yurkin}}, \citenamefont {{Zampa}},
  \citenamefont {{Zampa}},\ and\ \citenamefont {{Zverev}}}]{positrons.PAMELA}%
  \BibitemOpen
  \bibfield  {author} {\bibinfo {author} {O.~{Adriani}}, \bibinfo {author}
  {G.~C. {Barbarino}}, \bibinfo {author} {G.~A. {Bazilevskaya}} et~al.,\ }\href
  {https://doi.org/10.1103/PhysRevLett.111.081102} {\bibfield  {journal}
  {\bibinfo  {journal} {\prl}\ }\textbf {\bibinfo {volume} {111}},\ \bibinfo
  {eid} {081102} (\bibinfo {year} {2013})},\ \Eprint
  {https://arxiv.org/abs/1308.0133} {arXiv:1308.0133 [astro-ph.HE]}
  \BibitemShut {NoStop}%
\bibitem [{\citenamefont {{Aguilar}}\ \emph
  {et~al.}(2019{\natexlab{a}})\citenamefont {{Aguilar}}, \citenamefont {{Ali
  Cavasonza}}, \citenamefont {{Ambrosi}}, \citenamefont {{Arruda}},
  \citenamefont {{Attig}}, \citenamefont {{Azzarello}}, \citenamefont
  {{Bachlechner}}, \citenamefont {{Barao}}, \citenamefont {{Barrau}},
  \citenamefont {{Barrin}},\ and\ \citenamefont {et~al.}}]{positrons.AMS02}%
  \BibitemOpen
  \bibfield  {author} {\bibinfo {author} {M.~{Aguilar}}, \bibinfo {author}
  {L.~{Ali Cavasonza}}, \bibinfo {author} {G.~{Ambrosi}} et~al.,\ }\href
  {https://doi.org/10.1103/PhysRevLett.122.041102} {\bibfield  {journal}
  {\bibinfo  {journal} {\prl}\ }\textbf {\bibinfo {volume} {122}},\ \bibinfo
  {eid} {041102} (\bibinfo {year} {2019}{\natexlab{a}})}\BibitemShut {NoStop}%
\bibitem [{\citenamefont {{Hooper}}\ \emph {et~al.}(2009)\citenamefont
  {{Hooper}}, \citenamefont {{Blasi}},\ and\ \citenamefont
  {{Serpico}}}]{Hooper2009jcap}%
  \BibitemOpen
  \bibfield  {author} {\bibinfo {author} {D.~{Hooper}}, \bibinfo {author}
  {P.~{Blasi}}\ and\ \bibinfo {author} {P.~D. {Serpico}},\ }\href
  {https://doi.org/10.1088/1475-7516/2009/01/025} {\bibfield  {journal}
  {\bibinfo  {journal} {\jcap}\ }\textbf {\bibinfo {volume} {2009}},\ \bibinfo
  {eid} {025} (\bibinfo {year} {2009})},\ \Eprint
  {https://arxiv.org/abs/0810.1527} {arXiv:0810.1527 [astro-ph]} \BibitemShut
  {NoStop}%
\bibitem [{\citenamefont {{Grasso}}\ \emph {et~al.}(2009)\citenamefont
  {{Grasso}}, \citenamefont {{Profumo}}, \citenamefont {{Strong}},
  \citenamefont {{Baldini}}, \citenamefont {{Bellazzini}}, \citenamefont
  {{Bloom}}, \citenamefont {{Bregeon}}, \citenamefont {{Di Bernardo}},
  \citenamefont {{Gaggero}}, \citenamefont {{Giglietto}}, \citenamefont
  {{Kamae}}, \citenamefont {{Latronico}}, \citenamefont {{Longo}},
  \citenamefont {{Mazziotta}}, \citenamefont {{Moiseev}}, \citenamefont
  {{Morselli}}, \citenamefont {{Ormes}}, \citenamefont {{Pesce-Rollins}},
  \citenamefont {{Pohl}}, \citenamefont {{Razzano}}, \citenamefont {{Sgro}},
  \citenamefont {{Spandre}},\ and\ \citenamefont {{Stephens}}}]{Grasso2009aph}%
  \BibitemOpen
  \bibfield  {author} {\bibinfo {author} {D.~{Grasso}}, \bibinfo {author}
  {S.~{Profumo}}, \bibinfo {author} {A.~W. {Strong}} et~al.,\ }\href
  {https://doi.org/10.1016/j.astropartphys.2009.07.003} {\bibfield  {journal}
  {\bibinfo  {journal} {Astroparticle Physics}\ }\textbf {\bibinfo {volume}
  {32}},\ \bibinfo {pages} {140} (\bibinfo {year} {2009})},\ \Eprint
  {https://arxiv.org/abs/0905.0636} {arXiv:0905.0636 [astro-ph.HE]}
  \BibitemShut {NoStop}%
\bibitem [{\citenamefont {{Delahaye}}\ \emph {et~al.}(2010)\citenamefont
  {{Delahaye}}, \citenamefont {{Lavalle}}, \citenamefont {{Lineros}},
  \citenamefont {{Donato}},\ and\ \citenamefont {{Fornengo}}}]{Delahaye2010aa}%
  \BibitemOpen
  \bibfield  {author} {\bibinfo {author} {T.~{Delahaye}}, \bibinfo {author}
  {J.~{Lavalle}}, \bibinfo {author} {R.~{Lineros}} et~al.,\ }\href
  {https://doi.org/10.1051/0004-6361/201014225} {\bibfield  {journal} {\bibinfo
   {journal} {\aap}\ }\textbf {\bibinfo {volume} {524}},\ \bibinfo {eid} {A51}
  (\bibinfo {year} {2010})},\ \Eprint {https://arxiv.org/abs/1002.1910}
  {arXiv:1002.1910 [astro-ph.HE]} \BibitemShut {NoStop}%
\bibitem [{\citenamefont {{Linden}}\ and\ \citenamefont
  {{Profumo}}(2013)}]{Linden2013apj}%
  \BibitemOpen
  \bibfield  {author} {\bibinfo {author} {T.~{Linden}}\ and\ \bibinfo {author}
  {S.~{Profumo}},\ }\href {https://doi.org/10.1088/0004-637X/772/1/18}
  {\bibfield  {journal} {\bibinfo  {journal} {\apj}\ }\textbf {\bibinfo
  {volume} {772}},\ \bibinfo {eid} {18} (\bibinfo {year} {2013})},\ \Eprint
  {https://arxiv.org/abs/1304.1791} {arXiv:1304.1791 [astro-ph.HE]}
  \BibitemShut {NoStop}%
\bibitem [{\citenamefont {{Gaggero}}\ \emph {et~al.}(2013)\citenamefont
  {{Gaggero}}, \citenamefont {{Maccione}}, \citenamefont {{Di Bernardo}},
  \citenamefont {{Evoli}},\ and\ \citenamefont {{Grasso}}}]{Gaggero2013prl}%
  \BibitemOpen
  \bibfield  {author} {\bibinfo {author} {D.~{Gaggero}}, \bibinfo {author}
  {L.~{Maccione}}, \bibinfo {author} {G.~{Di Bernardo}} et~al.,\ }\href
  {https://doi.org/10.1103/PhysRevLett.111.021102} {\bibfield  {journal}
  {\bibinfo  {journal} {\prl}\ }\textbf {\bibinfo {volume} {111}},\ \bibinfo
  {eid} {021102} (\bibinfo {year} {2013})},\ \Eprint
  {https://arxiv.org/abs/1304.6718} {arXiv:1304.6718 [astro-ph.HE]}
  \BibitemShut {NoStop}%
\bibitem [{\citenamefont {{Di Mauro}}\ \emph {et~al.}(2014)\citenamefont {{Di
  Mauro}}, \citenamefont {{Donato}}, \citenamefont {{Fornengo}}, \citenamefont
  {{Lineros}},\ and\ \citenamefont {{Vittino}}}]{DiMauro2014jcap}%
  \BibitemOpen
  \bibfield  {author} {\bibinfo {author} {M.~{Di Mauro}}, \bibinfo {author}
  {F.~{Donato}}, \bibinfo {author} {N.~{Fornengo}} et~al.,\ }\href
  {https://doi.org/10.1088/1475-7516/2014/04/006} {\bibfield  {journal}
  {\bibinfo  {journal} {\jcap}\ }\textbf {\bibinfo {volume} {2014}},\ \bibinfo
  {eid} {006} (\bibinfo {year} {2014})},\ \Eprint
  {https://arxiv.org/abs/1402.0321} {arXiv:1402.0321 [astro-ph.HE]}
  \BibitemShut {NoStop}%
\bibitem [{\citenamefont {{Yuan}}\ \emph {et~al.}(2015)\citenamefont {{Yuan}},
  \citenamefont {{Bi}}, \citenamefont {{Chen}}, \citenamefont {{Guo}},
  \citenamefont {{Lin}},\ and\ \citenamefont {{Zhang}}}]{Yaun2015aph}%
  \BibitemOpen
  \bibfield  {author} {\bibinfo {author} {Q.~{Yuan}}, \bibinfo {author} {X.-J.
  {Bi}}, \bibinfo {author} {G.-M. {Chen}} et~al.,\ }\href
  {https://doi.org/10.1016/j.astropartphys.2014.05.005} {\bibfield  {journal}
  {\bibinfo  {journal} {Astroparticle Physics}\ }\textbf {\bibinfo {volume}
  {60}},\ \bibinfo {pages} {1} (\bibinfo {year} {2015})},\ \Eprint
  {https://arxiv.org/abs/1304.1482} {arXiv:1304.1482 [astro-ph.HE]}
  \BibitemShut {NoStop}%
\bibitem [{\citenamefont {{Cholis}}\ \emph {et~al.}(2018)\citenamefont
  {{Cholis}}, \citenamefont {{Karwal}},\ and\ \citenamefont
  {{Kamionkowski}}}]{Cholis2018prd}%
  \BibitemOpen
  \bibfield  {author} {\bibinfo {author} {I.~{Cholis}}, \bibinfo {author}
  {T.~{Karwal}}\ and\ \bibinfo {author} {M.~{Kamionkowski}},\ }\href
  {https://doi.org/10.1103/PhysRevD.98.063008} {\bibfield  {journal} {\bibinfo
  {journal} {\prd}\ }\textbf {\bibinfo {volume} {98}},\ \bibinfo {eid} {063008}
  (\bibinfo {year} {2018})},\ \Eprint {https://arxiv.org/abs/1807.05230}
  {arXiv:1807.05230 [astro-ph.HE]} \BibitemShut {NoStop}%
\bibitem [{\citenamefont {{Orusa}}\ \emph {et~al.}(2021)\citenamefont
  {{Orusa}}, \citenamefont {{Manconi}}, \citenamefont {{Donato}},\ and\
  \citenamefont {{Di Mauro}}}]{Orusa2021arxiv}%
  \BibitemOpen
  \bibfield  {author} {\bibinfo {author} {L.~{Orusa}}, \bibinfo {author}
  {S.~{Manconi}}, \bibinfo {author} {F.~{Donato}} et~al.,\ }\href@noop {}
  {\bibfield  {journal} {\bibinfo  {journal} {arXiv e-prints}\ ,\ \bibinfo
  {eid} {arXiv:2107.06300}} (\bibinfo {year} {2021})},\ \Eprint
  {https://arxiv.org/abs/2107.06300} {arXiv:2107.06300 [astro-ph.HE]}
  \BibitemShut {NoStop}%
\bibitem [{\citenamefont {{Evoli}}\ \emph {et~al.}(2021)\citenamefont
  {{Evoli}}, \citenamefont {{Amato}}, \citenamefont {{Blasi}},\ and\
  \citenamefont {{Aloisio}}}]{Evoli2021prd}%
  \BibitemOpen
  \bibfield  {author} {\bibinfo {author} {C.~{Evoli}}, \bibinfo {author}
  {E.~{Amato}}, \bibinfo {author} {P.~{Blasi}} et~al.,\ }\href
  {https://doi.org/10.1103/PhysRevD.103.083010} {\bibfield  {journal} {\bibinfo
   {journal} {\prd}\ }\textbf {\bibinfo {volume} {103}},\ \bibinfo {eid}
  {083010} (\bibinfo {year} {2021})},\ \Eprint
  {https://arxiv.org/abs/2010.11955} {arXiv:2010.11955 [astro-ph.HE]}
  \BibitemShut {NoStop}%
\bibitem [{\citenamefont {{Evoli}}\ \emph
  {et~al.}(2020{\natexlab{a}})\citenamefont {{Evoli}}, \citenamefont {{Blasi}},
  \citenamefont {{Amato}},\ and\ \citenamefont {{Aloisio}}}]{Evoli2020prl}%
  \BibitemOpen
  \bibfield  {author} {\bibinfo {author} {C.~{Evoli}}, \bibinfo {author}
  {P.~{Blasi}}, \bibinfo {author} {E.~{Amato}} et~al.,\ }\href
  {https://doi.org/10.1103/PhysRevLett.125.051101} {\bibfield  {journal}
  {\bibinfo  {journal} {\prl}\ }\textbf {\bibinfo {volume} {125}},\ \bibinfo
  {eid} {051101} (\bibinfo {year} {2020}{\natexlab{a}})},\ \Eprint
  {https://arxiv.org/abs/2007.01302} {arXiv:2007.01302 [astro-ph.HE]}
  \BibitemShut {NoStop}%
\bibitem [{\citenamefont {{Fang}}\ \emph {et~al.}(2021)\citenamefont {{Fang}},
  \citenamefont {{Bi}}, \citenamefont {{Lin}},\ and\ \citenamefont
  {{Yuan}}}]{Fang2021chphl}%
  \BibitemOpen
  \bibfield  {author} {\bibinfo {author} {K.~{Fang}}, \bibinfo {author} {X.-J.
  {Bi}}, \bibinfo {author} {S.-J. {Lin}} et~al.,\ }\href
  {https://doi.org/10.1088/0256-307X/38/3/039801} {\bibfield  {journal}
  {\bibinfo  {journal} {Chinese Physics Letters}\ }\textbf {\bibinfo {volume}
  {38}},\ \bibinfo {eid} {039801} (\bibinfo {year} {2021})},\ \Eprint
  {https://arxiv.org/abs/2007.15601} {arXiv:2007.15601 [astro-ph.HE]}
  \BibitemShut {NoStop}%
\bibitem [{\citenamefont {{Y{\"u}ksel}}\ \emph {et~al.}(2009)\citenamefont
  {{Y{\"u}ksel}}, \citenamefont {{Kistler}},\ and\ \citenamefont
  {{Stanev}}}]{Yuksel2009prl}%
  \BibitemOpen
  \bibfield  {author} {\bibinfo {author} {H.~{Y{\"u}ksel}}, \bibinfo {author}
  {M.~D. {Kistler}}\ and\ \bibinfo {author} {T.~{Stanev}},\ }\href
  {https://doi.org/10.1103/PhysRevLett.103.051101} {\bibfield  {journal}
  {\bibinfo  {journal} {\prl}\ }\textbf {\bibinfo {volume} {103}},\ \bibinfo
  {eid} {051101} (\bibinfo {year} {2009})},\ \Eprint
  {https://arxiv.org/abs/0810.2784} {arXiv:0810.2784 [astro-ph]} \BibitemShut
  {NoStop}%
\bibitem [{\citenamefont {{Fujita}}\ \emph {et~al.}(2009)\citenamefont
  {{Fujita}}, \citenamefont {{Kohri}}, \citenamefont {{Yamazaki}},\ and\
  \citenamefont {{Ioka}}}]{Fujita2009prd}%
  \BibitemOpen
  \bibfield  {author} {\bibinfo {author} {Y.~{Fujita}}, \bibinfo {author}
  {K.~{Kohri}}, \bibinfo {author} {R.~{Yamazaki}} et~al.,\ }\href
  {https://doi.org/10.1103/PhysRevD.80.063003} {\bibfield  {journal} {\bibinfo
  {journal} {\prd}\ }\textbf {\bibinfo {volume} {80}},\ \bibinfo {eid} {063003}
  (\bibinfo {year} {2009})},\ \Eprint {https://arxiv.org/abs/0903.5298}
  {arXiv:0903.5298 [astro-ph.HE]} \BibitemShut {NoStop}%
\bibitem [{\citenamefont {{Thoudam}}\ and\ \citenamefont
  {{H{\"o}randel}}(2012)}]{Thoudam2012mnras}%
  \BibitemOpen
  \bibfield  {author} {\bibinfo {author} {S.~{Thoudam}}\ and\ \bibinfo {author}
  {J.~R. {H{\"o}randel}},\ }\href
  {https://doi.org/10.1111/j.1365-2966.2011.20385.x} {\bibfield  {journal}
  {\bibinfo  {journal} {\mnras}\ }\textbf {\bibinfo {volume} {421}},\ \bibinfo
  {pages} {1209} (\bibinfo {year} {2012})},\ \Eprint
  {https://arxiv.org/abs/1112.3020} {arXiv:1112.3020 [astro-ph.HE]}
  \BibitemShut {NoStop}%
\bibitem [{\citenamefont {{Yin}}\ \emph {et~al.}(2013)\citenamefont {{Yin}},
  \citenamefont {{Yu}}, \citenamefont {{Yuan}},\ and\ \citenamefont
  {{Bi}}}]{Yin2013prd}%
  \BibitemOpen
  \bibfield  {author} {\bibinfo {author} {P.-F. {Yin}}, \bibinfo {author}
  {Z.-H. {Yu}}, \bibinfo {author} {Q.~{Yuan}} et~al.,\ }\href
  {https://doi.org/10.1103/PhysRevD.88.023001} {\bibfield  {journal} {\bibinfo
  {journal} {\prd}\ }\textbf {\bibinfo {volume} {88}},\ \bibinfo {eid} {023001}
  (\bibinfo {year} {2013})},\ \Eprint {https://arxiv.org/abs/1304.4128}
  {arXiv:1304.4128 [astro-ph.HE]} \BibitemShut {NoStop}%
\bibitem [{\citenamefont {{Boudaud}}\ \emph {et~al.}(2015)\citenamefont
  {{Boudaud}}, \citenamefont {{Aupetit}}, \citenamefont {{Caroff}},
  \citenamefont {{Putze}}, \citenamefont {{Belanger}}, \citenamefont
  {{Genolini}}, \citenamefont {{Goy}}, \citenamefont {{Poireau}}, \citenamefont
  {{Poulin}}, \citenamefont {{Rosier}}, \citenamefont {{Salati}}, \citenamefont
  {{Tao}},\ and\ \citenamefont {{Vecchi}}}]{Boudaud2015aa}%
  \BibitemOpen
  \bibfield  {author} {\bibinfo {author} {M.~{Boudaud}}, \bibinfo {author}
  {S.~{Aupetit}}, \bibinfo {author} {S.~{Caroff}} et~al.,\ }\href
  {https://doi.org/10.1051/0004-6361/201425197} {\bibfield  {journal} {\bibinfo
   {journal} {\aap}\ }\textbf {\bibinfo {volume} {575}},\ \bibinfo {eid} {A67}
  (\bibinfo {year} {2015})},\ \Eprint {https://arxiv.org/abs/1410.3799}
  {arXiv:1410.3799 [astro-ph.HE]} \BibitemShut {NoStop}%
\bibitem [{\citenamefont {{L{\'o}pez-Coto}}\ \emph {et~al.}(2018)\citenamefont
  {{L{\'o}pez-Coto}}, \citenamefont {{Parsons}}, \citenamefont {{Hinton}},\
  and\ \citenamefont {{Giacinti}}}]{LopezCoto2018prl}%
  \BibitemOpen
  \bibfield  {author} {\bibinfo {author} {R.~{L{\'o}pez-Coto}}, \bibinfo
  {author} {R.~D. {Parsons}}, \bibinfo {author} {J.~A. {Hinton}} et~al.,\
  }\href {https://doi.org/10.1103/PhysRevLett.121.251106} {\bibfield  {journal}
  {\bibinfo  {journal} {\prl}\ }\textbf {\bibinfo {volume} {121}},\ \bibinfo
  {eid} {251106} (\bibinfo {year} {2018})},\ \Eprint
  {https://arxiv.org/abs/1811.04123} {arXiv:1811.04123 [astro-ph.HE]}
  \BibitemShut {NoStop}%
\bibitem [{\citenamefont {{Fornieri}}\ \emph {et~al.}(2020)\citenamefont
  {{Fornieri}}, \citenamefont {{Gaggero}},\ and\ \citenamefont
  {{Grasso}}}]{Fornieri2020jcap}%
  \BibitemOpen
  \bibfield  {author} {\bibinfo {author} {O.~{Fornieri}}, \bibinfo {author}
  {D.~{Gaggero}}\ and\ \bibinfo {author} {D.~{Grasso}},\ }\href
  {https://doi.org/10.1088/1475-7516/2020/02/009} {\bibfield  {journal}
  {\bibinfo  {journal} {\jcap}\ }\textbf {\bibinfo {volume} {2020}},\ \bibinfo
  {eid} {009} (\bibinfo {year} {2020})},\ \Eprint
  {https://arxiv.org/abs/1907.03696} {arXiv:1907.03696 [astro-ph.HE]}
  \BibitemShut {NoStop}%
\bibitem [{\citenamefont {{Lee}}(1979)}]{Lee1979apj}%
  \BibitemOpen
  \bibfield  {author} {\bibinfo {author} {M.~A. {Lee}},\ }\href
  {https://doi.org/10.1086/156970} {\bibfield  {journal} {\bibinfo  {journal}
  {\apj}\ }\textbf {\bibinfo {volume} {229}},\ \bibinfo {pages} {424} (\bibinfo
  {year} {1979})}\BibitemShut {NoStop}%
\bibitem [{\citenamefont {{Blasi}}\ and\ \citenamefont
  {{Amato}}(2012{\natexlab{a}})}]{Blasi2012composition}%
  \BibitemOpen
  \bibfield  {author} {\bibinfo {author} {P.~{Blasi}}\ and\ \bibinfo {author}
  {E.~{Amato}},\ }\href {https://doi.org/10.1088/1475-7516/2012/01/010}
  {\bibfield  {journal} {\bibinfo  {journal} {\jcap}\ }\textbf {\bibinfo
  {volume} {2012}},\ \bibinfo {eid} {010} (\bibinfo {year}
  {2012}{\natexlab{a}})},\ \Eprint {https://arxiv.org/abs/1105.4521}
  {arXiv:1105.4521 [astro-ph.HE]} \BibitemShut {NoStop}%
\bibitem [{\citenamefont {{Ptuskin}}\ \emph {et~al.}(2006)\citenamefont
  {{Ptuskin}}, \citenamefont {{Jones}}, \citenamefont {{Seo}},\ and\
  \citenamefont {{Sina}}}]{Ptuskin2006adspr}%
  \BibitemOpen
  \bibfield  {author} {\bibinfo {author} {V.~S. {Ptuskin}}, \bibinfo {author}
  {F.~C. {Jones}}, \bibinfo {author} {E.~S. {Seo}} et~al.,\ }\href
  {https://doi.org/10.1016/j.asr.2005.08.036} {\bibfield  {journal} {\bibinfo
  {journal} {Advances in Space Research}\ }\textbf {\bibinfo {volume} {37}},\
  \bibinfo {pages} {1909} (\bibinfo {year} {2006})}\BibitemShut {NoStop}%
\bibitem [{\citenamefont {{Mertsch}}(2011)}]{Mertsch2011jcap}%
  \BibitemOpen
  \bibfield  {author} {\bibinfo {author} {P.~{Mertsch}},\ }\href
  {https://doi.org/10.1088/1475-7516/2011/02/031} {\bibfield  {journal}
  {\bibinfo  {journal} {\jcap}\ }\textbf {\bibinfo {volume} {2011}},\ \bibinfo
  {eid} {031} (\bibinfo {year} {2011})},\ \Eprint
  {https://arxiv.org/abs/1012.0805} {arXiv:1012.0805 [astro-ph.HE]}
  \BibitemShut {NoStop}%
\bibitem [{\citenamefont {{Bernard}}\ \emph {et~al.}(2012)\citenamefont
  {{Bernard}}, \citenamefont {{Delahaye}}, \citenamefont {{Salati}},\ and\
  \citenamefont {{Taillet}}}]{Bernard2012aa}%
  \BibitemOpen
  \bibfield  {author} {\bibinfo {author} {G.~{Bernard}}, \bibinfo {author}
  {T.~{Delahaye}}, \bibinfo {author} {P.~{Salati}} et~al.,\ }\href
  {https://doi.org/10.1051/0004-6361/201219502} {\bibfield  {journal} {\bibinfo
   {journal} {\aap}\ }\textbf {\bibinfo {volume} {544}},\ \bibinfo {eid} {A92}
  (\bibinfo {year} {2012})},\ \Eprint {https://arxiv.org/abs/1204.6289}
  {arXiv:1204.6289 [astro-ph.HE]} \BibitemShut {NoStop}%
\bibitem [{\citenamefont {{Mertsch}}(2018)}]{Mertsch2018jcap}%
  \BibitemOpen
  \bibfield  {author} {\bibinfo {author} {P.~{Mertsch}},\ }\href
  {https://doi.org/10.1088/1475-7516/2018/11/045} {\bibfield  {journal}
  {\bibinfo  {journal} {\jcap}\ }\textbf {\bibinfo {volume} {2018}},\ \bibinfo
  {eid} {045} (\bibinfo {year} {2018})},\ \Eprint
  {https://arxiv.org/abs/1809.05104} {arXiv:1809.05104 [astro-ph.HE]}
  \BibitemShut {NoStop}%
\bibitem [{\citenamefont {{Lagutin}}\ and\ \citenamefont
  {{Nikulin}}(1995)}]{Lagutin1995jetp}%
  \BibitemOpen
  \bibfield  {author} {\bibinfo {author} {A.~A. {Lagutin}}\ and\ \bibinfo
  {author} {Y.~A. {Nikulin}},\ }\href@noop {} {\bibfield  {journal} {\bibinfo
  {journal} {Soviet Journal of Experimental and Theoretical Physics}\ }\textbf
  {\bibinfo {volume} {81}},\ \bibinfo {pages} {825} (\bibinfo {year}
  {1995})}\BibitemShut {NoStop}%
\bibitem [{\citenamefont {{G{\'e}nolini}}\ \emph {et~al.}(2017)\citenamefont
  {{G{\'e}nolini}}, \citenamefont {{Serpico}}, \citenamefont {{Boudaud}},
  \citenamefont {{Caroff}}, \citenamefont {{Poulin}}, \citenamefont {{Derome}},
  \citenamefont {{Lavalle}}, \citenamefont {{Maurin}}, \citenamefont
  {{Poireau}}, \citenamefont {{Rosier}}, \citenamefont {{Salati}},\ and\
  \citenamefont {{Vecchi}}}]{Genolini2017prl}%
  \BibitemOpen
  \bibfield  {author} {\bibinfo {author} {Y.~{G{\'e}nolini}}, \bibinfo {author}
  {P.~D. {Serpico}}, \bibinfo {author} {M.~{Boudaud}} et~al.,\ }\href
  {https://doi.org/10.1103/PhysRevLett.119.241101} {\bibfield  {journal}
  {\bibinfo  {journal} {\prl}\ }\textbf {\bibinfo {volume} {119}},\ \bibinfo
  {eid} {241101} (\bibinfo {year} {2017})}\BibitemShut {NoStop}%
\bibitem [{\citenamefont {{Blasi}}\ and\ \citenamefont
  {{Amato}}(2012{\natexlab{b}})}]{Blasi2012anisotropy}%
  \BibitemOpen
  \bibfield  {author} {\bibinfo {author} {P.~{Blasi}}\ and\ \bibinfo {author}
  {E.~{Amato}},\ }\href {https://doi.org/10.1088/1475-7516/2012/01/011}
  {\bibfield  {journal} {\bibinfo  {journal} {\jcap}\ }\textbf {\bibinfo
  {volume} {2012}},\ \bibinfo {eid} {011} (\bibinfo {year}
  {2012}{\natexlab{b}})},\ \Eprint {https://arxiv.org/abs/1105.4529}
  {arXiv:1105.4529 [astro-ph.HE]} \BibitemShut {NoStop}%
\bibitem [{\citenamefont {{Ahlers}}(2016)}]{Ahlers2016prl}%
  \BibitemOpen
  \bibfield  {author} {\bibinfo {author} {M.~{Ahlers}},\ }\href
  {https://doi.org/10.1103/PhysRevLett.117.151103} {\bibfield  {journal}
  {\bibinfo  {journal} {\prl}\ }\textbf {\bibinfo {volume} {117}},\ \bibinfo
  {eid} {151103} (\bibinfo {year} {2016})},\ \Eprint
  {https://arxiv.org/abs/1605.06446} {arXiv:1605.06446 [astro-ph.HE]}
  \BibitemShut {NoStop}%
\bibitem [{\citenamefont {{Dragicevich}}\ \emph {et~al.}(1999)\citenamefont
  {{Dragicevich}}, \citenamefont {{Blair}},\ and\ \citenamefont
  {{Burman}}}]{Dragicevich1999mnras}%
  \BibitemOpen
  \bibfield  {author} {\bibinfo {author} {P.~M. {Dragicevich}}, \bibinfo
  {author} {D.~G. {Blair}}\ and\ \bibinfo {author} {R.~R. {Burman}},\ }\href
  {https://doi.org/10.1046/j.1365-8711.1999.02145.x} {\bibfield  {journal}
  {\bibinfo  {journal} {\mnras}\ }\textbf {\bibinfo {volume} {302}},\ \bibinfo
  {pages} {693} (\bibinfo {year} {1999})}\BibitemShut {NoStop}%
\bibitem [{\citenamefont {{Rozwadowska}}\ \emph {et~al.}(2021)\citenamefont
  {{Rozwadowska}}, \citenamefont {{Vissani}},\ and\ \citenamefont
  {{Cappellaro}}}]{Rozwadowska2021newa}%
  \BibitemOpen
  \bibfield  {author} {\bibinfo {author} {K.~{Rozwadowska}}, \bibinfo {author}
  {F.~{Vissani}}\ and\ \bibinfo {author} {E.~{Cappellaro}},\ }\href
  {https://doi.org/10.1016/j.newast.2020.101498} {\bibfield  {journal}
  {\bibinfo  {journal} {\na}\ }\textbf {\bibinfo {volume} {83}},\ \bibinfo
  {eid} {101498} (\bibinfo {year} {2021})},\ \Eprint
  {https://arxiv.org/abs/2009.03438} {arXiv:2009.03438 [astro-ph.HE]}
  \BibitemShut {NoStop}%
\bibitem [{\citenamefont {{Strong}}\ and\ \citenamefont
  {{Moskalenko}}(1998)}]{Moskalenko1998apja}%
  \BibitemOpen
  \bibfield  {author} {\bibinfo {author} {A.~W. {Strong}}\ and\ \bibinfo
  {author} {I.~V. {Moskalenko}},\ }\href {https://doi.org/10.1086/306470}
  {\bibfield  {journal} {\bibinfo  {journal} {\apj}\ }\textbf {\bibinfo
  {volume} {509}},\ \bibinfo {pages} {212} (\bibinfo {year} {1998})},\ \Eprint
  {https://arxiv.org/abs/astro-ph/9807150} {arXiv:astro-ph/9807150 [astro-ph]}
  \BibitemShut {NoStop}%
\bibitem [{\citenamefont {{Moskalenko}}\ and\ \citenamefont
  {{Strong}}(1998)}]{Moskalenko1998apjb}%
  \BibitemOpen
  \bibfield  {author} {\bibinfo {author} {I.~V. {Moskalenko}}\ and\ \bibinfo
  {author} {A.~W. {Strong}},\ }\href {https://doi.org/10.1086/305152}
  {\bibfield  {journal} {\bibinfo  {journal} {\apj}\ }\textbf {\bibinfo
  {volume} {493}},\ \bibinfo {pages} {694} (\bibinfo {year} {1998})},\ \Eprint
  {https://arxiv.org/abs/astro-ph/9710124} {arXiv:astro-ph/9710124 [astro-ph]}
  \BibitemShut {NoStop}%
\bibitem [{\citenamefont {{Evoli}}\ \emph {et~al.}(2008)\citenamefont
  {{Evoli}}, \citenamefont {{Gaggero}}, \citenamefont {{Grasso}},\ and\
  \citenamefont {{Maccione}}}]{Evoli2008jcap}%
  \BibitemOpen
  \bibfield  {author} {\bibinfo {author} {C.~{Evoli}}, \bibinfo {author}
  {D.~{Gaggero}}, \bibinfo {author} {D.~{Grasso}} et~al.,\ }\href
  {https://doi.org/10.1088/1475-7516/2008/10/018} {\bibfield  {journal}
  {\bibinfo  {journal} {\jcap}\ }\textbf {\bibinfo {volume} {2008}},\ \bibinfo
  {eid} {018} (\bibinfo {year} {2008})},\ \Eprint
  {https://arxiv.org/abs/0807.4730} {arXiv:0807.4730 [astro-ph]} \BibitemShut
  {NoStop}%
\bibitem [{\citenamefont {{Evoli}}\ \emph
  {et~al.}(2020{\natexlab{b}})\citenamefont {{Evoli}}, \citenamefont
  {{Morlino}}, \citenamefont {{Blasi}},\ and\ \citenamefont
  {{Aloisio}}}]{Evoli2020prd}%
  \BibitemOpen
  \bibfield  {author} {\bibinfo {author} {C.~{Evoli}}, \bibinfo {author}
  {G.~{Morlino}}, \bibinfo {author} {P.~{Blasi}} et~al.,\ }\href
  {https://doi.org/10.1103/PhysRevD.101.023013} {\bibfield  {journal} {\bibinfo
   {journal} {\prd}\ }\textbf {\bibinfo {volume} {101}},\ \bibinfo {eid}
  {023013} (\bibinfo {year} {2020}{\natexlab{b}})},\ \Eprint
  {https://arxiv.org/abs/1910.04113} {arXiv:1910.04113 [astro-ph.HE]}
  \BibitemShut {NoStop}%
\bibitem [{\citenamefont {{Weinrich}}\ \emph
  {et~al.}(2020{\natexlab{a}})\citenamefont {{Weinrich}}, \citenamefont
  {{Boudaud}}, \citenamefont {{Derome}}, \citenamefont {{G{\'e}nolini}},
  \citenamefont {{Lavalle}}, \citenamefont {{Maurin}}, \citenamefont
  {{Salati}}, \citenamefont {{Serpico}},\ and\ \citenamefont
  {{Weymann-Despres}}}]{Weinrich2020aaa}%
  \BibitemOpen
  \bibfield  {author} {\bibinfo {author} {N.~{Weinrich}}, \bibinfo {author}
  {M.~{Boudaud}}, \bibinfo {author} {L.~{Derome}} et~al.,\ }\href
  {https://doi.org/10.1051/0004-6361/202038064} {\bibfield  {journal} {\bibinfo
   {journal} {\aap}\ }\textbf {\bibinfo {volume} {639}},\ \bibinfo {eid} {A74}
  (\bibinfo {year} {2020}{\natexlab{a}})},\ \Eprint
  {https://arxiv.org/abs/2004.00441} {arXiv:2004.00441 [astro-ph.HE]}
  \BibitemShut {NoStop}%
\bibitem [{\citenamefont {{Weinrich}}\ \emph
  {et~al.}(2020{\natexlab{b}})\citenamefont {{Weinrich}}, \citenamefont
  {{G{\'e}nolini}}, \citenamefont {{Boudaud}}, \citenamefont {{Derome}},\ and\
  \citenamefont {{Maurin}}}]{Weinrich2020aab}%
  \BibitemOpen
  \bibfield  {author} {\bibinfo {author} {N.~{Weinrich}}, \bibinfo {author}
  {Y.~{G{\'e}nolini}}, \bibinfo {author} {M.~{Boudaud}} et~al.,\ }\href
  {https://doi.org/10.1051/0004-6361/202037875} {\bibfield  {journal} {\bibinfo
   {journal} {\aap}\ }\textbf {\bibinfo {volume} {639}},\ \bibinfo {eid} {A131}
  (\bibinfo {year} {2020}{\natexlab{b}})},\ \Eprint
  {https://arxiv.org/abs/2002.11406} {arXiv:2002.11406 [astro-ph.HE]}
  \BibitemShut {NoStop}%
\bibitem [{\citenamefont {{Atoyan}}\ \emph {et~al.}(1995)\citenamefont
  {{Atoyan}}, \citenamefont {{Aharonian}},\ and\ \citenamefont
  {{V{\"o}lk}}}]{Atoyan1995prd}%
  \BibitemOpen
  \bibfield  {author} {\bibinfo {author} {A.~M. {Atoyan}}, \bibinfo {author}
  {F.~A. {Aharonian}}\ and\ \bibinfo {author} {H.~J. {V{\"o}lk}},\ }\href
  {https://doi.org/10.1103/PhysRevD.52.3265} {\bibfield  {journal} {\bibinfo
  {journal} {\prd}\ }\textbf {\bibinfo {volume} {52}},\ \bibinfo {pages} {3265}
  (\bibinfo {year} {1995})}\BibitemShut {NoStop}%
\bibitem [{\citenamefont {{Pohl}}\ and\ \citenamefont
  {{Eichler}}(2013)}]{Pohl2013apj}%
  \BibitemOpen
  \bibfield  {author} {\bibinfo {author} {M.~{Pohl}}\ and\ \bibinfo {author}
  {D.~{Eichler}},\ }\href {https://doi.org/10.1088/0004-637X/766/1/4}
  {\bibfield  {journal} {\bibinfo  {journal} {\apj}\ }\textbf {\bibinfo
  {volume} {766}},\ \bibinfo {eid} {4} (\bibinfo {year} {2013})},\ \Eprint
  {https://arxiv.org/abs/1208.5338} {arXiv:1208.5338 [astro-ph.HE]}
  \BibitemShut {NoStop}%
\bibitem [{\citenamefont {{Manconi}}\ \emph {et~al.}(2020)\citenamefont
  {{Manconi}}, \citenamefont {{Di Mauro}},\ and\ \citenamefont
  {{Donato}}}]{Manconi2020prd}%
  \BibitemOpen
  \bibfield  {author} {\bibinfo {author} {S.~{Manconi}}, \bibinfo {author}
  {M.~{Di Mauro}}\ and\ \bibinfo {author} {F.~{Donato}},\ }\href
  {https://doi.org/10.1103/PhysRevD.102.023015} {\bibfield  {journal} {\bibinfo
   {journal} {\prd}\ }\textbf {\bibinfo {volume} {102}},\ \bibinfo {eid}
  {023015} (\bibinfo {year} {2020})},\ \Eprint
  {https://arxiv.org/abs/2001.09985} {arXiv:2001.09985 [astro-ph.HE]}
  \BibitemShut {NoStop}%
\bibitem [{\citenamefont {{Faucher-Gigu{\`e}re}}\ and\ \citenamefont
  {{Kaspi}}(2006)}]{FaucherGiguere2006apj}%
  \BibitemOpen
  \bibfield  {author} {\bibinfo {author} {C.-A. {Faucher-Gigu{\`e}re}}\ and\
  \bibinfo {author} {V.~M. {Kaspi}},\ }\href {https://doi.org/10.1086/501516}
  {\bibfield  {journal} {\bibinfo  {journal} {\apj}\ }\textbf {\bibinfo
  {volume} {643}},\ \bibinfo {pages} {332} (\bibinfo {year} {2006})},\ \Eprint
  {https://arxiv.org/abs/astro-ph/0512585} {arXiv:astro-ph/0512585 [astro-ph]}
  \BibitemShut {NoStop}%
\bibitem [{\citenamefont {{Berezinskii}}\ \emph {et~al.}(1990)\citenamefont
  {{Berezinskii}}, \citenamefont {{Bulanov}}, \citenamefont {{Dogiel}},\ and\
  \citenamefont {{Ptuskin}}}]{TheBible}%
  \BibitemOpen
  \bibfield  {author} {\bibinfo {author} {V.~S. {Berezinskii}}, \bibinfo
  {author} {S.~V. {Bulanov}}, \bibinfo {author} {V.~A. {Dogiel}} et~al.,\
  }\href@noop {} {\emph {\bibinfo {title} {{Astrophysics of cosmic rays}}}}\
  (\bibinfo  {publisher} {{North-Holland}},\ \bibinfo {year}
  {1990})\BibitemShut {NoStop}%
\bibitem [{\citenamefont {{Lorimer}}\ \emph {et~al.}(2006)\citenamefont
  {{Lorimer}}, \citenamefont {{Faulkner}}, \citenamefont {{Lyne}},
  \citenamefont {{Manchester}}, \citenamefont {{Kramer}}, \citenamefont
  {{McLaughlin}}, \citenamefont {{Hobbs}}, \citenamefont {{Possenti}},
  \citenamefont {{Stairs}}, \citenamefont {{Camilo}}, \citenamefont {{Burgay}},
  \citenamefont {{D'Amico}}, \citenamefont {{Corongiu}},\ and\ \citenamefont
  {{Crawford}}}]{Lorimer2006mnras}%
  \BibitemOpen
  \bibfield  {author} {\bibinfo {author} {D.~R. {Lorimer}}, \bibinfo {author}
  {A.~J. {Faulkner}}, \bibinfo {author} {A.~G. {Lyne}} et~al.,\ }\href
  {https://doi.org/10.1111/j.1365-2966.2006.10887.x} {\bibfield  {journal}
  {\bibinfo  {journal} {\mnras}\ }\textbf {\bibinfo {volume} {372}},\ \bibinfo
  {pages} {777} (\bibinfo {year} {2006})},\ \Eprint
  {https://arxiv.org/abs/astro-ph/0607640} {arXiv:astro-ph/0607640 [astro-ph]}
  \BibitemShut {NoStop}%
\bibitem [{\citenamefont {{Nogueras-Lara}}\ \emph {et~al.}(2021)\citenamefont
  {{Nogueras-Lara}}, \citenamefont {{Sch{\"o}del}},\ and\ \citenamefont
  {{Neumayer}}}]{NoguerasLara2021aa}%
  \BibitemOpen
  \bibfield  {author} {\bibinfo {author} {F.~{Nogueras-Lara}}, \bibinfo
  {author} {R.~{Sch{\"o}del}}\ and\ \bibinfo {author} {N.~{Neumayer}},\ }\href
  {https://doi.org/10.1051/0004-6361/202040073} {\bibfield  {journal} {\bibinfo
   {journal} {\aap}\ }\textbf {\bibinfo {volume} {653}},\ \bibinfo {eid} {A33}
  (\bibinfo {year} {2021})},\ \Eprint {https://arxiv.org/abs/2106.04529}
  {arXiv:2106.04529 [astro-ph.GA]} \BibitemShut {NoStop}%
\bibitem [{\citenamefont {{Phan}}\ \emph {et~al.}(2021)\citenamefont {{Phan}},
  \citenamefont {{Schulze}}, \citenamefont {{Mertsch}}, \citenamefont
  {{Recchia}},\ and\ \citenamefont {{Gabici}}}]{Phan2021prl}%
  \BibitemOpen
  \bibfield  {author} {\bibinfo {author} {V.~H.~M. {Phan}}, \bibinfo {author}
  {F.~{Schulze}}, \bibinfo {author} {P.~{Mertsch}} et~al.,\ }\href
  {https://doi.org/10.1103/PhysRevLett.127.141101} {\bibfield  {journal}
  {\bibinfo  {journal} {\prl}\ }\textbf {\bibinfo {volume} {127}},\ \bibinfo
  {eid} {141101} (\bibinfo {year} {2021})},\ \Eprint
  {https://arxiv.org/abs/2105.00311} {arXiv:2105.00311 [astro-ph.HE]}
  \BibitemShut {NoStop}%
\bibitem [{\citenamefont {{Evoli}}\ \emph
  {et~al.}(2018{\natexlab{b}})\citenamefont {{Evoli}}, \citenamefont
  {{Gaggero}}, \citenamefont {{Vittino}}, \citenamefont {{Di Mauro}},
  \citenamefont {{Grasso}},\ and\ \citenamefont {{Mazziotta}}}]{Evoli2018jcap}%
  \BibitemOpen
  \bibfield  {author} {\bibinfo {author} {C.~{Evoli}}, \bibinfo {author}
  {D.~{Gaggero}}, \bibinfo {author} {A.~{Vittino}} et~al.,\ }\href
  {https://doi.org/10.1088/1475-7516/2018/07/006} {\bibfield  {journal}
  {\bibinfo  {journal} {\jcap}\ }\textbf {\bibinfo {volume} {2018}},\ \bibinfo
  {eid} {006} (\bibinfo {year} {2018}{\natexlab{b}})},\ \Eprint
  {https://arxiv.org/abs/1711.09616} {arXiv:1711.09616 [astro-ph.HE]}
  \BibitemShut {NoStop}%
\bibitem [{\citenamefont {{Ferri{\`e}re}}(2001)}]{Ferriere2001review}%
  \BibitemOpen
  \bibfield  {author} {\bibinfo {author} {K.~M. {Ferri{\`e}re}},\ }\href
  {https://doi.org/10.1103/RevModPhys.73.1031} {\bibfield  {journal} {\bibinfo
  {journal} {Reviews of Modern Physics}\ }\textbf {\bibinfo {volume} {73}},\
  \bibinfo {pages} {1031} (\bibinfo {year} {2001})},\ \Eprint
  {https://arxiv.org/abs/astro-ph/0106359} {arXiv:astro-ph/0106359 [astro-ph]}
  \BibitemShut {NoStop}%
\bibitem [{\citenamefont {{Kafexhiu}}\ \emph {et~al.}(2014)\citenamefont
  {{Kafexhiu}}, \citenamefont {{Aharonian}}, \citenamefont {{Taylor}},\ and\
  \citenamefont {{Vila}}}]{Kafexhiu2014prd}%
  \BibitemOpen
  \bibfield  {author} {\bibinfo {author} {E.~{Kafexhiu}}, \bibinfo {author}
  {F.~{Aharonian}}, \bibinfo {author} {A.~M. {Taylor}} et~al.,\ }\href
  {https://doi.org/10.1103/PhysRevD.90.123014} {\bibfield  {journal} {\bibinfo
  {journal} {\prd}\ }\textbf {\bibinfo {volume} {90}},\ \bibinfo {eid} {123014}
  (\bibinfo {year} {2014})},\ \Eprint {https://arxiv.org/abs/1406.7369}
  {arXiv:1406.7369 [astro-ph.HE]} \BibitemShut {NoStop}%
\bibitem [{\citenamefont {{Coste}}\ \emph {et~al.}(2012)\citenamefont
  {{Coste}}, \citenamefont {{Derome}}, \citenamefont {{Maurin}},\ and\
  \citenamefont {{Putze}}}]{Coste2012aa}%
  \BibitemOpen
  \bibfield  {author} {\bibinfo {author} {B.~{Coste}}, \bibinfo {author}
  {L.~{Derome}}, \bibinfo {author} {D.~{Maurin}} et~al.,\ }\href
  {https://doi.org/10.1051/0004-6361/201117927} {\bibfield  {journal} {\bibinfo
   {journal} {\aap}\ }\textbf {\bibinfo {volume} {539}},\ \bibinfo {eid} {A88}
  (\bibinfo {year} {2012})},\ \Eprint {https://arxiv.org/abs/1108.4349}
  {arXiv:1108.4349 [astro-ph.GA]} \BibitemShut {NoStop}%
\bibitem [{\citenamefont {{Planck Collaboration}}\ \emph
  {et~al.}(2020)\citenamefont {{Planck Collaboration}}, \citenamefont
  {{Aghanim}}, \citenamefont {{Akrami}}, \citenamefont {{Ashdown}},
  \citenamefont {{Aumont}}, \citenamefont {{Baccigalupi}}, \citenamefont
  {{Ballardini}}, \citenamefont {{Banday}}, \citenamefont {{Barreiro}},
  \citenamefont {{Bartolo}}, \citenamefont {{Basak}}, \citenamefont {{Battye}},
  \citenamefont {{Benabed}}, \citenamefont {{Bernard}}, \citenamefont
  {{Bersanelli}}, \citenamefont {{Bielewicz}}, \citenamefont {{Bock}},
  \citenamefont {{Bond}}, \citenamefont {{Borrill}}, \citenamefont {{Bouchet}},
  \citenamefont {{Boulanger}}, \citenamefont {{Bucher}}, \citenamefont
  {{Burigana}}, \citenamefont {{Butler}}, \citenamefont {{Calabrese}},
  \citenamefont {{Cardoso}}, \citenamefont {{Carron}}, \citenamefont
  {{Challinor}}, \citenamefont {{Chiang}}, \citenamefont {{Chluba}},
  \citenamefont {{Colombo}}, \citenamefont {{Combet}}, \citenamefont
  {{Contreras}}, \citenamefont {{Crill}}, \citenamefont {{Cuttaia}},
  \citenamefont {{de Bernardis}}, \citenamefont {{de Zotti}}, \citenamefont
  {{Delabrouille}}, \citenamefont {{Delouis}}, \citenamefont {{Di Valentino}},
  \citenamefont {{Diego}}, \citenamefont {{Dor{\'e}}}, \citenamefont
  {{Douspis}}, \citenamefont {{Ducout}}, \citenamefont {{Dupac}}, \citenamefont
  {{Dusini}}, \citenamefont {{Efstathiou}}, \citenamefont {{Elsner}},
  \citenamefont {{En{\ss}lin}}, \citenamefont {{Eriksen}}, \citenamefont
  {{Fantaye}}, \citenamefont {{Farhang}}, \citenamefont {{Fergusson}},
  \citenamefont {{Fernandez-Cobos}}, \citenamefont {{Finelli}}, \citenamefont
  {{Forastieri}}, \citenamefont {{Frailis}}, \citenamefont {{Fraisse}},
  \citenamefont {{Franceschi}}, \citenamefont {{Frolov}}, \citenamefont
  {{Galeotta}}, \citenamefont {{Galli}}, \citenamefont {{Ganga}}, \citenamefont
  {{G{\'e}nova-Santos}}, \citenamefont {{Gerbino}}, \citenamefont {{Ghosh}},
  \citenamefont {{Gonz{\'a}lez-Nuevo}}, \citenamefont {{G{\'o}rski}},
  \citenamefont {{Gratton}}, \citenamefont {{Gruppuso}}, \citenamefont
  {{Gudmundsson}}, \citenamefont {{Hamann}}, \citenamefont {{Handley}},
  \citenamefont {{Hansen}}, \citenamefont {{Herranz}}, \citenamefont
  {{Hildebrandt}}, \citenamefont {{Hivon}}, \citenamefont {{Huang}},
  \citenamefont {{Jaffe}}, \citenamefont {{Jones}}, \citenamefont {{Karakci}},
  \citenamefont {{Keih{\"a}nen}}, \citenamefont {{Keskitalo}}, \citenamefont
  {{Kiiveri}}, \citenamefont {{Kim}}, \citenamefont {{Kisner}}, \citenamefont
  {{Knox}}, \citenamefont {{Krachmalnicoff}}, \citenamefont {{Kunz}},
  \citenamefont {{Kurki-Suonio}}, \citenamefont {{Lagache}}, \citenamefont
  {{Lamarre}}, \citenamefont {{Lasenby}}, \citenamefont {{Lattanzi}},
  \citenamefont {{Lawrence}}, \citenamefont {{Le Jeune}}, \citenamefont
  {{Lemos}}, \citenamefont {{Lesgourgues}}, \citenamefont {{Levrier}},
  \citenamefont {{Lewis}}, \citenamefont {{Liguori}}, \citenamefont {{Lilje}},
  \citenamefont {{Lilley}}, \citenamefont {{Lindholm}}, \citenamefont
  {{L{\'o}pez-Caniego}}, \citenamefont {{Lubin}}, \citenamefont {{Ma}},
  \citenamefont {{Mac{\'\i}as-P{\'e}rez}}, \citenamefont {{Maggio}},
  \citenamefont {{Maino}}, \citenamefont {{Mandolesi}}, \citenamefont
  {{Mangilli}}, \citenamefont {{Marcos-Caballero}}, \citenamefont {{Maris}},
  \citenamefont {{Martin}}, \citenamefont {{Martinelli}}, \citenamefont
  {{Mart{\'\i}nez-Gonz{\'a}lez}}, \citenamefont {{Matarrese}}, \citenamefont
  {{Mauri}}, \citenamefont {{McEwen}}, \citenamefont {{Meinhold}},
  \citenamefont {{Melchiorri}}, \citenamefont {{Mennella}}, \citenamefont
  {{Migliaccio}}, \citenamefont {{Millea}}, \citenamefont {{Mitra}},
  \citenamefont {{Miville-Desch{\^e}nes}}, \citenamefont {{Molinari}},
  \citenamefont {{Montier}}, \citenamefont {{Morgante}}, \citenamefont
  {{Moss}}, \citenamefont {{Natoli}}, \citenamefont {{N{\o}rgaard-Nielsen}},
  \citenamefont {{Pagano}}, \citenamefont {{Paoletti}}, \citenamefont
  {{Partridge}}, \citenamefont {{Patanchon}}, \citenamefont {{Peiris}},
  \citenamefont {{Perrotta}}, \citenamefont {{Pettorino}}, \citenamefont
  {{Piacentini}}, \citenamefont {{Polastri}}, \citenamefont {{Polenta}},
  \citenamefont {{Puget}}, \citenamefont {{Rachen}}, \citenamefont
  {{Reinecke}}, \citenamefont {{Remazeilles}}, \citenamefont {{Renzi}},
  \citenamefont {{Rocha}}, \citenamefont {{Rosset}}, \citenamefont {{Roudier}},
  \citenamefont {{Rubi{\~n}o-Mart{\'\i}n}}, \citenamefont {{Ruiz-Granados}},
  \citenamefont {{Salvati}}, \citenamefont {{Sandri}}, \citenamefont
  {{Savelainen}}, \citenamefont {{Scott}}, \citenamefont {{Shellard}},
  \citenamefont {{Sirignano}}, \citenamefont {{Sirri}}, \citenamefont
  {{Spencer}}, \citenamefont {{Sunyaev}}, \citenamefont {{Suur-Uski}},
  \citenamefont {{Tauber}}, \citenamefont {{Tavagnacco}}, \citenamefont
  {{Tenti}}, \citenamefont {{Toffolatti}}, \citenamefont {{Tomasi}},
  \citenamefont {{Trombetti}}, \citenamefont {{Valenziano}}, \citenamefont
  {{Valiviita}}, \citenamefont {{Van Tent}}, \citenamefont {{Vibert}},
  \citenamefont {{Vielva}}, \citenamefont {{Villa}}, \citenamefont
  {{Vittorio}}, \citenamefont {{Wandelt}}, \citenamefont {{Wehus}},
  \citenamefont {{White}}, \citenamefont {{White}}, \citenamefont {{Zacchei}},\
  and\ \citenamefont {{Zonca}}}]{Planck2020aa}%
  \BibitemOpen
  \bibfield  {author} {\bibinfo {author} {{Planck Collaboration}}, \bibinfo
  {author} {N.~{Aghanim}}, \bibinfo {author} {Y.~{Akrami}} et~al.,\ }\href
  {https://doi.org/10.1051/0004-6361/201833910} {\bibfield  {journal} {\bibinfo
   {journal} {\aap}\ }\textbf {\bibinfo {volume} {641}},\ \bibinfo {eid} {A6}
  (\bibinfo {year} {2020})},\ \Eprint {https://arxiv.org/abs/1807.06209}
  {arXiv:1807.06209 [astro-ph.CO]} \BibitemShut {NoStop}%
\bibitem [{\citenamefont {{Porter}}\ and\ \citenamefont
  {{Strong}}(2005)}]{Porter2005}%
  \BibitemOpen
  \bibfield  {author} {\bibinfo {author} {T.~A. {Porter}}\ and\ \bibinfo
  {author} {A.~W. {Strong}},\ }in\ \href@noop {} {\emph {\bibinfo {booktitle}
  {29th International Cosmic Ray Conference (ICRC29), Volume 4}}},\ \bibinfo
  {series} {International Cosmic Ray Conference}, Vol.~\bibinfo {volume} {4}\
  (\bibinfo {year} {2005})\ p.~\bibinfo {pages} {77},\ \Eprint
  {https://arxiv.org/abs/astro-ph/0507119} {arXiv:astro-ph/0507119 [astro-ph]}
  \BibitemShut {NoStop}%
\bibitem [{\citenamefont {{Popescu}}\ \emph {et~al.}(2017)\citenamefont
  {{Popescu}}, \citenamefont {{Yang}}, \citenamefont {{Tuffs}}, \citenamefont
  {{Natale}}, \citenamefont {{Rushton}},\ and\ \citenamefont
  {{Aharonian}}}]{Popescu2017mnras}%
  \BibitemOpen
  \bibfield  {author} {\bibinfo {author} {C.~C. {Popescu}}, \bibinfo {author}
  {R.~{Yang}}, \bibinfo {author} {R.~J. {Tuffs}} et~al.,\ }\href
  {https://doi.org/10.1093/mnras/stx1282} {\bibfield  {journal} {\bibinfo
  {journal} {\mnras}\ }\textbf {\bibinfo {volume} {470}},\ \bibinfo {pages}
  {2539} (\bibinfo {year} {2017})},\ \Eprint {https://arxiv.org/abs/1705.06652}
  {arXiv:1705.06652 [astro-ph.GA]} \BibitemShut {NoStop}%
\bibitem [{\citenamefont {{Zirakashvili}}\ and\ \citenamefont
  {{Aharonian}}(2007)}]{Zirakashvili2007aa}%
  \BibitemOpen
  \bibfield  {author} {\bibinfo {author} {V.~N. {Zirakashvili}}\ and\ \bibinfo
  {author} {F.~{Aharonian}},\ }\href
  {https://doi.org/10.1051/0004-6361:20066494} {\bibfield  {journal} {\bibinfo
  {journal} {\aap}\ }\textbf {\bibinfo {volume} {465}},\ \bibinfo {pages} {695}
  (\bibinfo {year} {2007})},\ \Eprint {https://arxiv.org/abs/astro-ph/0612717}
  {arXiv:astro-ph/0612717 [astro-ph]} \BibitemShut {NoStop}%
\bibitem [{\citenamefont {{Blasi}}(2010)}]{Blasi2010mnras}%
  \BibitemOpen
  \bibfield  {author} {\bibinfo {author} {P.~{Blasi}},\ }\href
  {https://doi.org/10.1111/j.1365-2966.2009.16110.x} {\bibfield  {journal}
  {\bibinfo  {journal} {\mnras}\ }\textbf {\bibinfo {volume} {402}},\ \bibinfo
  {pages} {2807} (\bibinfo {year} {2010})},\ \Eprint
  {https://arxiv.org/abs/0912.2053} {arXiv:0912.2053 [astro-ph.HE]}
  \BibitemShut {NoStop}%
\bibitem [{\citenamefont {{Vink}}(2012)}]{Vink2012aarv}%
  \BibitemOpen
  \bibfield  {author} {\bibinfo {author} {J.~{Vink}},\ }\href
  {https://doi.org/10.1007/s00159-011-0049-1} {\bibfield  {journal} {\bibinfo
  {journal} {\aapr}\ }\textbf {\bibinfo {volume} {20}},\ \bibinfo {eid} {49}
  (\bibinfo {year} {2012})},\ \Eprint {https://arxiv.org/abs/1112.0576}
  {arXiv:1112.0576 [astro-ph.HE]} \BibitemShut {NoStop}%
\bibitem [{\citenamefont {{Amato}}\ and\ \citenamefont
  {{Blasi}}(2018)}]{AmatoBlasi2018}%
  \BibitemOpen
  \bibfield  {author} {\bibinfo {author} {E.~{Amato}}\ and\ \bibinfo {author}
  {P.~{Blasi}},\ }\href {https://doi.org/10.1016/j.asr.2017.04.019} {\bibfield
  {journal} {\bibinfo  {journal} {Advances in Space Research}\ }\textbf
  {\bibinfo {volume} {62}},\ \bibinfo {pages} {2731} (\bibinfo {year}
  {2018})},\ \Eprint {https://arxiv.org/abs/1704.05696} {arXiv:1704.05696
  [astro-ph.HE]} \BibitemShut {NoStop}%
\bibitem [{\citenamefont {{Bykov}}\ \emph {et~al.}(2017)\citenamefont
  {{Bykov}}, \citenamefont {{Amato}}, \citenamefont {{Petrov}}, \citenamefont
  {{Krassilchtchikov}},\ and\ \citenamefont {{Levenfish}}}]{Bykov2017ssrv}%
  \BibitemOpen
  \bibfield  {author} {\bibinfo {author} {A.~M. {Bykov}}, \bibinfo {author}
  {E.~{Amato}}, \bibinfo {author} {A.~E. {Petrov}} et~al.,\ }\href
  {https://doi.org/10.1007/s11214-017-0371-7} {\bibfield  {journal} {\bibinfo
  {journal} {\ssr}\ }\textbf {\bibinfo {volume} {207}},\ \bibinfo {pages} {235}
  (\bibinfo {year} {2017})},\ \Eprint {https://arxiv.org/abs/1705.00950}
  {arXiv:1705.00950 [astro-ph.HE]} \BibitemShut {NoStop}%
\bibitem [{\citenamefont {{Arons}}(2003)}]{Arons2003apj}%
  \BibitemOpen
  \bibfield  {author} {\bibinfo {author} {J.~{Arons}},\ }\href
  {https://doi.org/10.1086/374776} {\bibfield  {journal} {\bibinfo  {journal}
  {\apj}\ }\textbf {\bibinfo {volume} {589}},\ \bibinfo {pages} {871} (\bibinfo
  {year} {2003})},\ \Eprint {https://arxiv.org/abs/astro-ph/0208444}
  {arXiv:astro-ph/0208444 [astro-ph]} \BibitemShut {NoStop}%
\bibitem [{\citenamefont {{Kotera}}\ \emph {et~al.}(2015)\citenamefont
  {{Kotera}}, \citenamefont {{Amato}},\ and\ \citenamefont
  {{Blasi}}}]{Kotera2015jcap}%
  \BibitemOpen
  \bibfield  {author} {\bibinfo {author} {K.~{Kotera}}, \bibinfo {author}
  {E.~{Amato}}\ and\ \bibinfo {author} {P.~{Blasi}},\ }\href
  {https://doi.org/10.1088/1475-7516/2015/08/026} {\bibfield  {journal}
  {\bibinfo  {journal} {\jcap}\ }\textbf {\bibinfo {volume} {2015}},\ \bibinfo
  {eid} {026} (\bibinfo {year} {2015})},\ \Eprint
  {https://arxiv.org/abs/1503.07907} {arXiv:1503.07907 [astro-ph.HE]}
  \BibitemShut {NoStop}%
\bibitem [{\citenamefont {{Yoon}}\ \emph {et~al.}(2017)\citenamefont {{Yoon}},
  \citenamefont {{Anderson}}, \citenamefont {{Barrau}}, \citenamefont
  {{Conklin}}, \citenamefont {{Coutu}}, \citenamefont {{Derome}}, \citenamefont
  {{Han}}, \citenamefont {{Jeon}}, \citenamefont {{Kim}}, \citenamefont
  {{Kim}}, \citenamefont {{Lee}}, \citenamefont {{Lee}}, \citenamefont {{Lee}},
  \citenamefont {{Lee}}, \citenamefont {{Link}}, \citenamefont
  {{Menchaca-Rocha}}, \citenamefont {{Mitchell}}, \citenamefont {{Mognet}},
  \citenamefont {{Nutter}}, \citenamefont {{Park}}, \citenamefont
  {{Picot-Clemente}}, \citenamefont {{Putze}}, \citenamefont {{Seo}},
  \citenamefont {{Smith}},\ and\ \citenamefont {{Wu}}}]{HHe.CREAM}%
  \BibitemOpen
  \bibfield  {author} {\bibinfo {author} {Y.~S. {Yoon}}, \bibinfo {author}
  {T.~{Anderson}}, \bibinfo {author} {A.~{Barrau}} et~al.,\ }\href
  {https://doi.org/10.3847/1538-4357/aa68e4} {\bibfield  {journal} {\bibinfo
  {journal} {\apj}\ }\textbf {\bibinfo {volume} {839}},\ \bibinfo {eid} {5}
  (\bibinfo {year} {2017})},\ \Eprint {https://arxiv.org/abs/1704.02512}
  {arXiv:1704.02512 [astro-ph.HE]} \BibitemShut {NoStop}%
\bibitem [{\citenamefont {{Aartsen}}\ \emph {et~al.}(2019)\citenamefont
  {{Aartsen}}, \citenamefont {{Ackermann}}, \citenamefont {{Adams}},
  \citenamefont {{Aguilar}}, \citenamefont {{Ahlers}}, \citenamefont
  {{Ahrens}}, \citenamefont {{Alispach}}, \citenamefont {{Andeen}},
  \citenamefont {{Anderson}}, \citenamefont {{Ansseau}},\ and\ \citenamefont
  {et~al.}}]{HHe.ICECUBE-ICETOP}%
  \BibitemOpen
  \bibfield  {author} {\bibinfo {author} {M.~G. {Aartsen}}, \bibinfo {author}
  {M.~{Ackermann}}, \bibinfo {author} {J.~{Adams}} et~al.,\ }\href
  {https://doi.org/10.1103/PhysRevD.100.082002} {\bibfield  {journal} {\bibinfo
   {journal} {\prd}\ }\textbf {\bibinfo {volume} {100}},\ \bibinfo {eid}
  {082002} (\bibinfo {year} {2019})},\ \Eprint
  {https://arxiv.org/abs/1906.04317} {arXiv:1906.04317 [astro-ph.HE]}
  \BibitemShut {NoStop}%
\bibitem [{\citenamefont {{Antoni}}\ \emph {et~al.}(2005)\citenamefont
  {{Antoni}}, \citenamefont {{Apel}}, \citenamefont {{Badea}}, \citenamefont
  {{Bekk}}, \citenamefont {{Bercuci}}, \citenamefont {{Bl{\"u}mer}},
  \citenamefont {{Bozdog}}, \citenamefont {{Brancus}}, \citenamefont
  {{Chilingarian}}, \citenamefont {{Daumiller}}, \citenamefont {{Doll}},
  \citenamefont {{Engel}}, \citenamefont {{Engler}}, \citenamefont
  {{Fe{\ss}ler}}, \citenamefont {{Gils}}, \citenamefont {{Glasstetter}},
  \citenamefont {{Haungs}}, \citenamefont {{Heck}}, \citenamefont
  {{H{\"o}randel}}, \citenamefont {{Kampert}}, \citenamefont {{Klages}},
  \citenamefont {{Maier}}, \citenamefont {{Mathes}}, \citenamefont {{Mayer}},
  \citenamefont {{Milke}}, \citenamefont {{M{\"u}ller}}, \citenamefont
  {{Obenland}}, \citenamefont {{Oehlschl{\"a}ger}}, \citenamefont
  {{Ostapchenko}}, \citenamefont {{Petcu}}, \citenamefont {{Rebel}},
  \citenamefont {{Risse}}, \citenamefont {{Risse}}, \citenamefont {{Roth}},
  \citenamefont {{Schatz}}, \citenamefont {{Schieler}}, \citenamefont
  {{Scholz}}, \citenamefont {{Thouw}}, \citenamefont {{Ulrich}}, \citenamefont
  {{van Buren}}, \citenamefont {{Vardanyan}}, \citenamefont {{Weindl}},
  \citenamefont {{Wochele}},\ and\ \citenamefont {{Zabierowski}}}]{H.KASCADE}%
  \BibitemOpen
  \bibfield  {author} {\bibinfo {author} {T.~{Antoni}}, \bibinfo {author}
  {W.~D. {Apel}}, \bibinfo {author} {A.~F. {Badea}} et~al.,\ }\href
  {https://doi.org/10.1016/j.astropartphys.2005.04.001} {\bibfield  {journal}
  {\bibinfo  {journal} {Astroparticle Physics}\ }\textbf {\bibinfo {volume}
  {24}},\ \bibinfo {pages} {1} (\bibinfo {year} {2005})},\ \Eprint
  {https://arxiv.org/abs/astro-ph/0505413} {arXiv:astro-ph/0505413 [astro-ph]}
  \BibitemShut {NoStop}%
\bibitem [{\citenamefont {{Grebenyuk}}\ \emph {et~al.}(2019)\citenamefont
  {{Grebenyuk}}, \citenamefont {{Karmanov}}, \citenamefont {{Kovalev}},
  \citenamefont {{Kudryashov}}, \citenamefont {{Kurganov}}, \citenamefont
  {{Panov}}, \citenamefont {{Podorozhny}}, \citenamefont {{Tkachenko}},
  \citenamefont {{Tkachev}}, \citenamefont {{Turundaevskiy}}, \citenamefont
  {{Vasiliev}},\ and\ \citenamefont {{Voronin}}}]{HHe.NUCLEON}%
  \BibitemOpen
  \bibfield  {author} {\bibinfo {author} {V.~{Grebenyuk}}, \bibinfo {author}
  {D.~{Karmanov}}, \bibinfo {author} {I.~{Kovalev}} et~al.,\ }\href
  {https://doi.org/10.1016/j.asr.2019.10.004} {\bibfield  {journal} {\bibinfo
  {journal} {Advances in Space Research}\ }\textbf {\bibinfo {volume} {64}},\
  \bibinfo {pages} {2546} (\bibinfo {year} {2019})}\BibitemShut {NoStop}%
\bibitem [{\citenamefont {{Mertsch}}\ and\ \citenamefont
  {{Funk}}(2015)}]{Mertsch2015prl}%
  \BibitemOpen
  \bibfield  {author} {\bibinfo {author} {P.~{Mertsch}}\ and\ \bibinfo {author}
  {S.~{Funk}},\ }\href {https://doi.org/10.1103/PhysRevLett.114.021101}
  {\bibfield  {journal} {\bibinfo  {journal} {\prl}\ }\textbf {\bibinfo
  {volume} {114}},\ \bibinfo {eid} {021101} (\bibinfo {year} {2015})},\ \Eprint
  {https://arxiv.org/abs/1408.3630} {arXiv:1408.3630 [astro-ph.HE]}
  \BibitemShut {NoStop}%
\bibitem [{\citenamefont {{Zirakashvili}}(2005)}]{Zirakashvili2005ijmpa}%
  \BibitemOpen
  \bibfield  {author} {\bibinfo {author} {V.~N. {Zirakashvili}},\ }\href
  {https://doi.org/10.1142/S0217751X05030314} {\bibfield  {journal} {\bibinfo
  {journal} {International Journal of Modern Physics A}\ }\textbf {\bibinfo
  {volume} {20}},\ \bibinfo {pages} {6858} (\bibinfo {year}
  {2005})}\BibitemShut {NoStop}%
\bibitem [{\citenamefont {{Bartoli}}\ \emph {et~al.}(2018)\citenamefont
  {{Bartoli}}, \citenamefont {{Bernardini}}, \citenamefont {{Bi}},
  \citenamefont {{Cao}}, \citenamefont {{Catalanotti}}, \citenamefont {{Chen}},
  \citenamefont {{Chen}}, \citenamefont {{Cui}}, \citenamefont {{Dai}},
  \citenamefont {{D'Amone}}, \citenamefont {{Danzengluobu}}, \citenamefont {{De
  Mitri}}, \citenamefont {{D'Ettorre Piazzoli}}, \citenamefont {{Di Girolamo}},
  \citenamefont {{Di Sciascio}}, \citenamefont {{Feng}}, \citenamefont
  {{Feng}}, \citenamefont {{Gao}}, \citenamefont {{Gou}}, \citenamefont
  {{Guo}}, \citenamefont {{He}}, \citenamefont {{Hu}}, \citenamefont {{Hu}},
  \citenamefont {{Iacovacci}}, \citenamefont {{Iuppa}}, \citenamefont {{Jia}},
  \citenamefont {{Labaciren}}, \citenamefont {{Li}}, \citenamefont {{Liu}},
  \citenamefont {{Liu}}, \citenamefont {{Liu}}, \citenamefont {{Lu}},
  \citenamefont {{Ma}}, \citenamefont {{Ma}}, \citenamefont {{Mancarella}},
  \citenamefont {{Mari}}, \citenamefont {{Marsella}}, \citenamefont
  {{Mastroianni}}, \citenamefont {{Montini}}, \citenamefont {{Ning}},
  \citenamefont {{Perrone}}, \citenamefont {{Pistilli}}, \citenamefont
  {{Ruffolo}}, \citenamefont {{Salvini}}, \citenamefont {{Santonico}},
  \citenamefont {{Shen}}, \citenamefont {{Sheng}}, \citenamefont {{Shi}},
  \citenamefont {{Surdo}}, \citenamefont {{Tan}}, \citenamefont {{Vallania}},
  \citenamefont {{Vernetto}}, \citenamefont {{Vigorito}}, \citenamefont
  {{Wang}}, \citenamefont {{Wu}}, \citenamefont {{Wu}}, \citenamefont {{Xue}},
  \citenamefont {{Yang}}, \citenamefont {{Yang}}, \citenamefont {{Yao}},
  \citenamefont {{Yuan}}, \citenamefont {{Zha}}, \citenamefont {{Zhang}},
  \citenamefont {{Zhang}}, \citenamefont {{Zhang}}, \citenamefont {{Zhang}},
  \citenamefont {{Zhao}}, \citenamefont {{Zhaxiciren}}, \citenamefont
  {{Zhaxisangzhu}}, \citenamefont {{Zhou}}, \citenamefont {{Zhu}},
  \citenamefont {{Zhu}},\ and\ \citenamefont {{ARGO-YBJ
  Collaboration}}}]{anisotropy.ARGO}%
  \BibitemOpen
  \bibfield  {author} {\bibinfo {author} {B.~{Bartoli}}, \bibinfo {author}
  {P.~{Bernardini}}, \bibinfo {author} {X.~J. {Bi}} et~al.,\ }\href
  {https://doi.org/10.3847/1538-4357/aac6cc} {\bibfield  {journal} {\bibinfo
  {journal} {\apj}\ }\textbf {\bibinfo {volume} {861}},\ \bibinfo {eid} {93}
  (\bibinfo {year} {2018})},\ \Eprint {https://arxiv.org/abs/1805.08980}
  {arXiv:1805.08980 [astro-ph.HE]} \BibitemShut {NoStop}%
\bibitem [{\citenamefont {{Aglietta}}\ \emph {et~al.}(2003)\citenamefont
  {{Aglietta}}, \citenamefont {{Alessandron}}, \citenamefont {{Antonioli}},
  \citenamefont {{Arneodo}}, \citenamefont {{Bergamasco}}, \citenamefont
  {{Bertaina}}, \citenamefont {{Castagnoli}}, \citenamefont {{Castellina}},
  \citenamefont {{Chiavassa}}, \citenamefont {{Castagnoli}}, \citenamefont
  {{D'Ettore Piazzoli}}, \citenamefont {{Di Sciascio}}, \citenamefont
  {{Fulgione}}, \citenamefont {{Galeotti}}, \citenamefont {{Ghia}},
  \citenamefont {{Iacovacci}}, \citenamefont {{Mannocchi}}, \citenamefont
  {{Morello}}, \citenamefont {{Navarra}}, \citenamefont {{Saavedra}},
  \citenamefont {{Trinchero}}, \citenamefont {{Valchierotti}}, \citenamefont
  {{Vallania}}, \citenamefont {{Vernetto}}, \citenamefont {{Vigorito}},\ and\
  \citenamefont {{EAS-TOP Collaboration}}}]{anisotropy.EASTOP}%
  \BibitemOpen
  \bibfield  {author} {\bibinfo {author} {M.~{Aglietta}}, \bibinfo {author}
  {B.~{Alessandron}}, \bibinfo {author} {P.~{Antonioli}} et~al.,\ }in\
  \href@noop {} {\emph {\bibinfo {booktitle} {International Cosmic Ray
  Conference}}},\ \bibinfo {series} {International Cosmic Ray Conference},
  Vol.~\bibinfo {volume} {1}\ (\bibinfo {year} {2003})\ p.\ \bibinfo {pages}
  {183}\BibitemShut {NoStop}%
\bibitem [{\citenamefont {{Abeysekara}}\ \emph {et~al.}(2018)\citenamefont
  {{Abeysekara}}, \citenamefont {{Alfaro}}, \citenamefont {{Alvarez}},
  \citenamefont {{{\'A}lvarez}}, \citenamefont {{Arceo}}, \citenamefont
  {{Arteaga-Vel{\'a}zquez}}, \citenamefont {{Avila Rojas}}, \citenamefont
  {{Ayala Solares}}, \citenamefont {{Becerril}}, \citenamefont
  {{Belmont-Moreno}}, \citenamefont {{BenZvi}}, \citenamefont {{Bernal}},
  \citenamefont {{Braun}}, \citenamefont {{Caballero-Mora}}, \citenamefont
  {{Capistr{\'a}n}}, \citenamefont {{Carrami{\~n}ana}}, \citenamefont
  {{Casanova}}, \citenamefont {{Castillo}}, \citenamefont {{Cotti}},
  \citenamefont {{Cotzomi}}, \citenamefont {{De Le{\'o}n}}, \citenamefont {{De
  la Fuente}}, \citenamefont {{Diaz Hernandez}}, \citenamefont {{Dichiara}},
  \citenamefont {{Dingus}}, \citenamefont {{DuVernois}}, \citenamefont
  {{D{\'\i}az-V{\'e}lez}}, \citenamefont {{Engel}}, \citenamefont {{Fiorino}},
  \citenamefont {{Fraija}}, \citenamefont {{Garc{\'\i}a-Gonz{\'a}lez}},
  \citenamefont {{Garfias}}, \citenamefont {{Gonz{\'a}lez Mu{\~n}oz}},
  \citenamefont {{Gonz{\'a}lez}}, \citenamefont {{Goodman}}, \citenamefont
  {{Hampel-Arias}}, \citenamefont {{Harding}}, \citenamefont {{Hernandez}},
  \citenamefont {{Hona}}, \citenamefont {{Hueyotl-Zahuantitla}}, \citenamefont
  {{Hui}}, \citenamefont {{H{\"u}ntemeyer}}, \citenamefont {{Iriarte}},
  \citenamefont {{Jardin-Blicq}}, \citenamefont {{Joshi}}, \citenamefont
  {{Kaufmann}}, \citenamefont {{Lara}}, \citenamefont {{Lauer}}, \citenamefont
  {{Lee}}, \citenamefont {{Le{\'o}n Vargas}}, \citenamefont {{Longinotti}},
  \citenamefont {{Luis-Raya}}, \citenamefont {{Luna-Garc{\'\i}a}},
  \citenamefont {{L{\'o}pez-C{\'a}mara}}, \citenamefont {{L{\'o}pez-Coto}},
  \citenamefont {{L{\'o}pez-C{\'a}mara}}, \citenamefont {{L{\'o}pez-Coto}},
  \citenamefont {{Malone}}, \citenamefont {{Marinelli}}, \citenamefont
  {{Martinez}}, \citenamefont {{Martinez-Castellanos}}, \citenamefont
  {{Mart{\'\i}nez-Castro}}, \citenamefont {{Mart{\'\i}nez-Huerta}},
  \citenamefont {{Matthews}}, \citenamefont {{Miranda-Romagnoli}},
  \citenamefont {{Moreno}}, \citenamefont {{Mostaf{\'a}}}, \citenamefont
  {{Nayerhoda}}, \citenamefont {{Nellen}}, \citenamefont {{Newbold}},
  \citenamefont {{Nisa}}, \citenamefont {{Noriega-Papaqui}}, \citenamefont
  {{Pelayo}}, \citenamefont {{Pretz}}, \citenamefont {{P{\'e}rez-P{\'e}rez}},
  \citenamefont {{Ren}}, \citenamefont {{Rho}}, \citenamefont {{Rivi{\`e}re}},
  \citenamefont {{Rosa-Gonz{\'a}lez}}, \citenamefont {{Rosenberg}},
  \citenamefont {{Ruiz-Velasco}}, \citenamefont {{Salesa Greus}}, \citenamefont
  {{Sandoval}}, \citenamefont {{Schneider}}, \citenamefont {{Schoorlemmer}},
  \citenamefont {{Seglar Arroyo}}, \citenamefont {{Sinnis}}, \citenamefont
  {{Smith}}, \citenamefont {{Springer}}, \citenamefont {{Surajbali}},
  \citenamefont {{Taboada}}, \citenamefont {{Tibolla}}, \citenamefont
  {{Tollefson}}, \citenamefont {{Torres}}, \citenamefont {{Vianello}},
  \citenamefont {{Villase{\~n}or}}, \citenamefont {{Weisgarber}}, \citenamefont
  {{Werner}}, \citenamefont {{Westerhoff}}, \citenamefont {{Wood}},
  \citenamefont {{Yapici}}, \citenamefont {{Zepeda}},\ and\ \citenamefont
  {{Zhou}}}]{anisotropy.HAWC}%
  \BibitemOpen
  \bibfield  {author} {\bibinfo {author} {A.~U. {Abeysekara}}, \bibinfo
  {author} {R.~{Alfaro}}, \bibinfo {author} {C.~{Alvarez}} et~al.,\ }\href
  {https://doi.org/10.3847/1538-4357/aad90c} {\bibfield  {journal} {\bibinfo
  {journal} {\apj}\ }\textbf {\bibinfo {volume} {865}},\ \bibinfo {eid} {57}
  (\bibinfo {year} {2018})},\ \Eprint {https://arxiv.org/abs/1805.01847}
  {arXiv:1805.01847 [astro-ph.HE]} \BibitemShut {NoStop}%
\bibitem [{\citenamefont {{Amenomori}}\ \emph {et~al.}(2005)\citenamefont
  {{Amenomori}}, \citenamefont {{Ayabe}}, \citenamefont {{Cui}}, \citenamefont
  {{Danzengluobu}}, \citenamefont {{Ding}}, \citenamefont {{Ding}},
  \citenamefont {{Feng}}, \citenamefont {{Feng}}, \citenamefont {{Gao}},
  \citenamefont {{Geng}}, \citenamefont {{Guo}}, \citenamefont {{He}},
  \citenamefont {{He}}, \citenamefont {{Hibino}}, \citenamefont {{Hotta}},
  \citenamefont {{Hu}}, \citenamefont {{Hu}}, \citenamefont {{Huang}},
  \citenamefont {{Huang}}, \citenamefont {{Jia}}, \citenamefont {{Kajino}},
  \citenamefont {{Kasahara}}, \citenamefont {{Katayose}}, \citenamefont
  {{Kato}}, \citenamefont {{Kawata}}, \citenamefont {{Labaciren}},
  \citenamefont {{Le}}, \citenamefont {{Li}}, \citenamefont {{Lu}},
  \citenamefont {{Lu}}, \citenamefont {{Meng}}, \citenamefont {{Mizutani}},
  \citenamefont {{Mu}}, \citenamefont {{Munakata}}, \citenamefont {{Nagai}},
  \citenamefont {{Nanjo}}, \citenamefont {{Nishizawa}}, \citenamefont
  {{Ohnishi}}, \citenamefont {{Ohta}}, \citenamefont {{Onuma}}, \citenamefont
  {{Ouchi}}, \citenamefont {{Ozawa}}, \citenamefont {{Ren}}, \citenamefont
  {{Saito}}, \citenamefont {{Sakata}}, \citenamefont {{Sasaki}}, \citenamefont
  {{Shibata}}, \citenamefont {{Shiomi}}, \citenamefont {{Shirai}},
  \citenamefont {{Sugimoto}}, \citenamefont {{Takita}}, \citenamefont {{Tan}},
  \citenamefont {{Tateyama}}, \citenamefont {{Torii}}, \citenamefont
  {{Tsuchiya}}, \citenamefont {{Udo}}, \citenamefont {{Utsugi}}, \citenamefont
  {{Wang}}, \citenamefont {{Wang}}, \citenamefont {{Wang}}, \citenamefont
  {{Wang}}, \citenamefont {{Wu}}, \citenamefont {{Xue}}, \citenamefont
  {{Yamamoto}}, \citenamefont {{Yan}}, \citenamefont {{Yang}}, \citenamefont
  {{Yasue}}, \citenamefont {{Ye}}, \citenamefont {{Yu}}, \citenamefont
  {{Yuan}}, \citenamefont {{Yuda}}, \citenamefont {{Zhang}}, \citenamefont
  {{Zhang}}, \citenamefont {{Zhang}}, \citenamefont {{Zhang}}, \citenamefont
  {{Zhang}}, \citenamefont {{Zhang}}, \citenamefont {{Zhaxisangzhu}},
  \citenamefont {{Zhou}},\ and\ \citenamefont {{Tibet As{\ensuremath{\gamma}}
  Collaboration}}}]{anisotropy.TIBET}%
  \BibitemOpen
  \bibfield  {author} {\bibinfo {author} {M.~{Amenomori}}, \bibinfo {author}
  {S.~{Ayabe}}, \bibinfo {author} {S.~W. {Cui}} et~al.,\ }\href
  {https://doi.org/10.1086/431582} {\bibfield  {journal} {\bibinfo  {journal}
  {\apjl}\ }\textbf {\bibinfo {volume} {626}},\ \bibinfo {pages} {L29}
  (\bibinfo {year} {2005})},\ \Eprint {https://arxiv.org/abs/astro-ph/0505114}
  {arXiv:astro-ph/0505114 [astro-ph]} \BibitemShut {NoStop}%
\bibitem [{\citenamefont {{Aguilar}}\ \emph
  {et~al.}(2019{\natexlab{b}})\citenamefont {{Aguilar}}, \citenamefont {{Ali
  Cavasonza}}, \citenamefont {{Alpat}}, \citenamefont {{Ambrosi}},
  \citenamefont {{Arruda}}, \citenamefont {{Attig}}, \citenamefont
  {{Azzarello}}, \citenamefont {{Bachlechner}}, \citenamefont {{Barao}},
  \citenamefont {{Barrau}},\ and\ \citenamefont {et~al.}}]{electrons.AMS02}%
  \BibitemOpen
  \bibfield  {author} {\bibinfo {author} {M.~{Aguilar}}, \bibinfo {author}
  {L.~{Ali Cavasonza}}, \bibinfo {author} {B.~{Alpat}} et~al.,\ }\href
  {https://doi.org/10.1103/PhysRevLett.122.101101} {\bibfield  {journal}
  {\bibinfo  {journal} {\prl}\ }\textbf {\bibinfo {volume} {122}},\ \bibinfo
  {eid} {101101} (\bibinfo {year} {2019}{\natexlab{b}})}\BibitemShut {NoStop}%
\bibitem [{\citenamefont {{Abdollahi}}\ \emph {et~al.}(2017)\citenamefont
  {{Abdollahi}}, \citenamefont {{Ackermann}}, \citenamefont {{Ajello}},
  \citenamefont {{Atwood}}, \citenamefont {{Baldini}}, \citenamefont
  {{Barbiellini}}, \citenamefont {{Bastieri}}, \citenamefont {{Bellazzini}},
  \citenamefont {{Bloom}}, \citenamefont {{Bonino}}, \citenamefont {{Brandt}},
  \citenamefont {{Bregeon}}, \citenamefont {{Bruel}}, \citenamefont
  {{Buehler}}, \citenamefont {{Cameron}}, \citenamefont {{Caputo}},
  \citenamefont {{Caragiulo}}, \citenamefont {{Castro}}, \citenamefont
  {{Cavazzuti}}, \citenamefont {{Cecchi}}, \citenamefont {{Chekhtman}},
  \citenamefont {{Ciprini}}, \citenamefont {{Cohen-Tanugi}}, \citenamefont
  {{Costanza}}, \citenamefont {{Cuoco}}, \citenamefont {{Cutini}},
  \citenamefont {{D'Ammando}}, \citenamefont {{de Palma}}, \citenamefont
  {{Desiante}}, \citenamefont {{Digel}}, \citenamefont {{Di Lalla}},
  \citenamefont {{Di Mauro}}, \citenamefont {{Di Venere}}, \citenamefont
  {{Drell}}, \citenamefont {{Drlica-Wagner}}, \citenamefont {{Favuzzi}},
  \citenamefont {{Focke}}, \citenamefont {{Funk}}, \citenamefont {{Fusco}},
  \citenamefont {{Gargano}}, \citenamefont {{Gasparrini}}, \citenamefont
  {{Giglietto}}, \citenamefont {{Giordano}}, \citenamefont {{Giroletti}},
  \citenamefont {{Green}}, \citenamefont {{Guillemot}}, \citenamefont
  {{Guiriec}}, \citenamefont {{Harding}}, \citenamefont {{Jogler}},
  \citenamefont {{J{\'o}hannesson}}, \citenamefont {{Kamae}}, \citenamefont
  {{Kuss}}, \citenamefont {{La Mura}}, \citenamefont {{Latronico}},
  \citenamefont {{Longo}}, \citenamefont {{Loparco}}, \citenamefont
  {{Lubrano}}, \citenamefont {{Maldera}}, \citenamefont {{Malyshev}},
  \citenamefont {{Manfreda}}, \citenamefont {{Mazziotta}}, \citenamefont
  {{Michelson}}, \citenamefont {{Mirabal}}, \citenamefont {{Mitthumsiri}},
  \citenamefont {{Mizuno}}, \citenamefont {{Moiseev}}, \citenamefont
  {{Monzani}}, \citenamefont {{Morselli}}, \citenamefont {{Moskalenko}},
  \citenamefont {{Negro}}, \citenamefont {{Nuss}}, \citenamefont {{Orlando}},
  \citenamefont {{Paneque}}, \citenamefont {{Perkins}}, \citenamefont
  {{Pesce-Rollins}}, \citenamefont {{Piron}}, \citenamefont {{Pivato}},
  \citenamefont {{Porter}}, \citenamefont {{Principe}}, \citenamefont
  {{Rain{\`o}}}, \citenamefont {{Rando}}, \citenamefont {{Razzano}},
  \citenamefont {{Reimer}}, \citenamefont {{Reimer}}, \citenamefont
  {{Sgr{\`o}}}, \citenamefont {{Simone}}, \citenamefont {{Siskind}},
  \citenamefont {{Spada}}, \citenamefont {{Spandre}}, \citenamefont
  {{Spinelli}}, \citenamefont {{Tajima}}, \citenamefont {{Thayer}},
  \citenamefont {{Tibaldo}}, \citenamefont {{Torres}}, \citenamefont {{Troja}},
  \citenamefont {{Wood}}, \citenamefont {{Worley}}, \citenamefont
  {{Zaharijas}}, \citenamefont {{Zimmer}},\ and\ \citenamefont {{Fermi-LAT
  Collaboration}}}]{leptons.FERMI}%
  \BibitemOpen
  \bibfield  {author} {\bibinfo {author} {S.~{Abdollahi}}, \bibinfo {author}
  {M.~{Ackermann}}, \bibinfo {author} {M.~{Ajello}} et~al.,\ }\href
  {https://doi.org/10.1103/PhysRevD.95.082007} {\bibfield  {journal} {\bibinfo
  {journal} {\prd}\ }\textbf {\bibinfo {volume} {95}},\ \bibinfo {eid} {082007}
  (\bibinfo {year} {2017})},\ \Eprint {https://arxiv.org/abs/1704.07195}
  {arXiv:1704.07195 [astro-ph.HE]} \BibitemShut {NoStop}%
\bibitem [{\citenamefont {{Archer}}\ \emph {et~al.}(2018)\citenamefont
  {{Archer}}, \citenamefont {{Benbow}}, \citenamefont {{Bird}}, \citenamefont
  {{Brose}}, \citenamefont {{Buchovecky}}, \citenamefont {{Buckley}},
  \citenamefont {{Bugaev}}, \citenamefont {{Connolly}}, \citenamefont {{Cui}},
  \citenamefont {{Daniel}}, \citenamefont {{Feng}}, \citenamefont {{Finley}},
  \citenamefont {{Fortson}}, \citenamefont {{Furniss}}, \citenamefont
  {{Gillanders}}, \citenamefont {{H{\"u}tten}}, \citenamefont {{Hanna}},
  \citenamefont {{Hervet}}, \citenamefont {{Holder}}, \citenamefont {{Hughes}},
  \citenamefont {{Humensky}}, \citenamefont {{Johnson}}, \citenamefont
  {{Kaaret}}, \citenamefont {{Kar}}, \citenamefont {{Kelley-Hoskins}},
  \citenamefont {{Kertzman}}, \citenamefont {{Kieda}}, \citenamefont
  {{Krause}}, \citenamefont {{Krennrich}}, \citenamefont {{Kumar}},
  \citenamefont {{Lang}}, \citenamefont {{Lin}}, \citenamefont {{Maier}},
  \citenamefont {{McArthur}}, \citenamefont {{Moriarty}}, \citenamefont
  {{Mukherjee}}, \citenamefont {{O'Brien}}, \citenamefont {{Ong}},
  \citenamefont {{Otte}}, \citenamefont {{Petrashyk}}, \citenamefont {{Pohl}},
  \citenamefont {{Pueschel}}, \citenamefont {{Quinn}}, \citenamefont {{Ragan}},
  \citenamefont {{Reynolds}}, \citenamefont {{Richards}}, \citenamefont
  {{Roache}}, \citenamefont {{Rulten}}, \citenamefont {{Sadeh}}, \citenamefont
  {{Santander}}, \citenamefont {{Sembroski}}, \citenamefont {{Staszak}},
  \citenamefont {{Sushch}}, \citenamefont {{Wakely}}, \citenamefont {{Wells}},
  \citenamefont {{Wilcox}}, \citenamefont {{Wilhelm}}, \citenamefont
  {{Williams}}, \citenamefont {{Williamson}}, \citenamefont {{Zitzer}},\ and\
  \citenamefont {{VERITAS Collaboration}}}]{leptons.VERITAS}%
  \BibitemOpen
  \bibfield  {author} {\bibinfo {author} {A.~{Archer}}, \bibinfo {author}
  {W.~{Benbow}}, \bibinfo {author} {R.~{Bird}} et~al.,\ }\href
  {https://doi.org/10.1103/PhysRevD.98.062004} {\bibfield  {journal} {\bibinfo
  {journal} {\prd}\ }\textbf {\bibinfo {volume} {98}},\ \bibinfo {eid} {062004}
  (\bibinfo {year} {2018})},\ \Eprint {https://arxiv.org/abs/1808.10028}
  {arXiv:1808.10028 [astro-ph.HE]} \BibitemShut {NoStop}%
\bibitem [{\citenamefont {{Diesing}}\ and\ \citenamefont
  {{Caprioli}}(2019)}]{Diesing2019prl}%
  \BibitemOpen
  \bibfield  {author} {\bibinfo {author} {R.~{Diesing}}\ and\ \bibinfo {author}
  {D.~{Caprioli}},\ }\href {https://doi.org/10.1103/PhysRevLett.123.071101}
  {\bibfield  {journal} {\bibinfo  {journal} {\prl}\ }\textbf {\bibinfo
  {volume} {123}},\ \bibinfo {eid} {071101} (\bibinfo {year} {2019})},\ \Eprint
  {https://arxiv.org/abs/1905.07414} {arXiv:1905.07414 [astro-ph.HE]}
  \BibitemShut {NoStop}%
\bibitem [{\citenamefont {{Cristofari}}\ \emph {et~al.}(2021)\citenamefont
  {{Cristofari}}, \citenamefont {{Blasi}},\ and\ \citenamefont
  {{Caprioli}}}]{Cristofari2021aa}%
  \BibitemOpen
  \bibfield  {author} {\bibinfo {author} {P.~{Cristofari}}, \bibinfo {author}
  {P.~{Blasi}}\ and\ \bibinfo {author} {D.~{Caprioli}},\ }\href
  {https://doi.org/10.1051/0004-6361/202140448} {\bibfield  {journal} {\bibinfo
   {journal} {\aap}\ }\textbf {\bibinfo {volume} {650}},\ \bibinfo {eid} {A62}
  (\bibinfo {year} {2021})},\ \Eprint {https://arxiv.org/abs/2103.02375}
  {arXiv:2103.02375 [astro-ph.HE]} \BibitemShut {NoStop}%
\bibitem [{\citenamefont {{Morlino}}\ and\ \citenamefont
  {{Celli}}(2021)}]{Morlino2021mnras}%
  \BibitemOpen
  \bibfield  {author} {\bibinfo {author} {G.~{Morlino}}\ and\ \bibinfo {author}
  {S.~{Celli}},\ }\href {https://doi.org/10.1093/mnras/stab2972} {\bibfield
  {journal} {\bibinfo  {journal} {\mnras}\ }\textbf {\bibinfo {volume} {508}},\
  \bibinfo {pages} {6142} (\bibinfo {year} {2021})},\ \Eprint
  {https://arxiv.org/abs/2106.06488} {arXiv:2106.06488 [astro-ph.HE]}
  \BibitemShut {NoStop}%
\bibitem [{\citenamefont {{Ackermann}}\ \emph {et~al.}(2010)\citenamefont
  {{Ackermann}}, \citenamefont {{Ajello}}, \citenamefont {{Atwood}},
  \citenamefont {{Baldini}}, \citenamefont {{Ballet}}, \citenamefont
  {{Barbiellini}}, \citenamefont {{Bastieri}}, \citenamefont {{Bechtol}},
  \citenamefont {{Bellazzini}}, \citenamefont {{Berenji}}, \citenamefont
  {{Bloom}}, \citenamefont {{Bonamente}}, \citenamefont {{Borgland}},
  \citenamefont {{Bouvier}}, \citenamefont {{Bregeon}}, \citenamefont {{Brez}},
  \citenamefont {{Brigida}}, \citenamefont {{Bruel}}, \citenamefont
  {{Buehler}}, \citenamefont {{Burnett}}, \citenamefont {{Buson}},
  \citenamefont {{Caliandro}}, \citenamefont {{Cameron}}, \citenamefont
  {{Caraveo}}, \citenamefont {{Carrigan}}, \citenamefont {{Casandjian}},
  \citenamefont {{Cecchi}}, \citenamefont {{{\c{C}}elik}}, \citenamefont
  {{Charles}}, \citenamefont {{Chekhtman}}, \citenamefont {{Cheung}},
  \citenamefont {{Chiang}}, \citenamefont {{Ciprini}}, \citenamefont {{Claus}},
  \citenamefont {{Cohen-Tanugi}}, \citenamefont {{Conrad}}, \citenamefont
  {{Cuoco}}, \citenamefont {{Dermer}}, \citenamefont {{de Angelis}},
  \citenamefont {{de Palma}}, \citenamefont {{Digel}}, \citenamefont {{di
  Bernardo}}, \citenamefont {{Do Couto E Silva}}, \citenamefont {{Drell}},
  \citenamefont {{Dubois}}, \citenamefont {{Favuzzi}}, \citenamefont {{Fegan}},
  \citenamefont {{Focke}}, \citenamefont {{Frailis}}, \citenamefont
  {{Fukazawa}}, \citenamefont {{Funk}}, \citenamefont {{Fusco}}, \citenamefont
  {{Gaggero}}, \citenamefont {{Gargano}}, \citenamefont {{Germani}},
  \citenamefont {{Giglietto}}, \citenamefont {{Giommi}}, \citenamefont
  {{Giordano}}, \citenamefont {{Giroletti}}, \citenamefont {{Glanzman}},
  \citenamefont {{Godfrey}}, \citenamefont {{Grasso}}, \citenamefont
  {{Grenier}}, \citenamefont {{Grove}}, \citenamefont {{Guiriec}},
  \citenamefont {{Gustafsson}}, \citenamefont {{Hadasch}}, \citenamefont
  {{Harding}}, \citenamefont {{Hayashi}}, \citenamefont {{Hays}}, \citenamefont
  {{Hughes}}, \citenamefont {{J{\'o}hannesson}}, \citenamefont {{Johnson}},
  \citenamefont {{Johnson}}, \citenamefont {{Kamae}}, \citenamefont
  {{Katagiri}}, \citenamefont {{Kataoka}}, \citenamefont {{Kerr}},
  \citenamefont {{Kn{\"o}dlseder}}, \citenamefont {{Kuss}}, \citenamefont
  {{Lande}}, \citenamefont {{Latronico}}, \citenamefont {{Lee}}, \citenamefont
  {{Lemoine-Goumard}}, \citenamefont {{Llena Garde}}, \citenamefont {{Longo}},
  \citenamefont {{Loparco}}, \citenamefont {{Lovellette}}, \citenamefont
  {{Lubrano}}, \citenamefont {{Makeev}}, \citenamefont {{Mazziotta}},
  \citenamefont {{McEnery}}, \citenamefont {{Mehault}}, \citenamefont
  {{Michelson}}, \citenamefont {{Mizuno}}, \citenamefont {{Moiseev}},
  \citenamefont {{Monte}}, \citenamefont {{Monzani}}, \citenamefont
  {{Moretti}}, \citenamefont {{Morselli}}, \citenamefont {{Moskalenko}},
  \citenamefont {{Murgia}}, \citenamefont {{Nakamori}}, \citenamefont
  {{Naumann-Godo}}, \citenamefont {{Nolan}}, \citenamefont {{Nuss}},
  \citenamefont {{Ohsugi}}, \citenamefont {{Okumura}}, \citenamefont
  {{Omodei}}, \citenamefont {{Orlando}}, \citenamefont {{Ormes}}, \citenamefont
  {{Paneque}}, \citenamefont {{Panetta}}, \citenamefont {{Parent}},
  \citenamefont {{Pelassa}}, \citenamefont {{Pepe}}, \citenamefont
  {{Pesce-Rollins}}, \citenamefont {{Piron}}, \citenamefont {{Porter}},
  \citenamefont {{Profumo}}, \citenamefont {{Rain{\`o}}}, \citenamefont
  {{Rando}}, \citenamefont {{Razzano}}, \citenamefont {{Reimer}}, \citenamefont
  {{Reimer}}, \citenamefont {{Reposeur}}, \citenamefont {{Ripken}},
  \citenamefont {{Ritz}}, \citenamefont {{Roth}}, \citenamefont
  {{Sadrozinski}}, \citenamefont {{Sander}}, \citenamefont {{Schalk}},
  \citenamefont {{Sgr{\`o}}}, \citenamefont {{Siegal-Gaskins}}, \citenamefont
  {{Siskind}}, \citenamefont {{Smith}}, \citenamefont {{Smith}}, \citenamefont
  {{Spandre}}, \citenamefont {{Spinelli}}, \citenamefont {{Strickman}},
  \citenamefont {{Strong}}, \citenamefont {{Suson}}, \citenamefont
  {{Takahashi}}, \citenamefont {{Takahashi}}, \citenamefont {{Tanaka}},
  \citenamefont {{Thayer}}, \citenamefont {{Thayer}}, \citenamefont
  {{Thompson}}, \citenamefont {{Tibaldo}}, \citenamefont {{Torres}},
  \citenamefont {{Tosti}}, \citenamefont {{Tramacere}}, \citenamefont
  {{Uchiyama}}, \citenamefont {{Usher}}, \citenamefont {{Vandenbroucke}},
  \citenamefont {{Vasileiou}}, \citenamefont {{Vilchez}}, \citenamefont
  {{Vitale}}, \citenamefont {{Waite}}, \citenamefont {{Wang}}, \citenamefont
  {{Winer}}, \citenamefont {{Wood}}, \citenamefont {{Yang}}, \citenamefont
  {{Ylinen}}, \citenamefont {{Zaharijas}},\ and\ \citenamefont
  {{Ziegler}}}]{Ackermann2010prd}%
  \BibitemOpen
  \bibfield  {author} {\bibinfo {author} {M.~{Ackermann}}, \bibinfo {author}
  {M.~{Ajello}}, \bibinfo {author} {W.~B. {Atwood}} et~al.,\ }\href
  {https://doi.org/10.1103/PhysRevD.82.092003} {\bibfield  {journal} {\bibinfo
  {journal} {\prd}\ }\textbf {\bibinfo {volume} {82}},\ \bibinfo {eid} {092003}
  (\bibinfo {year} {2010})},\ \Eprint {https://arxiv.org/abs/1008.5119}
  {arXiv:1008.5119 [astro-ph.HE]} \BibitemShut {NoStop}%
\bibitem [{\citenamefont {{Kobayashi}}\ \emph {et~al.}(2004)\citenamefont
  {{Kobayashi}}, \citenamefont {{Komori}}, \citenamefont {{Yoshida}},\ and\
  \citenamefont {{Nishimura}}}]{Kobayashi2004apj}%
  \BibitemOpen
  \bibfield  {author} {\bibinfo {author} {T.~{Kobayashi}}, \bibinfo {author}
  {Y.~{Komori}}, \bibinfo {author} {K.~{Yoshida}} et~al.,\ }\href
  {https://doi.org/10.1086/380431} {\bibfield  {journal} {\bibinfo  {journal}
  {\apj}\ }\textbf {\bibinfo {volume} {601}},\ \bibinfo {pages} {340} (\bibinfo
  {year} {2004})},\ \Eprint {https://arxiv.org/abs/astro-ph/0308470}
  {arXiv:astro-ph/0308470 [astro-ph]} \BibitemShut {NoStop}%
\bibitem [{\citenamefont {{Manconi}}\ \emph {et~al.}(2017)\citenamefont
  {{Manconi}}, \citenamefont {{Di Mauro}},\ and\ \citenamefont
  {{Donato}}}]{Manconi2017jcap}%
  \BibitemOpen
  \bibfield  {author} {\bibinfo {author} {S.~{Manconi}}, \bibinfo {author}
  {M.~{Di Mauro}}\ and\ \bibinfo {author} {F.~{Donato}},\ }\href
  {https://doi.org/10.1088/1475-7516/2017/01/006} {\bibfield  {journal}
  {\bibinfo  {journal} {\jcap}\ }\textbf {\bibinfo {volume} {2017}},\ \bibinfo
  {eid} {006} (\bibinfo {year} {2017})},\ \Eprint
  {https://arxiv.org/abs/1611.06237} {arXiv:1611.06237 [astro-ph.HE]}
  \BibitemShut {NoStop}%
\bibitem [{\citenamefont {{Ferrand}}\ and\ \citenamefont
  {{Safi-Harb}}(2012)}]{ATNF.cat}%
  \BibitemOpen
  \bibfield  {author} {\bibinfo {author} {G.~{Ferrand}}\ and\ \bibinfo {author}
  {S.~{Safi-Harb}},\ }\href {https://doi.org/10.1016/j.asr.2012.02.004}
  {\bibfield  {journal} {\bibinfo  {journal} {Advances in Space Research}\
  }\textbf {\bibinfo {volume} {49}},\ \bibinfo {pages} {1313} (\bibinfo {year}
  {2012})},\ \Eprint {https://arxiv.org/abs/1202.0245} {arXiv:1202.0245
  [astro-ph.HE]} \BibitemShut {NoStop}%
\bibitem [{\citenamefont {{Faherty}}\ \emph {et~al.}(2007)\citenamefont
  {{Faherty}}, \citenamefont {{Walter}},\ and\ \citenamefont
  {{Anderson}}}]{Faherty2007apss}%
  \BibitemOpen
  \bibfield  {author} {\bibinfo {author} {J.~{Faherty}}, \bibinfo {author}
  {F.~M. {Walter}}\ and\ \bibinfo {author} {J.~{Anderson}},\ }\href
  {https://doi.org/10.1007/s10509-007-9368-0} {\bibfield  {journal} {\bibinfo
  {journal} {\apss}\ }\textbf {\bibinfo {volume} {308}},\ \bibinfo {pages}
  {225} (\bibinfo {year} {2007})}\BibitemShut {NoStop}%
\bibitem [{\citenamefont {{Cristofari}}\ and\ \citenamefont
  {{Blasi}}(2019)}]{Cristofari2019mnras}%
  \BibitemOpen
  \bibfield  {author} {\bibinfo {author} {P.~{Cristofari}}\ and\ \bibinfo
  {author} {P.~{Blasi}},\ }\href {https://doi.org/10.1093/mnras/stz2126}
  {\bibfield  {journal} {\bibinfo  {journal} {\mnras}\ }\textbf {\bibinfo
  {volume} {489}},\ \bibinfo {pages} {108} (\bibinfo {year} {2019})},\ \Eprint
  {https://arxiv.org/abs/1907.13591} {arXiv:1907.13591 [astro-ph.HE]}
  \BibitemShut {NoStop}%
\end{thebibliography}%

\end{document}